\documentclass[11pt]{article}
\usepackage{jheppub}

\usepackage{amssymb}
\usepackage{amsmath}
\usepackage{amsfonts}

\usepackage{graphicx, subfigure, placeins, float}

\usepackage{array} 
\usepackage{longtable} 
\usepackage{colortab} 
\usepackage{colortbl}
\usepackage{arydshln}

\usepackage{dsfont,bbm}
\usepackage{mathrsfs}  
\usepackage{xfrac}
\usepackage{slashed}
\usepackage{mathabx}
\usepackage{enumitem}
\usepackage{shuffle}
\usepackage{stmaryrd}

\makeatletter
\newcommand\xleftrightarrow[2][]{%
  \ext@arrow 9999{\longleftrightarrowfill@}{#1}{#2}}
\newcommand\longleftrightarrowfill@{%
  \arrowfill@\leftarrow\relbar\rightarrow}
\makeatother

\newcommand{\Rome}[1]{\uppercase\expandafter{\romannumeral#1}}

\newcommand{\itbf}[1]{\textbf{\textit{#1}}}

\newcolumntype{C}[1]{>{\centering\arraybackslash}m{#1}}

\makeatletter
\newcommand{\mathleft}{\@fleqntrue\@mathmargin0pt}
\newcommand{\mathcenter}{\@fleqnfalse}
\makeatother

\newcommand{\sign}[1]{\text{sgn}(#1)}
\newcommand{\pl}{\text{PL}}
\newcommand{\np}{\text{NP}}
\newcommand{\SU}[1]{\text{SU}(#1)}

\usepackage{extarrows}

\allowdisplaybreaks[4]

\title{Full-color three-loop three-point form factors in ${\cal N}=4$ SYM}

\author[a,b]{Guanda Lin,}
\emailAdd{linguandak@pku.edu.cn}
\author[a,c,d,e,f]{Gang Yang,}
\emailAdd{yangg@itp.ac.cn}
\author[a,c]{and Siyuan Zhang}
\emailAdd{zhangsiyuan@itp.ac.cn}
\affiliation[a]{CAS Key Laboratory of Theoretical Physics, Institute of Theoretical Physics, \\Chinese Academy of Sciences, Beijing 100190, China}
\affiliation[b]{School of Physics, Peking University, Beijing 100871, China}
\affiliation[c]{School of Physical Sciences, University of Chinese Academy of Sciences, Beijing 100049, China}
\affiliation[d]{School of Fundamental Physics and Mathematical Sciences, Hangzhou Institute for Advanced Study, UCAS, Hangzhou 310024, China}
\affiliation[e]{International Centre for Theoretical Physics Asia-Pacific, Beijing/Hangzhou, China}
\affiliation[f]{Peng Huanwu Center for Fundamental Theory, Xian 710127, China}

\abstract{
We present the detailed computation of full-color three-loop three-point form factors of both the stress-tensor supermultiplet and a length-three BPS operator in ${\cal N}=4$ SYM. 
The integrands are constructed based on the color-kinematics (CK) duality and generalized unitarity method. 
An interesting observation is that the CK-dual integrands contain a large number of free parameters.
We discuss the origin of these free parameters in detail and check that they cancel in the simplified integrands.
We further perform the numerical evaluation of the integrals at a special kinematics point using public packages FIESTA and pySecDec based on the sector-decomposition approach.
We find that the numerical computation can be significantly simplified by expressing the integrals in terms of uniformly transcendental basis,
although the final three-loop computations still require large computational resources.
Having the full-color numerical results, we verify that the non-planar infrared divergences reproduce the non-dipole structures, which firstly appear at three loops. 
As for the finite remainder functions, we check that the numerical planar remainder for the stress-tensor supermultiplet is consistent with the known result of the bootstrap computation. 
We also obtain for the first time the numerical results of the three-loop non-planar remainder for the stress-tensor supermultiplet as well as the three-loop remainder for the length-three operator.
}

\begin{document}

\maketitle

\setcounter{footnote}{0}

\section{Introduction}

Significant progress has been made in the last thirty years in our understanding of scattering amplitudes, in which the modern on-shell methods have played important roles, see \emph{e.g.}~\cite{Elvang:2013cua, Henn:2014yza} for an introduction. 
Form factors, as quantities that encode the information of both on-shell asymptotic states and off-shell local operators, have also attracted increasing attentions in recent years.
Specifically, an $n$-point form factor can be defined as the matrix element between $n$ on-shell asymptotic states and a local gauge invariant operator $\mathcal{O}$ as
\begin{equation}
   \mathcal{F}_{\mathcal{O}}(1, \ldots, n ; q)\equiv\int d^{D} x e^{-i q \cdot x}\langle 1 \ldots n|\mathcal{O}(x)| 0\rangle=(2 \pi)^{D} \delta^{(D)}\Big(q-\sum_{i=1}^{n} p_{i}\Big)\langle 1 \ldots n|\mathcal{O}(0)| 0\rangle\,.
\end{equation}

Since form factors involve both on-shell states and local operators, 
they may be taken as a bridge between on-shell amplitudes and off-shell correlation functions,
in particular, it is possible to apply powerful modern techniques originally designed for amplitudes to study them.
Indeed, much advance has been made in the studies of form factors in ${\cal N}=4$ SYM, including the strong-coupling picture via AdS/CFT correspondence \cite{Alday:2007he, Maldacena:2010kp, Gao:2013dza}, MHV structure and supersymmetric formalisms \cite{Brandhuber:2010ad, Bork:2010wf, Brandhuber:2011tv, Bork:2011cj}, Grassmannian and polytope pictures \cite{Frassek:2015rka, Bork:2016hst, Bork:2016xfn, Bork:2017qyh, Bork:2014eqa}, twistor formalism \cite{Koster:2016ebi, Koster:2016loo, Chicherin:2016qsf, Koster:2016fna}, and the connected description of form factors \cite{He:2016dol, Brandhuber:2016xue, He:2016jdg}.
Generalized unitarity method has been applied to high-loop form factors of BPS operators \cite{Gehrmann:2011xn, Brandhuber:2012vm, Brandhuber:2014ica} and non-protected operators \cite{Wilhelm:2014qua, Nandan:2014oga, Loebbert:2015ova, Brandhuber:2016fni, Loebbert:2016xkw, Caron-Huot:2016cwu}.
Bootstrap methods have also proved to be successful for form factors: the symbol bootstrap method has been applied to form factors at two loops \cite{Brandhuber:2012vm} and recently at much higher loops in \cite{Dixon:2020bbt} with the help of the form factor operator product expansion (OPE) \cite{Sever:2020jjx, Sever:2021nsq}; besides, a new bootstrap strategy based on the master-integral ansatz has been developed to compute a two-loop four-point form factor in \cite{Guo:2021bym}.
See also some other studies in \cite{Henn:2011by, Bork:2012tt, Engelund:2012re, Johansson:2012zv, Huang:2016bmv, Ahmed:2016vgl, Bolshov:2018eos, Bianchi:2018peu,Bianchi:2018rrj}.
A recent introduction and review of form factors in ${\cal N}=4$ SYM can be found in \cite{Yang:2019vag}. 

In this paper, we study a special class of three-point form factors in ${\cal N}=4$ SYM at three loops with complete color dependence. We focus on the form factors of the BPS operators, $\operatorname{tr}(\phi^2)$ and $\operatorname{tr}(\phi^3)$.
Although we consider form factors in ${\cal N}=4$ SYM, they are closely related to Higgs amplitudes in QCD.
In particular, scattering of Higgs and gluons can be understood as form factors of operators appearing in the Higgs effective theory, where Higgs-gluons interaction vertices are obtained by integrating out the heavy top quark loop \cite{Ellis:1975ap, Georgi:1977gs, Wilczek:1977zn, Shifman:1979eb}.
Many evidences from form factor results at two loops \cite{Gehrmann:2011aa, Brandhuber:2012vm, Brandhuber:2014ica, Loebbert:2015ova, Brandhuber:2016fni, Loebbert:2016xkw, Banerjee:2016kri, Brandhuber:2017bkg, Banerjee:2017faz, Jin:2018fak, Brandhuber:2018xzk, Brandhuber:2018kqb, Jin:2019ile, Jin:2019opr, Jin:2020pwh} support the intriguing maximal transcendentality principle (MTP) \cite{Kotikov:2002ab,Kotikov:2004er} which conjectures that $\mathcal{N}=4$ SYM captures the maximally transcendental part of QCD. 
The new results obtained in this paper are also expected to provide the maximally transcendental pieces of three-loop QCD results.

Our strategy for calculating these three-loop form factors, in particular to get the full-color integrands, is to apply the color-kinematics (CK) duality \cite{Bern:2008qj, Bern:2010ue}.
The CK duality implies a direct connection between the color and kinematics of gauge amplitudes or form factors, and thus provides a powerful practical tool to construct loop integrands.
Together with the generalized unitarity method \cite{Bern:1994zx, Bern:1994cg, Britto:2004nc},
many non-trivial high-loop constructions have been achieved with the help of CK duality, see \emph{e.g.~}high-loop amplitudes in SYM \cite{Carrasco:2011mn, Bern:2012uf, Bern:2013qca, Bern:2014sna, Johansson:2017bfl, Bern:2017ucb, Kalin:2018thp} and pure YM \cite{Boels:2013bi, Bern:2013yya, Bern:2015ooa, Mogull:2015adi}, and also form factors in ${\cal N}=4$ SYM \cite{Boels:2012ew,Yang:2016ear, Lin:2020dyj}.
Apart from its significance in gauge theories,  the duality also make it possible to compute gravity amplitudes via the double-copy of gauge amplitudes, which has played an important role in studying the (super)gravity theories and their ultraviolet divergences, see \emph{e.g.}~\cite{Bern:2017ucb,Bern:2017yxu,Bern:2012gh,Bern:2012cd,Bern:2012uf,Bern:2010ue}. 
See \cite{Bern:2019prr} for an extensive review of the duality and its applications.

Although the CK duality has been understood at tree level \cite{BjerrumBohr:2009rd, Stieberger:2009hq, Feng:2010my}, it is still a conjecture at loop level.
Thus, it is important to explore more explicit constructions and see to what extent the duality applies.
The CK-dual integrands constructed in this paper offer new non-trivial three-loop examples which suggest that the CK-duality structure should exist in  more general situations.
Interestingly, the new form factor integrands show a novel feature compared with the amplitude cases: the CK-dual solutions contain a large number of free parameters which originate from a new type of generalized gauge transformations induced by the insertion of operator in form factors. 
We would also like to stress the results take the ``simplest" form, in the sense that they have the minimal power counting of loop momenta expected in $\mathcal{N}=4$ SYM and manifest all diagrammatic symmetries.

The form factors after integration are also expected to possess interesting structures. 
Starting from three loops, infrared (IR) divergences receive contributions beyond the dipole form  \cite{Almelid:2015jia}, which should be observed in our three-point form factors as well. 
To study these structures, we also compute the three-loop integrals using the numerical approach based on sector decomposition strategy \cite{Binoth:2000ps} with pubic programs FIESTA  \cite{Smirnov:2015mct} and pySecDec \cite{Borowka:2017idc}. To tackle the complicated three-loop non-planar integrals, an improved representation based on the $d\log$-form integrals is also employed \cite{Arkani-Hamed:2014via, Bern:2014kca}. Such $d\log$ integrals have simple pole structures and empirical observations show that it is much more efficient to evaluate them numerically than ordinary Feynman integrals.  
A similar observation was also made for the four-loop Sudakov form factor in \cite{Boels:2017ftb}.
We obtain the numerical results of full-color form factors for both $\operatorname{tr}(\phi^2)$ and $\operatorname{tr}(\phi^3)$ operators up to the $O(\epsilon^0)$ order at a special phase space point. 
Our three-loop results consistently reproduce the IR structure with non-dipole corrections \cite{Almelid:2015jia} and also verify the planar remainder of the operator $\operatorname{tr}(\phi^2)$ obtained from the bootstrap method \cite{Dixon:2020bbt}. 

Finally, we mention that the integrands of the above three-loop form factors have been reported in the short letter \cite{Lin:2021kht}, while in this paper we present more details, including the non-planar unitarity cuts and the parameter cancellation in the solution space, as well as new results on the integral simplification and evaluation.
The organization of this paper is as follows. In Section~\ref{sec:strategy}, we provide a detailed description of our strategy for constructing integrands in the CK-dual representation. The three-loop integrands satisfying CK duality are discussed in Section~\ref{sec:ckintegrand}, where the two-loop three-point form factor for $\operatorname{tr}(\phi^2)$ is also revisited as a warm-up example. Section~\ref{sec:ckintegrand} also includes  an analysis on the aforementioned parameters in the integrands, reflecting new features for form factors.  Then we describe the method and results for simplifying and reorganizing the integrands in Section~\ref{sec:simintegrand}, so that the integrands are suitable for subsequent numerical evaluations. The numerical integration, IR subtraction and remainders are discussed in Section~\ref{sec:ir}. 
Some details and data appearing in all these discussions are presented in Appendix \ref{ap:ck2loop}--\ref{ap:ir}.

\section{Strategy for loop-integrand construction}\label{sec:strategy}

In this section, we describe the general strategy for constructing loop integrand by color-kinematics duality and unitarity cuts. 
First, we give a brief review of color-kinematics duality and describe how to use it to construct an ansatz for the loop integrand of form factors.
Then we explain how to apply physical constraints to fix the ansatz, including  diagrammatic symmetries and unitarity cuts, where a particular emphasis will be on the application of non-planar cuts.
Readers are also referred to \cite{Bern:2012uf, Carrasco:2015iwa, Yang:2019vag} for some further details of the strategy.

\subsection{CK duality and the construction of ansatz}
\label{ssec:ckansatz}

Color-kinematics duality refers to the statement that  it is possible to rearrange the perturbative amplitudes in gauge theories, such that the Jacobi relations satisfied by color factors of cubic graphs also apply to the kinematics numerators of the same graphs \cite{Bern:2008qj, Bern:2010ue}.

The simplest example to illustrate this duality should be the four-gluon tree amplitude. 
The CK-dual representation of this amplitude takes the following form:
\begin{equation}
\label{eq:4g-stu}
    \itbf{A}^{(0)}_{4}=\frac{C_s N_s}{s}+\frac{C_t N_t}{t}+\frac{C_u N_u}{u}\,,
\end{equation}
where $C_{s,t,u}$ and $N_{s,t,u}$ are color factors and numerators of the $s$-, $t$- and $u$-channel cubic graphs respectively:
\begin{equation}\label{eq:A4tree}
\begin{aligned}
        \includegraphics[height=0.11\linewidth]{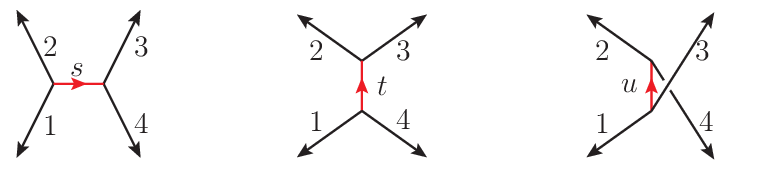}
\end{aligned} .
\end{equation}
The color factors are defined by attaching a structure constant $\tilde{f}^{abc}$ to  each trivalent vertex, namely
\begin{equation}
    C_{s}=\tilde{f}^{a_1 a_2 \rm{x} }\tilde{f}^{a_3 a_4 \rm{x} },\qquad C_{t}=\tilde{f}^{a_4 a_1 \rm{x} }\tilde{f}^{a_2 a_3 \rm{x} }, \qquad C_{u}=\tilde{f}^{a_3 a_1 \rm{x} }\tilde{f}^{a_4 a_2 \rm{x} }\,,
\end{equation}
where 
\begin{equation}
        \tilde{f}^{abc}=\operatorname{tr}(T^aT^bT^c)-\operatorname{tr}(T^aT^cT^b)\,,
\end{equation}
and the color generators $T^a $ are normalized as $\operatorname{tr}(T^{a}T^{b})=\delta^{ab}$.
For concreteness, we will often consider  the  $\SU{N_c}$ gauge group in this paper.

As for the numerators, the requirement of color-kinematics duality means that the kinematics numerators satisfy the same algebraic relation as the color factors:
\begin{equation}\label{eq:basicckdual}
    C_{s}=C_{t}+C_{u}\quad \Rightarrow \quad  N_{s}=N_{t}+N_{u}\,.
\end{equation}
It is easy to obtain such a numerator solution for the four-gluon amplitude:
\begin{align}\label{eq:4ptcksolution}
\nonumber
& N_{s}=\left(\mathcal{E}_{12} {\bf p}_{12}^{\mu}+2\mathcal{P}_{21} \varepsilon_{2}^{\mu}-2 \mathcal{P}_{12} \varepsilon_{1}^{\mu}\right) \left(\mathcal{E}_{34} {\bf p}_{34,\mu}+2\mathcal{P}_{43} \varepsilon_{4,\mu}-2\mathcal{P}_{34} \varepsilon_{3,\mu}\right)+  s_{12}\left(\mathcal{E}_{13}\mathcal{E}_{24}-\mathcal{E}_{14}\mathcal{E}_{23}\right) ,\\
& N_{t}=\left.N_{s}\right|_{1 \leftrightarrow 3}, \qquad N_{u}=\left.N_{s}\right|_{2 \leftrightarrow 3} ,
\end{align}
where $\mathcal{E}_{ij}=\varepsilon_i\cdot \varepsilon_j$, $\mathcal{P}_{ij}=p_i\cdot \varepsilon_j$,
and ${\bf p}_{ij}=p_i-p_j$, 
with $p_i$ and $\varepsilon_i$ the momentum and polarization vector of the $i$-th particle. 
We also remark that the on-shell conditions for the four external legs are needed to show that \eqref{eq:basicckdual} is indeed a CK-dual solution. 

The highly non-trivial observation is that the color-kinematics duality can be generalized at loop level.
For the form factors considered in this paper, the loop integrand in a representation satisfying CK duality can be expressed in the following form:
\begin{equation}\label{eq:generalckform}
    \itbf{F}^{(\ell)}_{\mathcal{O},n}=  \sum_{\sigma} \sum_{\Gamma_i} \int \prod_{j=1}^{\ell}\frac{ \mathrm{d}^{D}l_{j}}{(2\pi i)^{D}} \frac{1}{S_i} \sigma\cdot\frac{C_i {\tilde N}_i }{\prod_{\alpha_i}P^2_{\alpha_i}}\,,
\end{equation}
where 
\begin{enumerate}
    \item $\Gamma_i$ represent topological distinct trivalent graphs, with $i$ labelling of the graphs;
    \item $S_i$ are symmetry factors determined by the graph symmetry of the corresponding $\Gamma_i$;
    \item $P^2_{\alpha_i}$ are Feynman propagators corresponding to $\Gamma_i$;
    \item $C_i$ are color factors of diagram $\Gamma_i$ composed of the following two parts: each trivalent vertex contributes a structure constant $\tilde{f}^{abc}$; as for the operator vertex, the associated color factor is determined by the  operator $\mathcal{O}$;
    \item The kinematics numerators are ${\tilde N}_i$ and can be factorized as ${\cal F}_{\mathcal{O},n}^{(0)} N_i$,  where the tree-level form factor ${\cal F}_{\mathcal{O},n}^{(0)} $ carries all helicity weights, and $N_i$ are functions of loop and external momenta, which will be the central object of the discussions below;
    \item The sum over $\sigma$ refers to summing all permutations of external legs, as well as possible inequivalent permutations of legs directly connected to the operator; and $\sigma$ acts on ${\tilde N}_i, P^2_{\alpha_i} $ and $C_i $. 
\end{enumerate}

\begin{figure}[t!]
    \centering
    \subfigure[s-channel]{
        \includegraphics[width=0.25\linewidth]{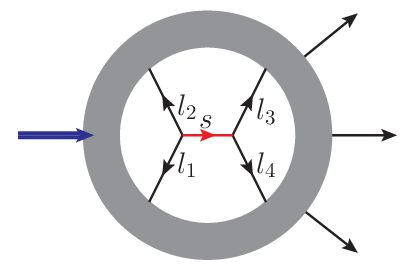}
         \label{fig:cks}
        }
    \centering
    \subfigure[t-channel]{
        \includegraphics[width=0.25\linewidth]{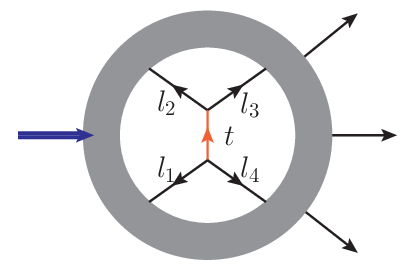}
        \label{fig:ckt}
        }
    \centering
    \subfigure[u-channel]{
        \includegraphics[width=0.25\linewidth]{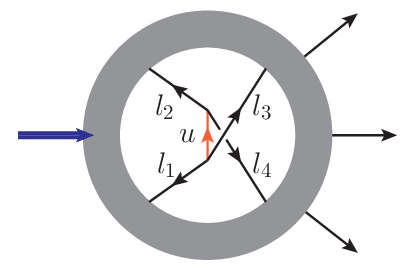}
        \label{fig:cku}
        }
    \caption{$s,t,u$-channel graphs related by Jacobi relations. }
    \label{fig:ckstu}
\end{figure}

The crucial point of the representation is that the color factors $C$ and kinematics factors $\tilde{N}$ are required to satisfy the same set of Jacobi relations. As a concrete example, let us consider Figure~\ref{fig:cks} which represents a cubic graph with a four-point tree-level sub-graph shown explicitly with external legs $\{l_1,l_2,l_3,l_4\}$. By changing the four-point tree sub-topology (keeping other edges of the graph unchanged), one can get the other two $t$- and $u$-channel graphs as shown in Figure~\ref{fig:ckt} and Figure~\ref{fig:cku}. 
The Jacobi relation $C_s=C_t+C_u$ satisfied by the color factors of these three graphs indicates the following dual Jacobi relation:
\begin{align} 
\label{eq:dualjacobi}
{\tilde N}_s(\{l_1,l_2,l_s\},\{-l_s,l_3,l_4\}) = {\tilde N}_t(\{l_4,l_1,l_t\},\{-l_t,l_2,l_3\})  + {\tilde N}_u(\{l_3,l_1,l_u\},\{-l_u,l_4,l_2\}) \,, 
\end{align}
where $l_i$ label the momenta, and each $\{l_a,l_b,l_c\}$ specifies a trivalent vertex in the graphs. 
Similar relations should also be valid when applying above transformation between $s$-, $t$- and $u$-channel graphs to other propagators and other topologies. Note that unlike the above four-gluon tree amplitude, the momenta $\{l_1, l_2, l_3, l_4\}$ are generally off-shell for loop diagrams, and it is non-trivial that the duality relations such as \eqref{eq:dualjacobi} still hold at loop level. 

The practical construction of an ansatz as in \eqref{eq:generalckform} can be given in the following steps:
\subsubsection*{1) Generating trivalent topologies}
We first generate topologically distinct cubic graphs. 
A useful selection criterion can be used here: the good UV behavior of BPS form factors in $\mathcal{N}=4$ SYM can help to constrain the type of graphs appearing in the ansatz, similar to the amplitude cases \cite{Bern:1994cg,Bern:2010ue,Bern:2012uf}. 
In particular, we exclude any graphs with one-loop sub-tadpole, sub-bubble and sub-triangle diagrams, unless the sub-triangle is directly connected to $q^2$ vertex.%
\footnote{For the graphs with sub-bubble connected to the off-shell vertex, they are generally related to integrals with UV divergences. In some cases, these integrals can be zero with special numerators so they bypass the constraint based on UV behavior, see \cite{Lin:2020dyj}. For the results in this paper,  diagrams containing sub-bubble are not necessary, \emph{i.e.}~there exists a solution with numerators corresponding to these diagrams zero. } 

\subsubsection*{2) Generating dual Jacobi relations and master topologies}
Given the topologies, it is straightforward to generate all dual Jacobi relations as  \eqref{eq:dualjacobi}. 
Note that to generate dual Jacobi relations, we consider only the propagators that are not directly connected to the operator vertex. We mention that considering the excluded propagators leads to some interesting relations with algebraic structures different from Jacobi relations, and this point will be discussed in Section~\ref{ssec:parameters}. 

The set of dual Jacobi relations relate the numerators between different topologies, and 
a crucial consequence is that all numerators can be deduced from a small number of \emph{master numerators}. 
The corresponding topologies will be called master topologies. We point out that the choice of master topologies is not unique, and we will  choose masters as planar topologies for the convenience of ansatz construction.\footnote{Choosing non-planar topologies may give a smaller set of masters, see examples of two-loop three-point $\operatorname{tr}(\phi^2) $ form factor and three-loop three-point $\operatorname{tr}(\phi^3) $ form factor in Section~\ref{sec:ckintegrand}.}

\subsubsection*{3) Constructing numerator Ansatz}

To construct an ansatz for the full integrand, it is enough to focus on the master numerators.
A few useful conditions can be imposed on the ansatz:
\begin{itemize}
    \item The numerators $N_i$ are local, namely, the numerators have no poles and are polynomials of Lorentz products. 
In practice, since we choose all the masters to be planar, one can use zone variables, denoted as $x_{ij}^2$ in the dual momentum space \cite{Drummond:2006rz}, to express Lorentz products.
    \item For $\mathcal{N}=4$ SYM, one can restrict the power of loop momenta in the ansatz: for any $n$-point sub-graph, if it is an amplitude, the numerator should contain no more than $(n-4)$ powers of the loop momentum of this loop; if it is a form factor, the upper bound becomes $(n-3)$. 
    These constraints stem from the good UV behavior of amplitudes or BPS form factors in ${\cal N}=4$ SYM \cite{Bern:2012uf, Boels:2012ew}.
\end{itemize}
These conditions allow us to write down an ansatz as a polynomial with correct mass dimensions for each master numerator
\begin{equation}
N^{\rm ansatz} = \sum_a c_a m_a(x_{ij}^2) \,,
\end{equation}
where $m_a(x_{ij}^2) $ are monomials of zone variables and $c_a $ are the free parameters in the ansatz.

Finally, using dual Jacobi relations, all other numerators can be obtained as certain linear combinations of master numerators. In this way, one construct the form factor ansatz in the form of \eqref{eq:generalckform} which are linear functions of parameters $c_a $.
The next task is to solve for these parameters by using various constraints, and two major types are the graph symmetries and unitarity cuts, which will be discussed in the following two subsections.

\subsection{Graph symmetry constraints}\label{ssec:cksym}
We first require the numerator of any topology shares the same symmetry of the diagram. 
Schematically, the condition from diagrammatic symmetry reads
\begin{equation}\label{eq:symmety}
C {\tilde N} = \mathbf{s} [ C {\tilde N} ]\,,
\end{equation}
where $\mathbf{s}$ is a symmetry operator (due to certain diagrammatic isomorphism) acting on both the color factor and the kinematics numerator. 
For the three-point form factors, under the symmetry actions, the color factor $C$ and the tree-level form factor $\mathcal{F}_3^{(0)} \propto (\langle 12\rangle\langle23\rangle\langle31\rangle)^{-1}$ in $\tilde{N}$ (recall that $\tilde{N}=\mathcal{F}^{(0)}N$) can only give a possible minus sign. Thus the main goal here is to find the relation between $\mathbf{s}[N(l,p)]$ and $N(l,p)$.
Below we explain the operation by explicit examples. 

\begin{figure}[t]
    \centering
    \subfigure[]{
        \includegraphics[width=0.32\linewidth]{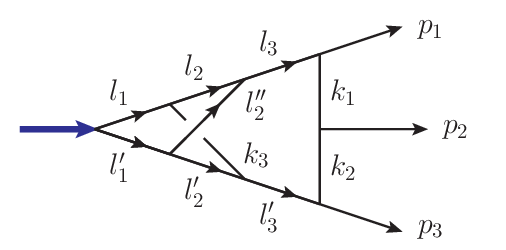}
         \label{fig:symConstraintExample}
        }
    \centering
    \subfigure[]{
        \includegraphics[width=0.32\linewidth]{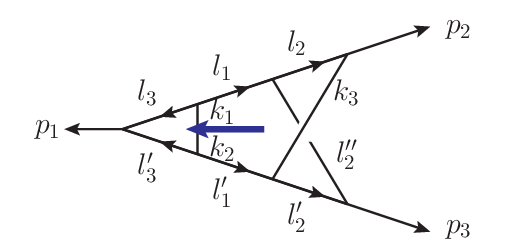}
        \label{fig:symConstraintExample1}
        }
    \caption{Example diagrams for diagrammatic symmetry constraints. The color factor $C_a$ of (a) is proportional to $\tilde f^{a_1 a_2 a_3}$ and non-vanishing while the color factor $C_b$ of (b) is zero.}
\end{figure}

First we consider the topology shown in Figure~\ref{fig:symConstraintExample} from the two-loop $\operatorname{tr}(\phi^2)$ form factor.
This topology has a reflection symmetry 
\begin{equation}
\mathbf{s}_1: \quad \{p_1\leftrightarrow p_3, \ \  k_1\leftrightarrow k_2, \ \ l_2^{\prime\prime}\leftrightarrow k_3, \ \ l_m\leftrightarrow l_m^\prime \textrm{~for~} m=1,2,3 \} \,.
\end{equation}
From \eqref{eq:symmety}, the diagrammatic symmetry requires
\begin{equation}\label{eq:symmetryeg1}
\begin{aligned}
   C_a {\cal F}^{(0)}(p_1,p_2,p_3) &N_a(l_1,l_2,l_3,p_1,p_2,p_3)
   =\mathbf{s}_1\big[C_{a} \, {\cal F}^{(0)}(p_1,p_2,p_3) \, N_a(l_1,l_2,l_3,p_1,p_2,p_3)\big]\,.
\end{aligned}
\end{equation}
The symmetry action of $\mathbf{s}_1$ on color factors is
\begin{equation}\label{eq:symeg1ca}
 \begin{aligned}
       C_a  &= \delta^{l_1 l_1^\prime} \tilde{f}^{l_1 l_2 k_3} \tilde{f}^{l_1^\prime l_2^\prime l_2^{\prime\prime}} \tilde{f}^{l_2 l_3 l_2^{\prime\prime}} \tilde{f}^{l_2^\prime l_3^\prime k_3} \tilde{f}^{l_3 a_1 k_1} \tilde{f}^{l_3^\prime a_3 k_2} \tilde{f}^{k_1 a_2 k_2}\,, \\ 
      \Rightarrow \quad \mathbf{s}_1\left[C_{a}\right]  &= \delta^{l_1^\prime l_1} \tilde{f}^{l_1^\prime l_2^\prime l_2^{\prime\prime}} \tilde{f}^{l_1 l_2 k_3} \tilde{f}^{l_2^\prime l_3^\prime k_3} \tilde{f}^{l_2 l_3 l_2^{\prime\prime}} \tilde{f}^{l_3^\prime a_3 k_2} \tilde{f}^{l_3 a_1 k_1} \tilde{f}^{k_2 a_2 k_1}=-C_a\,,
\end{aligned}
\end{equation} 
where the last identity comes from the fact the only difference between $\mathbf{s}_1\left[C_{a}\right] $  and $C_{a} $ is the ordering of indicies in the last $\tilde{f}$. For the kinematics numerators, one has
\begin{align}
&\mathbf{s}_1\big[{\cal F}^{(0)}(p_1,p_2,p_3)\big]={\cal F}^{(0)}(p_3,p_2,p_1)=-{\cal F}^{(0)}(p_1,p_2,p_3)\,,  \\
&\mathbf{s}_1\left[N_a(l_1,l_2,l_3,p_1,p_2,p_3)\right]=N_a(l_1^\prime,l_2^\prime,l_3^\prime,p_3,p_2,p_1) \,.
\end{align}
As a result,  we get the constraint on $N_{a}$ as
\begin{equation}\label{eq:symmetryeg1n}
    N_a(l_1,l_2,l_3,p_1,p_2,p_3)=N_a(l_1^\prime,l_2^\prime,l_3^\prime,p_3,p_2,p_1).
\end{equation}

In this example, it is actually not necessary to write down the lengthy product of $\tilde{f}$ as  in \eqref{eq:symeg1ca}, and one can directly consider the simpler form of color factor with all internal indices contracted, expressed as\footnote{For simplicity, we use $\tilde{f}^{a_1 a_2 a_3}$ as a short notation for $\tilde{f}^{a_1a_2a_3}$}
\begin{equation}\label{eq:symeg1ca2}
    C_a=-N_c^2\tilde{f}^{a_1 a_2 a_3}, \quad \quad \mathbf{s}_1\left[C_{a}\right]=-N_c^2\tilde{f}^{a_3 a_2 a_1}=-C_a\,.
\end{equation}
Practically, such an analysis as \eqref{eq:symeg1ca2} is easier because the (simplified) color factor $C$ equals to $P\left(N_c\right)\tilde{f}^{a_1 a_2 a_3}$ ($P(N_c)$ is a polynomial of $N_c$) and the combination $\tilde{f}^{a_1 a_2 a_3}\mathcal{F}^{(0)}(p_1,p_2,p_3)$ is always invariant under any symmetry operation. As a result,  $N_a=\mathbf{s}\left[N_a\right]$ for any $\mathbf{s}$, bypassing detailed analyses on $C$ and $\mathcal{F}^{(0)}$. 

However, there are exceptional situations if $P\left(N_c\right)=0$, such as  the topology shown in Figure~\ref{fig:symConstraintExample1}. 
Consider the graph symmetry of this topology:
\begin{equation}
\mathbf{s}_2: \quad \{l_2\leftrightarrow l_2^{\prime\prime}, \ \  k_3\leftrightarrow l_2^{\prime}, \ \  p_2\leftrightarrow p_3 \} \,.
\end{equation}
Since $C_{b}=0$, both $\mathbf{s}_2\big[C_{b}]=\pm C_{b}$ are mathematically allowed, and an analysis as \eqref{eq:symeg1ca2} is no longer possible. 
One can consider the trivalent form of the color factor, after following a similar path from \eqref{eq:symmetryeg1} to \eqref{eq:symmetryeg1n},one gets
\begin{equation}\label{eq:symeg2cb}
 \begin{aligned}
       \mathbf{s_2}\left[ C_{b}\right] =\mathbf{s_2}\big[\tilde{f}^{l_3^\prime a_1 l_3} \delta^{k_1 k_2} \tilde{f}^{l_3 l_1 k_1} \tilde{f}^{l_3^\prime l_1^\prime k_2} \tilde{f}^{l_1 l_2 l_2^{\prime\prime}} \tilde{f}^{l_1^\prime l_2^\prime k_3} \tilde{f}^{l_2 a_2 k_3} \tilde{f}^{l_2^\prime a_3 l_2^{\prime\prime}}\big]=C_b\,.
    \end{aligned}
\end{equation}
Since $\mathbf{s}_2\big[{\cal F}^{(0)}(p_1,p_2,p_3)\big]=-{\cal F}^{(0)}(p_1,p_2,p_3)$, we obtain
\begin{equation} \label{eq:symmetryeg2n}
     N_b(l_1,l_2,l_3,p_1,p_2,p_3)=-N_b(l_1,l_2^{\prime\prime},l_3,p_1,p_3,p_2)\,.
\end{equation} 
One should be particularly careful about the minus sign on the RHS.

As a comparison, Figure~\ref{fig:symConstraintExample1} has  another symmetry $\mathbf{s}_3$:
\begin{equation}
 \mathbf{s}_3:\qquad \{p_2 \leftrightarrow p_3, \, k_1\leftrightarrow k_2, \, l_2^{\prime\prime}\leftrightarrow k_3, \, l_m\leftrightarrow l_m^\prime \textrm{~for~} m=1,2,3 \}\,.
\end{equation}
In this case, one has 
\begin{equation}\label{eq:symeg2cb2}
    \begin{aligned}
       \mathbf{s_3}\left[C_b\right] &=\mathbf{s}_3\big[\tilde{f}^{l_3^\prime a_1 l_3} \delta^{k_1 k_2} \tilde{f}^{l_3 l_1 k_1} \tilde{f}^{l_3^\prime l_1^\prime k_2} \tilde{f}^{l_1 l_2 l_2^{\prime\prime}} \tilde{f}^{l_1^\prime l_2^\prime k_3} \tilde{f}^{l_2 a_2 k_3} \tilde{f}^{l_2^\prime a_3 l_2^{\prime\prime}}\big]=-C_b\,,
    \end{aligned}
\end{equation}
and thus 
\begin{equation}
    N_b(l_1,l_2,l_3,p_1,p_2,p_3)=N_b(l_1^\prime,l_2^\prime,l_3^\prime,p_1,p_3,p_2)\,.
\end{equation}
Note that there is no minus sign, different from \eqref{eq:symmetryeg2n}, which indicates that for those topologies with zero color factors, one should be cautious about the symmetry conditions.

As we will see in the construction in Section~\ref{sec:ckintegrand}, the application of these symmetry constraints can usually fix a large number of parameters in the numerator ansatz.

\subsection{Full-color unitarity-cut constraints}\label{ssec:ckunitarity}

To fix the remaining parameters and to ensure that the ansatz provides a physical integrand, we apply generalized unitarity cuts \cite{Bern:1994zx, Bern:1994cg, Britto:2004nc}. 
The idea of the generalized unitarity method is that when performing unitarity cuts, \emph{i.e.}~setting certain internal lines to be on-shell
\begin{equation}
    \frac{i}{l^{2}} \stackrel{\text { cut }}{\longrightarrow} 2 \pi \delta_{+}\left(l^{2}\right)\,,
\end{equation}
the loop amplitude or form factor will be factorized as products of simpler on-shell building blocks, such as tree-level amplitudes and form factors. 
If a integrand is consistent with a spanning set of cuts (thus all possible cuts), the integrand is then guaranteed to be the correct physical result.

Below we first explain how we perform the full-color cuts by the color decomposition of the cut integrands. Then we discuss the general color-stripped cuts such as the simplest maximal cuts as well as the most complicated quadruple cuts at three loops.

\subsection*{1) The color decomposition of cut integrand}

When performing a full-color cut, the integrand under the cut should reproduce the product of color-dressed blocks as
\begin{equation}\label{eq:generalcutform}
\sum_{\sigma} \sum_{\Gamma_i} \int \prod_{j=1}^{\ell}\frac{ \mathrm{d}^{D}l_{j}}{(2\pi i)^{D}} \frac{1}{S_i} \sigma\cdot\frac{ C_i {\cal F}_{\mathcal{O},n}^{(0)}   N_i}{\prod_{\alpha_i}P^2_{\alpha_i}}    \bigg|_{\{l_{\rm c}^2 \shortrightarrow 0\}} = \int \mathrm{dPS}_{\{l_{\rm c}\}} \int \mathrm{d}\eta_{\{l_{\rm c}\}} \itbf{F}^{(0)}\prod_{I} \itbf{A}^{(0)}_{I}\,,
\end{equation} 
where $\{l_c\}$ is the list of internal cut lines, $\mathrm{dPS}$ represents the phase space measure, and the Grassman integration over $\mathrm{d}\eta$ serves to sum over internal states. 

\begin{figure}
    \centering
    \subfigure[Full-color $\itbf{F}_3$-$\itbf{A}_4$ cut]{\includegraphics[width=0.32\linewidth]{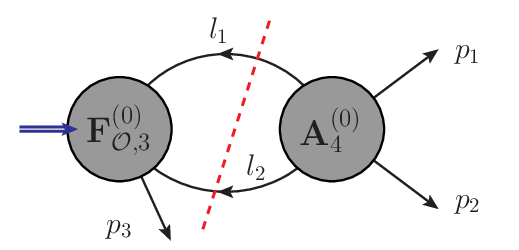}}
    \subfigure[Color-stripped  $\mathcal{F}_3$-$\mathcal{A}_4$ cut]{\includegraphics[width=0.32\linewidth]{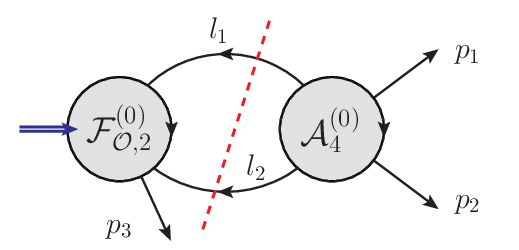}}
    \subfigure[Color-stripped  $\mathcal{F}_{3}$-$\mathcal{A}_{4}$ cut]{\includegraphics[width=0.32\linewidth]{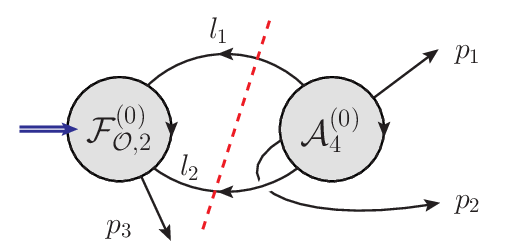}}
    \caption{Cuts for one-loop three-point form factors, where $\mathcal{O}=\operatorname{tr}(\phi^2) $. The dark gray blobs in (a) represent full-color blocks, while the light gray ones in (b) and (c) are color-stripped amplitudes and form factors. }
    \label{fig:PartialF2A4cut}
\end{figure}

In practice, it is convenient to perform color decomposition for \eqref{eq:generalcutform}, such that only color-ordered tree amplitudes or form factors are needed as blocks.
To illustrate this, we consider a double cut in the $s_{12}$ channel for full-color form factor $\itbf{F}^{(1)}_{\operatorname{tr}(\phi^2),3}$, as shown in Figure~\ref{fig:PartialF2A4cut}(a).

We first consider the tree product on the RHS of \eqref{eq:generalcutform}, which can be explicitly given as
\begin{equation}\label{eq:1loopfullcolorcut}
    \text{Full-color cut}:\ 
    \int \prod_{i=1}^{2} \mathrm{d}^{4} \eta_{l_{i}}\itbf{F}^{(0),\mathrm{MHV}}_{\operatorname{tr}(\phi^2),3}(-l_1,-l_2,p_3)\itbf{A}^{(0),\mathrm{MHV}}_{4}(l_1,l_2,p_1,p_2)\,,
\end{equation}
where we omit the phase space measure for simplicity.
The tree blocks take the standard single-trace color-decomposition form \cite{Mangano:1990by}:
\begin{align}
    \itbf{F}_{\operatorname{tr}(\phi^2),3}^{(0)}(-l_1,-l_2,p_3)&=\sum_{\sigma_1\in S_3/\mathbb{Z}_3}\operatorname{tr}(\sigma_1(a_{l_1}a_{l_2}a_{3}))\mathcal{F}^{(0)}_{\operatorname{tr}(\phi^2),3}(\sigma_1\{-l_1,-l_2,p_3\} ) \,, \\
    \itbf{A}_{4}^{(0)}(l_2,l_1,p_1,p_2)&=\sum_{\sigma_2\in S_4/\mathbb{Z}_4}\operatorname{tr}(\sigma_2(a_{l_2}a_{l_1}a_{1}a_{2}))\mathcal{A}^{(0)}_{4}(\sigma_2\{l_2,l_1,p_1,p_2\} ) \,,
\end{align}
where we use the short notation $\operatorname{tr}(abc)$ for $\operatorname{tr}(T^a T^b T^c)$.
As a result, the product in \eqref{eq:1loopfullcolorcut} can also be decomposed according to trace-color factor as
\begin{equation}
    \text{Full-color Cut}:\ \sum_{\sigma_1\times \sigma_2} \operatorname{tr}(\sigma_1(a_{l_1}a_{l_2}a_{3}))\operatorname{tr}(\sigma_2(a_{l_2}a_{l_1}a_{1}a_{2}))\mathcal{K}(\sigma_1,\sigma_2) \,, 
\end{equation}
with $\mathcal{K}$ representing the product of  color-stripped building blocks:
\begin{equation}
 \mathcal{K}(\sigma_1,\sigma_2)=\int \prod_{i=1}^{2} \mathrm{d}^{4} \eta_{l_{i}} \mathcal{F}^{(0)}_{\operatorname{tr}(\phi^2),3}(\sigma_1\{-l_1,-l_2,p_3\} ) \mathcal{A}^{(0)}_{4}(\sigma_2\{l_2,l_1,p_1,p_2\} ) \,. \nonumber
\end{equation}
Figure~\ref{fig:PartialF2A4cut}(b) and \ref{fig:PartialF2A4cut}(c) show two configurations of such color-stripped cuts, corresponding to color factors $\operatorname{tr}(a_{l_1}a_{l_2}a_{3}) \operatorname{tr}(a_{l_2}a_{l_1}a_{1}a_{2}) $ and $\operatorname{tr}(a_{l_1}a_{l_2}a_{3}) \operatorname{tr}(a_{l_2}a_{2}a_{l_1}a_{1}) $, respectively.

\begin{figure}
    \centering
    \subfigure[ ]{\includegraphics[width=0.20\linewidth]{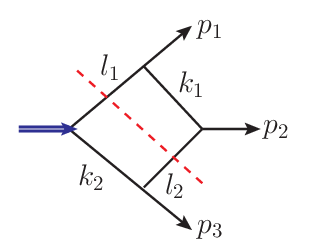}}
    \subfigure[ ]{\includegraphics[width=0.20\linewidth]{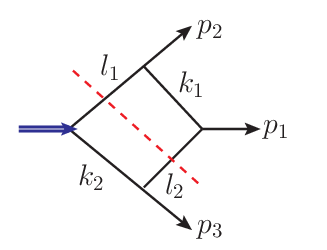}}
    \subfigure[ ]{\includegraphics[width=0.19\linewidth]{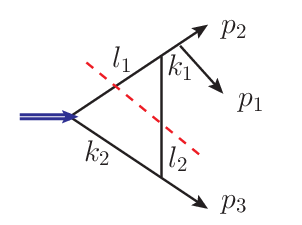}}
    \caption{Trivalent graphs contributing to the full-color $s_{12}$ cut for one-loop three-point form factors.}
    \label{fig:oneloopdiagram}
\end{figure}

Next we can consider the LHS of \eqref{eq:generalcutform} given by the ansatz. Similarly, the color factor $C_i$ in the ansatz can also be expanded in terms of the same trace color factors $\operatorname{tr}(\sigma_1)\operatorname{tr}(\sigma_2)$ as mentioned above. Concretely, there are three trivalent graphs contributing in this cut example, as shown in Figure~\ref{fig:oneloopdiagram}. 
 The color factors of the three trivalent graphs can be naturally factorized under the cuts as the product of color factors of sub-tree graphs as
    \begin{equation}
        \begin{aligned}
            C^{\rm (a)}=C^{\rm (a)}_{\itbf{F}_3}C^{\rm (a)}_{\itbf{A}_4}=(\tilde{f}^{a_3 k_2 l_2}\delta^{ k_2 l_1})(\tilde{f}^{l_1 a_1 k_1 }\tilde{f}^{ k_1 a_2 l_2})\,, \\
            C^{\rm (b)}=C^{\rm (b)}_{\itbf{F}_3}C^{\rm (b)}_{\itbf{A}_4}=(\tilde{f}^{a_3 k_2 l_2}\delta^{ k_2 l_1})(\tilde{f}^{l_1 a_2 k_1 }\tilde{f}^{ k_1 a_1 l_2})\,, \\
            C^{\rm (c)}=C^{\rm (c)}_{\itbf{F}_3}C^{\rm (c)}_{\itbf{A}_4}=(\tilde{f}^{a_3 k_2 l_2}\delta^{ k_2 l_1})(\tilde{f}^{a_2 a_1 k_1 }\tilde{f}^{ k_1 l_2 l_1})\,.
        \end{aligned}
    \end{equation}    
    For the color factors of trivalent sub-tree graphs, such as $C^{\rm (a)}_{\itbf{F}_3}$ and $C^{\rm (a)}_{\itbf{A}_4}$, one can also extract proper trace color components.
    For example, to reproduce the contribution to the product of trace color factors $\mathcal{T}_*=\operatorname{tr}(a_{l_1}a_{l_2}a_{p_3}) \operatorname{tr}(a_{l_2}a_{l_1}a_{p_1}a_{p_2}) $ as for Figure~\ref{fig:PartialF2A4cut}(b), we have
    \begin{equation}
    \label{eq:colordecompC}
        \begin{aligned}
            & C^{\rm (a)}_{\itbf{F}_3}\rightarrow(+1)\operatorname{tr}(a_{l_1}a_{l_2}a_{3}), \qquad C^{\rm (a)}_{\itbf{A}_4}\rightarrow(+1)\operatorname{tr}(a_{l_2}a_{l_1}a_{1}a_{2})\,, \\
            & C^{\rm (b)}_{\itbf{F}_3}\rightarrow(+1)\operatorname{tr}(a_{l_1}a_{l_2}a_{3}), \qquad C^{\rm (b)}_{\itbf{A}_4}\rightarrow(0)\operatorname{tr}(a_{l_2}a_{l_1}a_{1}a_{2})\,, \\
            & C^{\rm (c)}_{\itbf{F}_3}\rightarrow(+1)\operatorname{tr}(a_{l_1}a_{l_2}a_{3}), \qquad C^{\rm (c)}_{\itbf{A}_4}\rightarrow(-1)\operatorname{tr}(a_{l_2}a_{l_1}a_{1}a_{2})\,. 
        \end{aligned}
    \end{equation}
From \eqref{eq:colordecompC}, one can get: 
\begin{equation}
\label{eq:signCabc}
\{C^{\rm (a)},C^{\rm (b)},C^{\rm (c)}\}\big|_{\mathcal{T}_*}=\{1,0,-1\} \,.
\end{equation}
 
With the above decompositions, we obtain a color-stripped version of \eqref{eq:generalcutform} by selecting certain color orderings.
For example, for Figure~\ref{fig:PartialF2A4cut}(b), selecting the component associated to color factor $\mathcal{T}_*$, one has the color-stripped unitarity relation:
\begin{align}
    {\cal F}_{\operatorname{tr}(\phi^2),3}^{(0)}
&    \Bigl(
\hskip -.2cm
\begin{tabular}{c}{\includegraphics[width=0.14\linewidth]{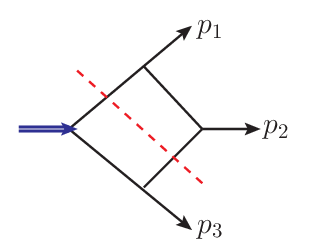}} \end{tabular} \hskip -.4cm \times (+1)N^{\rm(a)}
+\hskip -.2cm
\begin{tabular}{c}{\includegraphics[width=0.13\linewidth]{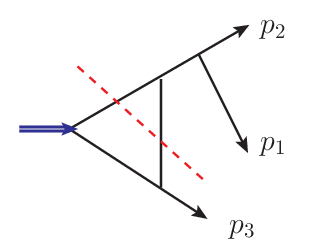}} \end{tabular} \hskip -.4cm \times (-1)N^{\rm(c)}
    \Bigr) \nonumber\\
& =\int \prod_{i=1}^{2} \mathrm{d}^{4} \eta_{l_{i}} \mathcal{F}^{(0)}_{\operatorname{tr}(\phi^2),3}(-l_1,-l_2,p_3 ) \mathcal{A}^{(0)}_{4}(l_2,l_1,p_1,p_2)\,,
\label{eq:oneloopcutsimp}
\end{align}
where the signs in front of $N^{\rm (a)}$ and $N^{\rm (c)}$ are given by \eqref{eq:signCabc}.
The same consideration can be easily applied to other orderings, including the  non-planar cut in Figure~\ref{fig:PartialF2A4cut}(c).
The generalization to generic high-loop cases is also straightforward.

\subsubsection*{2) Maximal cuts}
A special class of cuts is the maximal cuts where all possible propagators are put on-shell \cite{Britto:2004nc, Bern:2007ct}. 
In this case, only one topology contributes to the cut and one has 
\begin{equation}\label{eq:generalmc}
   {\cal F}_{\mathcal{O},n}^{(0)} \, N_i \Big|_{P^2_{\alpha_i}\shortrightarrow 0}=\int \left(\mathrm{d}\eta_{\{l_{\rm c}\}}\right)^{4} \mathcal{F}^{(0)}_{{\cal O},{\rm min}} \prod_{\rm I} \mathcal{A}^{(0)}_{3,\mathrm{I}}\,,
\end{equation}
where $\mathcal{F}^{(0)}_{{\cal O},{\rm min}}$ is the minimal form factor and $\mathcal{A}_{3}^{(0)}$ are three-point amplitudes.
Sometimes it is possible to apply the maximal-cut without actually calculating the product of three-point tree amplitudes. 
In such cases, one can apply the rung-rule \cite{Bern:1998ug} to guess the numerators and compare them with the ansatz under maximal cuts. 

Maximal cuts can be regarded as a property of each individual diagram, similar to the graph symmetry constraint in Section~\ref{ssec:cksym}. In practice, it is often used before applying more complicated cuts.

\subsubsection*{3) General non-maximal cuts}

\begin{figure}
    \centering
    \subfigure[$\mathcal{F}_2$-$\mathcal{A}_5$-$\mathcal{A}_6$ planar cut]{\includegraphics[width=0.4\linewidth]{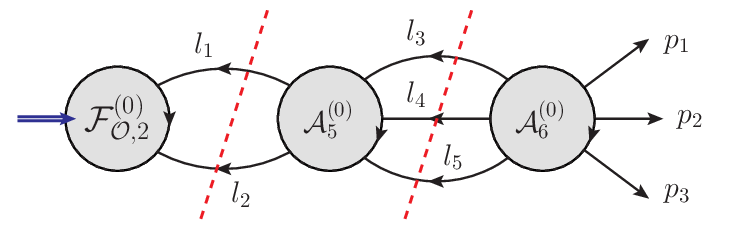}}
    \subfigure[$\mathcal{F}_2$-$\mathcal{A}_5$-$\mathcal{A}_6$ non-planar cut]{\includegraphics[width=0.4\linewidth]{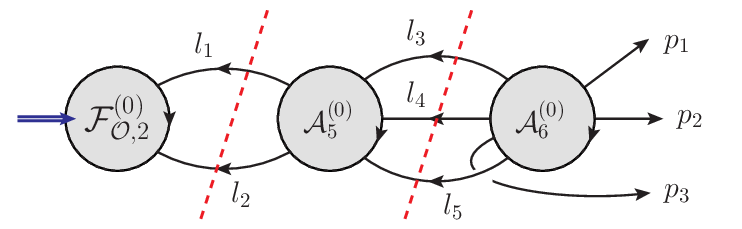}}
    \subfigure[$\mathcal{F}_4$-$\mathcal{A}_7$ cut]{\includegraphics[width=0.3\linewidth]{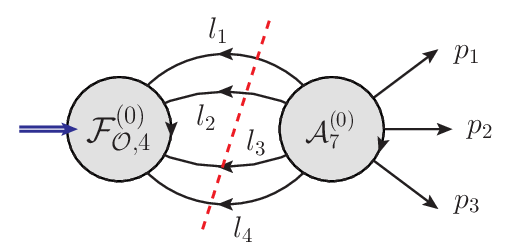}}
    \subfigure[$\mathcal{F}_{5}$-$\mathcal{A}_{6}$ cut]{\includegraphics[width=0.3\linewidth]{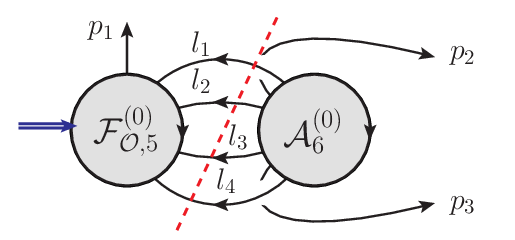}}
    \caption{Quintuple and quadruple cuts for three-loop three-point form factors.}
    \label{fig:quinquadcuts}
\end{figure}

A complete set of unitarity condition involves more general unitarity cuts, in which less propagators are put on-shell such as examples given in Figure~\ref{fig:quinquadcuts}.
For the three-loop form factors, the most complicated cuts are the quadruple cuts, where only four propagators are cut and the full integrand factorizes as the product of two tree blocks.
Two such cuts are shown in Figure~\ref{fig:quinquadcuts}(c) and~\ref{fig:quinquadcuts}(d).  
Non-trivial tree building blocks are involved in such cases,
for example, the tree product for the case of Figure~\ref{fig:quinquadcuts}(d) can be given as
\begin{equation}
\begin{aligned}
\int \prod_{i=1}^{4} \left(d \eta_{l_{i}}\right)^{4}\Big[& \mathcal{F}_{\mathcal{O},5}^{(0), \mathrm{MHV}}\left(p_{1},-l_{1},-l_{2},-l_{3},-l_{4}\right) \mathcal{A}_{6}^{(0), \mathrm{NNMHV}}\left(p_{2},l_{1},l_{4},p_{3},l_{3},l_{2}\right) \\
+&\mathcal{F}_{\mathcal{O},5}^{(0), \mathrm{NMHV}}\left(p_{1},-l_{1},-l_{2},-l_{3},-l_{4}\right) \mathcal{A}_{6}^{(0), \mathrm{NMHV}}\left(p_{2},l_{1},l_{4},p_{3},l_{3},l_{2}\right)\\
+&\mathcal{F}_{\mathcal{O},5}^{(0), \mathrm{NNMHV}}\left(p_{1},-l_{1},-l_{2},-l_{3},-l_{4}\right) \mathcal{A}_{6}^{(0), \mathrm{MHV}}\left(p_{2},l_{1},l_{4},p_{3},l_{3},l_{2}\right)\Big]\,.
\end{aligned}
\end{equation}
The MHV tree-level amplitudes are standard Parke-Taylor form \cite{Parke:1986gb}, and MHV form factors of $\mathcal{O}=\operatorname{tr}(\phi^L)$ take simple Parke-Taylor-like form (see \emph{e.g.}~\cite{Yang:2019vag}). As for the N$^{k}$MHV amplitudes and form factors, they can be computed via BCFW on-shell recursion \cite{Britto:2005fq} or CSW vertex expansion \cite{Cachazo:2004kj} methods. 

The tree products can be compared with the cut of CK-dual ansatz, similar to \eqref{eq:oneloopcutsimp}.  
For three-loop cases, the ansatz cut-diagrams have typically a total number of order $\sim O(100)$. 
Thus the quadruple cuts provide very non-trivial constraints on the results as well as the most stringent checks.

\subsubsection*{4) On $D$-dimensional cuts}

Since we use dimensional regularization and the loop momenta are $D$-dimensional, one should  in principle apply unitarity cuts in $D$ dimensions. 
In this respect, we have checked our results by applying $D$-dimensional cuts that contain four-point sub-amplitudes, following the strategy in \cite{Bern:2010tq}. 
Although it would be interesting to consider further for a complete set of $D$-dimensional cuts, we would like to make the following two remarks that strongly support the completeness of our results:
\begin{itemize}
\item 
First, the dual Jacobi relations that relate numerators of different diagrams are expected to hold in any dimension, and the ansatz of the numerators are given in terms of Lorentz products (not spinor products) which are in a form valid in general dimensions.
\item
Second, we have evaluated the full-color form factors numerically, as will be discussed later in Section~\ref{sec:ir}. Our results give consistent full-color IR divergences as well as the planar finite remainder function from a bootstrap computation \cite{Dixon:2020bbt}. A possibly missing $D$-dimensional contribution (such as $\mu^2$-term integral) would easily modify (the divergent or finite part of) the results and thus would lead to inconsistency. 
\end{itemize}

\section{Constructing three-loop form factor integrands}\label{sec:ckintegrand}

In this section, we apply the strategy described in the previous section to construct the full-color integrands of three-loop three-point form factors:
\begin{equation}
    \mathcal{F}_{\mathcal{O}}(1, 2, 3 ; q) = \int d^{D} x e^{-i q \cdot x}\langle \Phi_1(p_1) \Phi_2(p_2) \Phi_3(p_3)|\mathcal{O}(x)| 0\rangle \,,
\end{equation}
where $p_i^2=0, i = 1,2,3,$ and $q^2 = (p_1+p_2+p_3)^2 \neq0$.
We consider two types of operators: one is $\operatorname{tr}(\phi^2)$ which is also a component of the stress-tensor supermultiplet, and the other is the length-three half-BPS operator ${\rm tr}(\phi^3)$.

\subsection{Warm-up: Two-loop three-point form factor of $\operatorname{tr}(\phi^2)$}\label{ssec:resphi22loop}

As a warm-up example, let us briefly revisit the simpler case|the two-loop three-point form factor of the stress-tensor supermultiplet \cite{Boels:2012ew}. 
Here we will obtain a more general CK-dual solution that contains four free parameters.

\subsubsection*{Constructing ansatz}
\begin{enumerate}
    \item[(1)]\textbf{Trivalent topologies}

Following the criteria of selecting trivalent topology in Section~\ref{ssec:ckansatz}, 
one finds that there are six contributed topologies, given by (1)--(6) in Figure~\ref{fig:phi2tops2loop}.

\begin{figure}[t]
    \centering
  \includegraphics[width=0.99\linewidth]{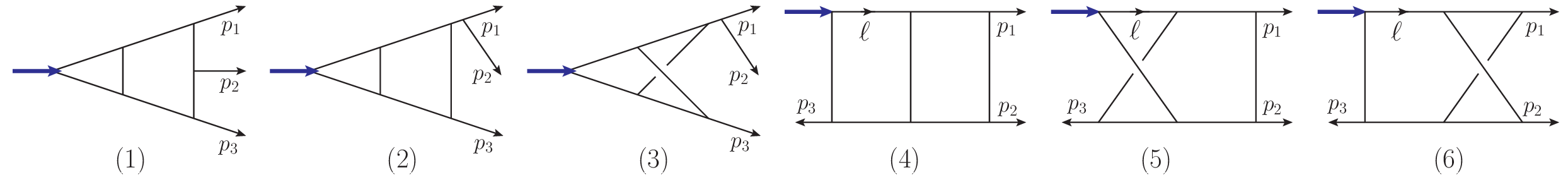}
\caption{Trivalent topologies for the two-loop form factor of  $\operatorname{tr}(\phi^2)$. }
\label{fig:phi2tops2loop}
\end{figure}

\begin{figure}
    \centering
    \includegraphics[width=0.55\linewidth]{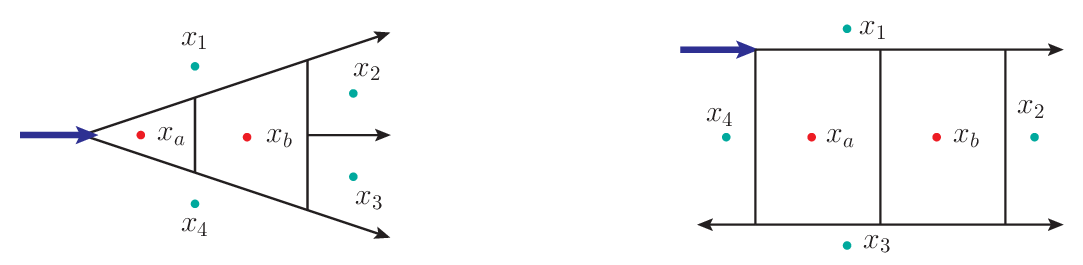}
    \caption{Master graphs for two-loop $\operatorname{tr}(\phi^2)$ form factor. }
    \label{fig:phi2master2loop}
\end{figure}

    \item[(2)]\textbf{Jacobi relations and master numerators}

    By generating the dual Jacobi relations, one can find it is possible to choose two planar topologies as master topologies,  as  shown in Figure~\ref{fig:phi2master2loop}.\footnote{Note that in this case it is also possible to choose one non-planar graph, that is Figure~\ref{fig:phi2tops2loop}(5), as the single master topology.}
    Other numerators can be determined from master numerators $N_1$ and $N_4$ by the following dual Jacobi relations:
    \begin{align}
    &N_2(p_1,p_2,p_3) = N_1(p_2,p_3,p_1) + N_1(p_3,p_2,p_1) \,, \nonumber\\ 
    &N_5(p_1,p_2,p_3,\ell) = N_1(p_1,p_2,p_3) - N_4(p_1,p_2,p_3,\ell) \,, \nonumber\\  
    &N_3(p_1,p_2,p_3) = N_2(p_1,p_2,p_3) \,, \qquad N_6(p_1,p_2,p_3,\ell) = N_4(p_1,p_2,p_3,\ell) \,.
    \end{align}

    \item[(3)]\textbf{Master numerators and the Ansatz}

    Taking advantage of the planarity of master graphs, we apply zone variables  to parametrize the momenta, which are defined explicitly in Figure~\ref{fig:phi2master2loop}. 
    The master numerators are polynomials of zone variables with degree three.
    Moreover, as described in Section~\ref{ssec:ckansatz}, the good UV property of ${\cal N}=4$ SYM provides strong power-counting constraints on the loop momenta dependence in master numerators: 
    for $N_1 $ of the first master in Figure~\ref{fig:phi2master2loop}, $x_b $ can appear at most once in Lorentz products, so $(x_{ib}^2)^1 $, with $i=1,2,3,4 $, is allowed, and any term containing $x_a $ is forbidden; while for the other master numerator $N_4 $, $x_a $ can appear at most once.
    With these conditions, we obtain an ansatz for two master numerators with 68 parameters in total.
   
\end{enumerate}

\subsubsection*{Solving ansatz}

Next, to solve the ansatz, we impose various conditions as follows.

\begin{figure}[t]
    \centering
    \subfigure[]{
        \includegraphics[width=0.25\linewidth]{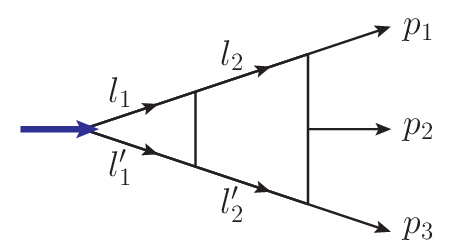}
         \label{fig:2loopSymConstraintExample}
        }
    \centering
    \subfigure[]{
        \includegraphics[width=0.25\linewidth]{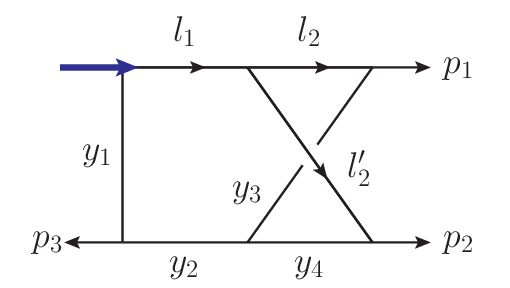}
        \label{fig:2loopSymConstraintExample1}
        }
    \caption{Symmetry constraint examples for the two-loop form factor of  $\operatorname{tr}(\phi^2)$.}
\end{figure}

\begin{enumerate}
\item[(1)]
First we apply symmetry constraints on all topologies. As mentioned in Section~\ref{ssec:cksym}, if the color factor is non-zero, it is straightforward to write down conditions from diagrammatic symmetries for three-point form factors; otherwise, a more detailed analysis is required. 

Specifically, color factors of (1)--(5) in Figure~\ref{fig:phi2tops2loop} are not zero, whereas the color factor of Figure~\ref{fig:phi2tops2loop}(6) is zero, and below we choose (1) and (6) as examples.
For Figure~\ref{fig:phi2tops2loop}(1), one has a symmetry $\mathbf{s}_1:\{p_1 \leftrightarrow p_3, l_{1,2}\leftrightarrow l_{1,2}^{\prime}\}$.
Such a symmetry leads to
\begin{equation}
	C_1\mathcal{F}^{(0)}N_1(l_1,l_2,p_1,p_2,p_3)=\mathbf{s}_1\big[C_1\mathcal{F}^{(0)}N_1(l_1,l_2,p_1,p_2,p_3)\big]\,.
\end{equation}
Here we have 
\begin{equation}
\begin{aligned}
&\mathbf{s}_1\big[C_1] = \mathbf{s}_1[2N_c^2\tilde{f}^{a_1 a_2 a_3}]=2N_c^2\tilde{f}^{a_3 a_2 a_1}=-C_1;\\
&  \mathbf{s}_1[\mathcal{F}^{(0)}(p_1,p_2,p_3)]=\mathcal{F}^{(0)}(p_3,p_2,p_1)=-\mathcal{F}^{(0)}(p_1,p_2,p_3)\,,
\end{aligned}
\end{equation}
resulting in 
\begin{equation}
   N_1(l_1,l_2,p_1,p_2,p_3)=N_1(l_1^\prime,l_2^\prime,p_3,p_2,p_1)\,.
\end{equation}

For Figure~\ref{fig:phi2tops2loop}(6), since its color factor is zero, it is necessary to consider the full trivalent product form.
Consider the symmetry $\mathbf{s}_2:\{p_1\leftrightarrow p_2,l_2\leftrightarrow l_2^\prime, y_3\leftrightarrow y_4\}$, one gets 
\begin{equation}
\begin{aligned}
      \mathbf{s}_2\left[  C_{6}\right] &= 
      \mathbf{s}_2\left[  \delta^{l_1 k_1} \tilde{f}^{k_1 k_2 a_3} \tilde{f}^{l_1 l_2 l_2^\prime} \tilde{f}^{k_2 k_3 k_4} \tilde{f}^{l_2 a_1 k_3} \tilde{f}^{l_2^\prime a_2 k_4} \right]=C_6\,,
    \end{aligned}
\end{equation}
so that 
\begin{equation}
     N_6(l_1,l_2,p_1,p_2,p_3)=-N_6(l_1,l_2^\prime,p_2,p_1,p_3)\,.
\end{equation}

After considering all symmetries, 68 free parameters are reduced to only 16.

\begin{figure}
    \centering
    \subfigure[$\mathcal{F}_3$-$\mathcal{A}_6$ cut]{\includegraphics[width=0.3\linewidth]{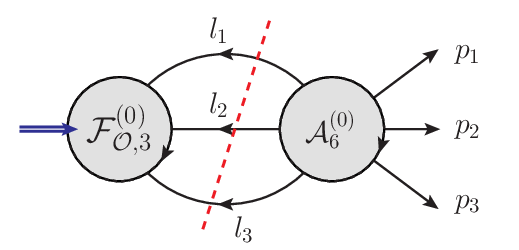}
    \label{fig:tricuts2loop}}
    \subfigure[$\mathcal{F}_{4}$-$\mathcal{A}_{5}$ cut]{\includegraphics[width=0.3\linewidth]{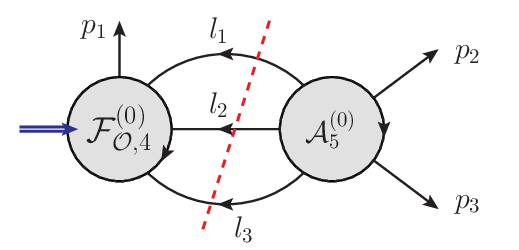}
    \label{fig:tricuts2loop1}}
    \caption{Triple cuts for two-loop three-point form factors.}
\end{figure}

\item[(2)]
Next we apply generalized unitarity cuts, 
where two most constraining planar cuts are shown in Figure~\ref{fig:tricuts2loop}--\ref{fig:tricuts2loop1}. The expressions of tree products are
\begin{equation}
\begin{aligned}
\hskip -2.5pt 
\text{Cut-(a)}:\ \int \prod_{i=1}^{3} \left(d \eta_{l_{i}}\right)^{4}\Big[& \mathcal{F}_{\operatorname{tr}(\phi^2),3}^{(0), \mathrm{MHV}}\left(-l_{1},-l_{2},-l_{3}\right) \mathcal{A}_{6}^{(0), \mathrm{NMHV}}\left(p_{1}, p_{2},p_{3}, l_{3}, l_{2}, l_{1}\right) \\
+&\mathcal{F}_{\operatorname{tr}(\phi^2),3}^{(0), \mathrm{NMHV}}\left(-l_{1},-l_{2},-l_{3}\right) \mathcal{A}_{6}^{(0), \mathrm{MHV}}\left(p_{1}, p_{2}, p_{3}, l_{3}, l_{2}, l_{1}\right)\Big]\,,
\end{aligned}
\end{equation}
\begin{equation}
\begin{aligned}
\hskip -2.5pt 
\text{Cut-(b)}:\ \int \prod_{i=1}^{3} \left(d \eta_{l_{i}}\right)^{4}\Big[& \mathcal{F}_{\operatorname{tr}(\phi^2),4}^{(0), \mathrm{MHV}}\left(p_{1},-l_{1},-l_{2},-l_{3}\right) \mathcal{A}_{5}^{(0), \mathrm{NMHV}}\left(p_{2},p_{3},l_{3},l_{2},l_{1}\right) \\
+&\mathcal{F}_{\operatorname{tr}(\phi^2),4}^{(0), \mathrm{NMHV}}\left(p_{1},-l_{1},-l_{2},-l_{3}\right) \mathcal{A}_{5}^{(0), \mathrm{MHV}}\left(p_{2},p_{3},l_{3},l_{2},l_{1}\right)\Big]\,.
\end{aligned}
\end{equation}
These two cuts can fix 12 parameters among the 16 parameters. 

\item[(3)]
Finally, one can check the integrand solution with four free parameters satisfies all other cuts and all the dual Jacobi relations. 
Hence, this four-parameter solution is the final solution that manifests CK-dual structure and is consistent with all unitarity conditions. 

\end{enumerate}

The master-numerator solution with four parameters takes the following form:
\begin{equation}\label{eq:phi2master2loop}
\begin{aligned}
           N_1 =& x_{14}^2 x_{13}^2 x_{23}^2/2 \\
      &  -  c_1\bigl(S_{\rm p,3}\big(x_{b3}^2-x_{b4}^2\big)+ S_{\rm p,2}\big(x_{b2}^2-x_{b3}^2\big) +S_{\rm p,1}(x_{b1}^2-x_{b2}^2\big)\bigr) \\
      &  -  c_2\bigl(S_{\rm s,3}\big(x_{b3}^2-x_{b4}^2\big)+ S_{\rm s,2}\big(x_{b2}^2-x_{b3}^2\big) +S_{\rm s,1}(x_{b1}^2-x_{b2}^2\big)\bigr)\\
        N_4 =& x_{13}^2(x_{13}^2  x_{24}^2 + (x_{13}^2 -x_{14}^2)(x_{1a}^2-x_{2a}^2) +  x_{24}^2 ( x_{1a}^2 -  x_{3a}^2))/2 \\
      &  +  c_1 S_{\rm p,3} \bigl(x_{a3}^2 - x_{a4}^2 \bigr) +  c_2 S_{\rm s,3} \bigl(x_{a3}^2  - x_{a4}^2 \bigr)\\
      &  +  c_3 S_{\rm p,3} \bigl( x_{a4}^2 -  x_{a1}^2 \bigr)+  c_4 S_{\rm s,3} \bigl( x_{a4}^2 -  x_{a1}^2 \bigr)
\end{aligned} 
\end{equation}
where $c_{1,2,3,4}$ denote the four parameters and $S_{\mathrm{p},i}$, $S_{\mathrm{s},i}$ are simple functions of Mandelstam variables $s_{ij}$, for example $ S_{\rm p,3}=s_{12}(s_{23}-s_{13})$. See more details in Appendix~\ref{ap:ck2loop}.

We mention that our result reproduces the result in \cite{Boels:2012ew} when setting $c_{1,2,3,4}=0$. Of course, there is no physical requirement that forces $c_i$ to be 0. We will show in Section~\ref{ssec:parameters} that these parameters come from a special deformation of the integrand and they cancel at the integrand level. 

\subsubsection*{Final solution}

The final two-loop full-color integrand reads
\begin{equation}  
\label{eq:ff2loop}
{\cal F}_{{\rm tr}(\phi^2), 3}^{(2)} = {\cal F}_{{\rm tr}(\phi^2), 3}^{(0)}   \sum_{\sigma_e} {\sign{\sigma_e} } \sum_{i=1}^{6} \int \prod_{j=1}^{3} d^D \ell_j {1\over S_i} \,\sigma_e\cdot{C_i \, N_i \over \prod_{\alpha_i} P^2_{\alpha_i}} \,,
\end{equation}
where the various factors are given in Appendix~\ref{ap:ck2loop}. 
The permutation operator $\sigma_e \in S_3$ acts on the three external momenta and color indices in $N_i, P^2_{\alpha_i} $ and $C_i $. 
Notice that the tree-level form factor is factorized out and not permuted by $\sigma_e$, which is different from \eqref{eq:generalckform}; this requires a compensation which is responsible for the extra $\sign{\sigma_e}$ factor.
The $\sign{\sigma_e} $ is +1/-1 for even/odd permutation $\sigma_e $ respectively, rising from the fact that the tree-level form factor gives a $\sign{\sigma_e}$ when permuting external lines. .

Finally, we comment that as we will see in the three-loop cases, the existence of free parameters in final CK-dual solutions appears to be a more general feature.

\subsection{Three-loop three-point form factor of $\operatorname{tr}(\phi^2)$}\label{ssec:resphi2}

In this subsection, we construct the full-color three-loop integrand for the three-point form factor of the stress-tensor supermultiplet. 

\begin{figure}[t]
    \centering
    \includegraphics[width=0.99\textwidth]{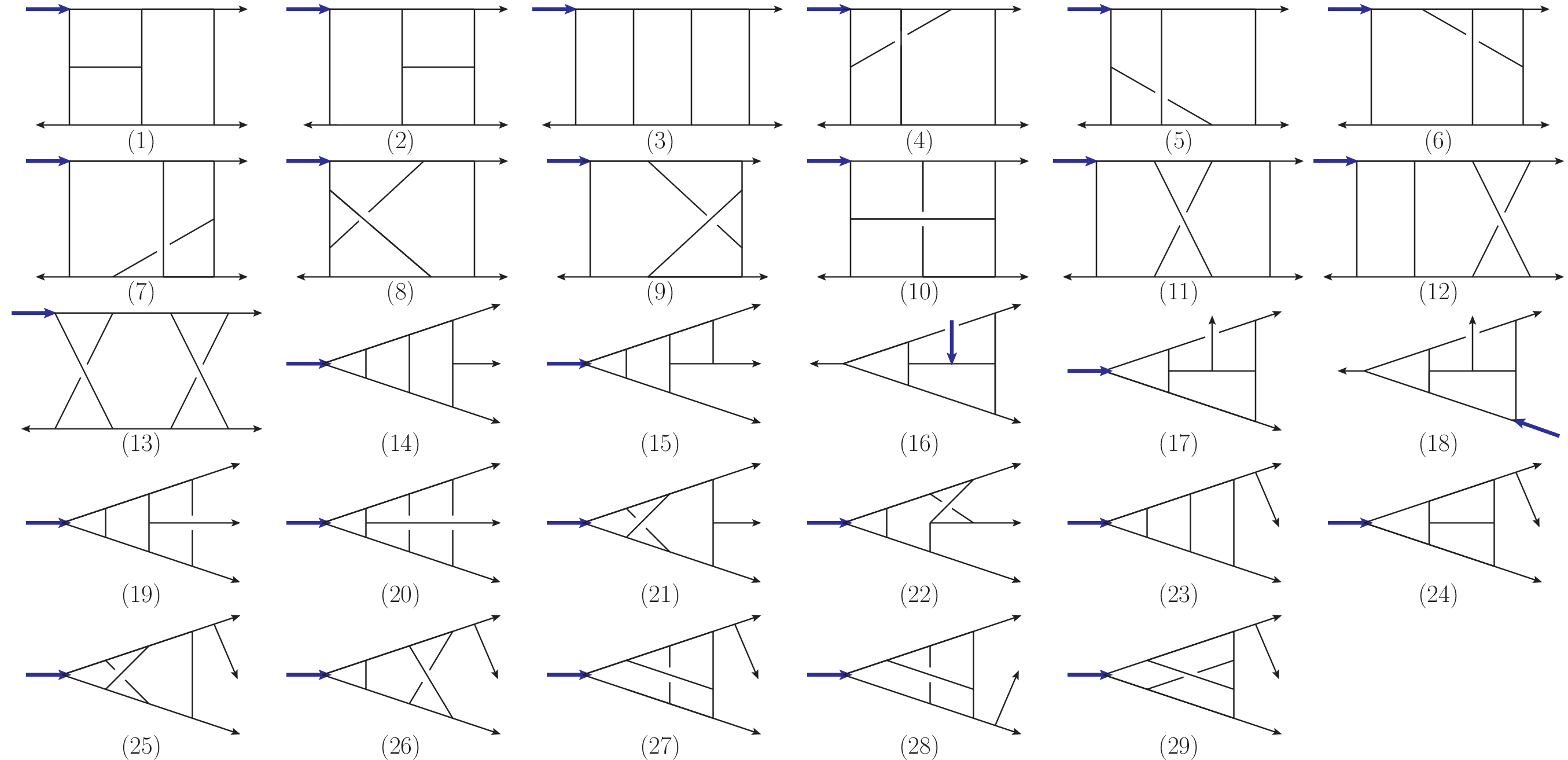}
\caption{Trivalent topologies for the three-loop form factor of  $\operatorname{tr}(\phi^2)$.}
\label{fig:phi2tops}
\end{figure}

\subsubsection*{CK-dual ansatz}
\begin{figure}[t]
    \centering
    \includegraphics[width=0.65\linewidth]{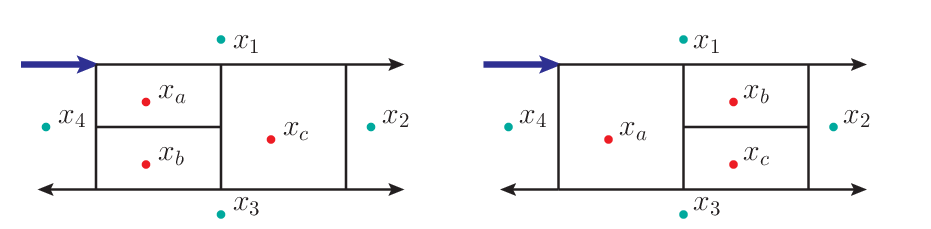}
    \caption{Master graphs for three-loop $\operatorname{tr}(\phi^2)$ form factor. }
    \label{fig:phi2master}
\end{figure}

\begin{enumerate}
    \item[(1)]\textbf{Trivalent topologies}
   The first step is to construct three-loop trivalent topologies. 
	Using the same selection rule mentioned in Section~\ref{ssec:ckansatz}, we find that there are in total 29 topologies, which are listed in Figure~\ref{fig:phi2tops}. 

    \item[(2)]\textbf{Jacobi relations and master numerators}

    We assume the form factor can be expressed in the form of \eqref{eq:generalckform}, where the numerators $N_i, i=1,...,29,$ satisfy dual Jacobi relations.
    By inspecting the dual Jacobi relations for all 29 trivalent topologies, we find that the minimal number of master graphs is two.
    A convenient choice is to select $N_1, N_2 $ (both are planar) as master numerators, and their topologies are shown in Figure~\ref{fig:phi2master} respectively. 
    \item[(3)]\textbf{Ansatz for master numerators}
    
	 As for the ansatz for master numerators, similar to the two-loop case,  we define zone variables   as in Figure~\ref{fig:phi2master}. 
    The ansatz should be a degree-four polynomial of zone variables, and the power-counting constraints are as follows.
    For $N_1 $,   $x_a,x_c $ can appear at most once, so both $(x^2_{ai})^1,(x^2_{ci})^1 $, with $i=1,2,3,4 $, and $(x^2_{ac})^1 $ are allowed; any term containing $x_b $ or  more than one $x_a $ or $x_c $, such as $(x^2_{ac})^2,(x^2_{a1})^2 $ are forbidden. 
    For  $N_2 $, only $x_b $ can appear and can appear at most twice, so only $x^2_{bi} $ can appear, with possible power $2,1,0 $. 
    With these requirements, the total number of parameters is 316: 201 for the first numerator and 115 for the second. 
\end{enumerate}

\subsubsection*{Solving ansatz}
\
Given the integrand ansatz, we now  consider diagrammatic symmetries and unitarity conditions to solve for  the parameters. 
\begin{enumerate}
\item[(1)]
We first apply the symmetries. Examples of applying symmetry conditions for three-loop topologies have been given in Section~\ref{ssec:cksym}, and we skip details here. 
After applying symmetry conditions of all topologies, an ansatz with only 105 parameters is obtained. 
One can see that the symmetry conditions provide strong constraints and can substantially reduce the number of parameters.

\item[(2)]
Next we apply unitarity-cut constraints.
In this case, we directly apply the most constraining cuts, \emph{i.e.}~the quadruple cuts discussed in Section~\ref{ssec:ckunitarity}, such as  (c) and (d) in Figure~\ref{fig:quinquadcuts}. 
With these cuts, only 24 parameters remains unsolved. 

\item[(3)]
Finally, one can check that the solution with 24 free parameters satisfies all the dual Jacobi relations as well as other unitarity cuts, thus the 24-parameter solution is indeed physical.
It is interesting and a little surprising that the final solution still contains a quite large number of free parameters, which we will discuss more in Section~\ref{ssec:parameters}.

\end{enumerate}

\subsubsection*{Final solutions}

The full-color integrand for $\operatorname{tr}(\phi^2)$ form factor at three loops can be given as follows:
\begin{equation}  
\label{eq:fullcolor-3loop-L2}
 \itbf{F}_{{\rm tr}(\phi^2), 3}^{(3)} = {\cal F}_{{\rm tr}(\phi^2), 3}^{(0)}   \sum_{\sigma_e} {\sign{\sigma_e} } \sum_{i=1}^{29} \int \prod_{j=1}^{3} d^D \ell_j {1\over S_i} \,\sigma_e \cdot {C_i \, N_i \over \prod_{\alpha_i} P^2_{\alpha_i}} \,,
\end{equation}
where the notations have been explained in the previous two-loop case, and the explicit expressions of various factors can be found in the ancillary files in \cite{Lin:2021kht}.\footnote{In the ancillary files in \cite{Lin:2021kht}, we explain how the permutation acts on the numerators. In particular, the $\sigma_e \in S_3$ acting on $N_i$ may give an extra sign in \cite{Lin:2021kht}. Here in \eqref{eq:fullcolor-3loop-L2}, we explicitly extract this possible sign by introducing the signature function $\sign{\sigma_e}$.}

\subsection{Three-loop three-point form factor of ${\rm tr}(\phi^3)$}\label{ssec:resphi3}

In this subsection, we construct the full-color three-loop integrand for the three-point form factor of the half-BPS operator ${\rm tr}(\phi^3)$. 
Compared with the previous construction for the form factor of $\operatorname{tr}(\phi^2)$, the $\operatorname{tr}(\phi^3)$ case is simpler, partly because of the smaller number of propagators for each cubic graph.

\subsubsection*{Constructing ansatz}
We now provide some details of CK-ansatz construction of the three-point form factor of $\operatorname{tr}(\phi^3)$. 
\begin{enumerate}
    \item[(1)]\textbf{Trivalent topologies}
    
    Also following the topology selection rule described in Section~\ref{ssec:ckansatz},     there are 26 topologies. Twenty of them, (1)--(20) in Figure~\ref{fig:phi3tops}, the operator vertex is connected to three internal lines; they will be called range-three topologies.
    While in the remaining six topologies, (21)--(26) \emph{i.e.}~the last row in Figure~\ref{fig:phi3tops}, the operator vertex is connected to only two internal lines,  and one of the external lines is decoupled in the interaction; they will be called range-two topologies.

\begin{figure}[t]
\centering
\includegraphics[width=0.9\textwidth]{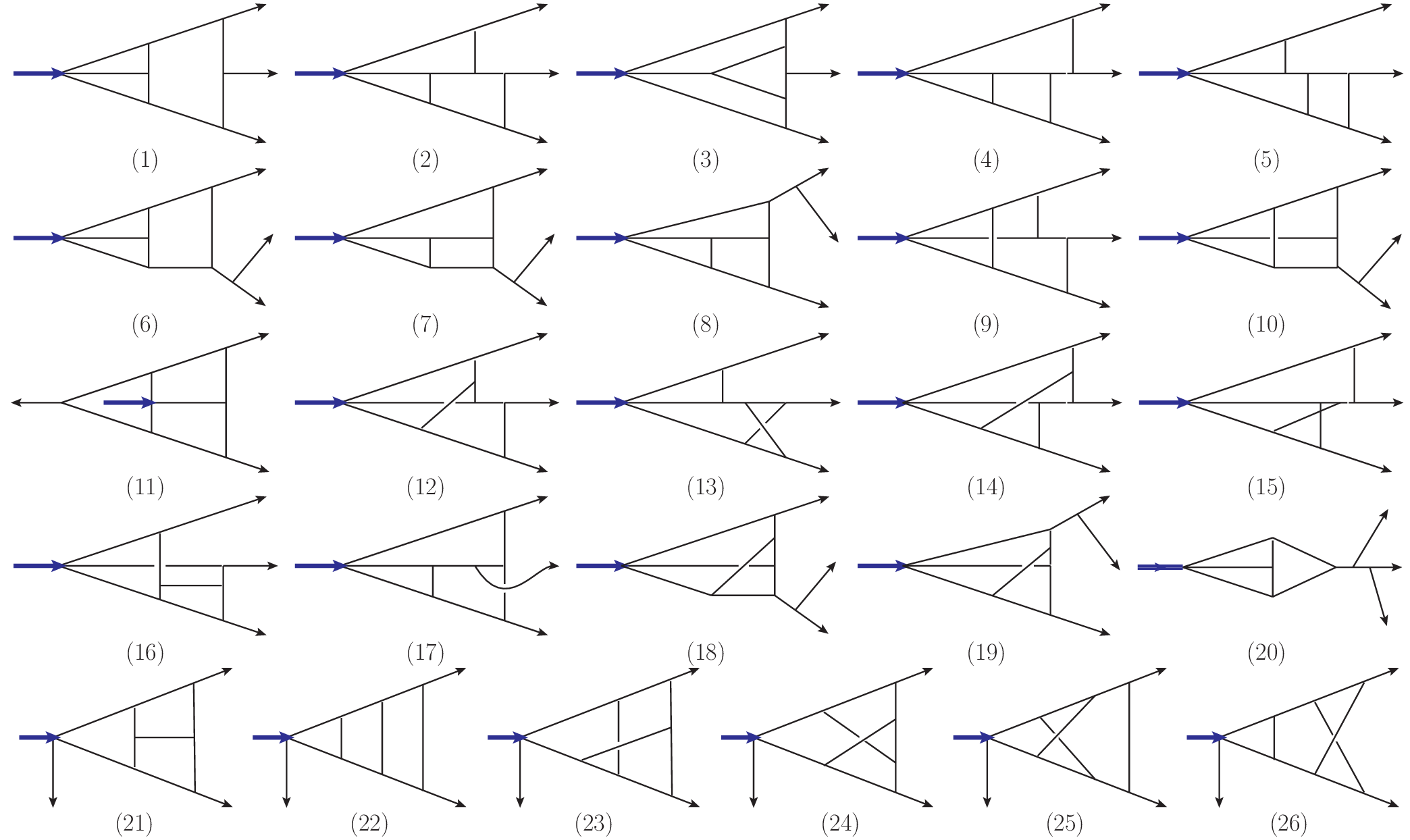}
\caption{Trivalent topologies for the three-loop form factor of  $\operatorname{tr}(\phi^3)$.}
\label{fig:phi3tops}
\end{figure}

    \item[(2)]\textbf{Jacobi relations and master numerators}
    
    Since we generate dual Jacobi relations from propagators not directly attached to the $q^2 $ vertex, for high-length operator (length$\geq$3) the Jacobi relations for diagrams with different interacting ranges are decoupled.
    Thus, numerators $N_{i}$ with $i=1,\ldots,20$ and $i=21,\ldots,26$  satisfy two sets of decoupled dual Jacobi relations. Accordingly, the master numerators should be selected separately.
    
    For the range-three topologies, we choose $N_{1},N_{2},N_{3}$ as planar master numerators; while for the range-two graphs, we choose $N_{21}$ as a master.  Although the choice of masters is not unique, we choose above topologies mainly because they are planar and mostly symmetric, which are nice properties in the ansatz construction.\footnote{It is possible to choose only three masters but involving non-planar graphs, such as $N_{9}, N_{11} $ and $N_{21} $.}
    
    \item[(3)]\textbf{Ansatz for master numerators}

\begin{figure}[t]
    \centering
    \includegraphics[width=0.8\linewidth]{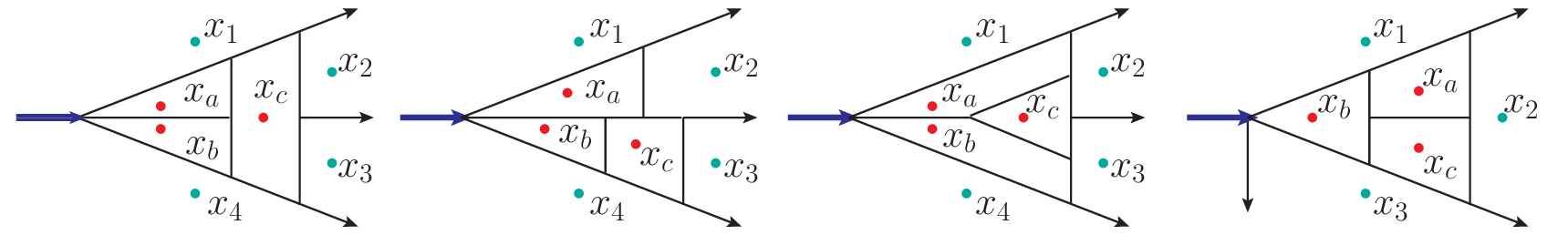}
    \caption{Master graphs for three-loop $\operatorname{tr}(\phi^3)$ form factor.}
    \label{fig:phi3master}
\end{figure}

The master graphs and the defintion of zone variables are shown in Figure~\ref{fig:phi3master}. The power-counting constraints of these four masters are: 
for $N_1 $,  only $x^2_{ci} $ can appear, with possible power $2,1,0 $; for $N_2$,  both $(x^2_{ai})^1,(x^2_{ci})^1 $, and $(x^2_{ac})^1 $ are allowed; for  $N_3$,  both $(x^2_{ai})^1,(x^2_{bi})^1 $, and $(x^2_{ab})^1 $ are allowed; in the end we have only $x^2_{bi} $ with possible power $1,0 $.
Given these power-counting constraints, we can write down an ansatz as a linear combination of monomial with undetermined coefficients. The total number of parameters to fix is 295.
    
\end{enumerate}

\subsubsection*{Solving ansatz}

\begin{enumerate}
    \item[(1)] In this case, we first impose conditions for each master graphs individually, including the maximal cuts and symmetry constraints.
    
    For maximal cuts, we can use rung-rule \cite{Bern:1998ug,Bern:2010tq} to determine the ``irreducible" part (\emph{i.e.}~the maximal-cut-detectable part) of each master numerator, reading
    \begin{equation}
       0, \quad  -x_{24}^2 x_{a4}^2 x_{1c}^2, \quad -x_{2b}^2x_{3a}x_{14}^2+x_{3a}x_{1b}^2x_{14}^2+x_{4a}x_{2b}^2x_{14}^2, \quad  x_{2b}^2 x_{13}\,,
    \end{equation}
    and 19 parameters can be fixed after applying maximal cuts.
    
    Furthermore, the symmetries of master topologies themselves can be considered, solving 92 more parameters and resulting in a simpler master ansatz.
    
    \item[(2)] Since we can derive all other numerators based on dual Jacobi relations and master numerators, we can then apply the constraints|symmetries, dual Jacobi relations and maximal cuts|on diagrams other than masters, as described in Section~\ref{ssec:ckunitarity}.

    Concretely, the symmetries of graphs other than master graphs   are considered and 138 more parameters are fixed.
    Next all the possible dual Jacobi relations are checked, and 13 more parameters are solved. 
    Then all the remaining maximal cuts can be considered which further fix 7 parameters. 
    The dimension of the parameter space is thus reduced to only 26. 
    
    \item[(3)] 
	  Finally, the constraints from other unitarity cuts are  considered, involving the two quadruple cuts stated in Section~\ref{ssec:ckunitarity} and 16 parameters can be solved.
    With this 10-parameter integrand, we have checked a spanning set of cuts and no further parameters can be solved.
    Thus, we reach the final solution for master numerators. 
\end{enumerate}

\subsubsection*{Final solution}

The full-color integrand for $\operatorname{tr}(\phi^3)$ form factor at three loops can be given as follows:\footnote{As a side remark, no extra signature like ${\sign{\sigma_e} }$ appears here because tree-level form factor ${\cal F}_{{\rm tr}(\phi^3), 3}^{(0)}=1$ is invariant under permutation.}
\begin{equation}  
\label{eq:phi3res2}
 \itbf{F}_{{\rm tr}(\phi^3), 3}^{(3)} = {\cal F}_{{\rm tr}(\phi^3), 3}^{(0)}   \sum_{\sigma} \sum_{i=1}^{26} \int \prod_{j=1}^{3} d^D \ell_j {1\over S_i} \,\sigma \cdot {C_i \, N_i \over \prod_{\alpha_i} P^2_{\alpha_i}} \,,
\end{equation}
where the various factors can also be found in anciliary files in \cite{Lin:2021kht}.
We only need to point out that unlike the form factor of ${\rm tr}(\phi^2)$, here the permutation $\sigma$ has a more complicated structure. We write $\sigma=\sigma_i\times\sigma_{e} $:  $\sigma_e\in S_3$ acts on external color indices and momenta as before and $\sigma_i\in S_3/\mathbb{Z}_3$ acts on lines directly connected to the operator vertex, because there are two inequivalent ways of connecting three internal lines to a single-trace length-three operator. About this point, further examples and explanations can be found in Section~\ref{ssec:color}.

\subsection{Free parameters and generalized gauge transformations}\label{ssec:parameters}

One very interesting finding about the integrand solutions obtained in previous subsections is that they contain a large number of free parameters. 
In other words, the form factor integrand lies in a large solution space that manifests CK-duality. 
Since we have considered a spanning set of unitarity cuts, all the solutions in the solution space should be equivalent and can be deformed from one another.\footnote{In Section~\ref{sec:simintegrand}, we will show that all free parameters indeed cancel by simplifying the integrand properly.}
As we will see, such CK-reserving deformations have an origin of generalized gauge transformations and reflect new features due to the operator insertion for form factors.

\subsubsection*{From traditional gauge transformation (GT) to generalized GT}

It is well-known that the traditional gauge transformations (GTs) correspond to the change of external polarization vectors. 
For example, amplitudes and form factors should be invariant under the following transformation of gluon polarization vectors:
\begin{equation}\label{eq:tragauge}
\varepsilon_i^\mu \rightarrow \varepsilon_i^\mu + \alpha p_i^\mu \,,
\end{equation}
where $\alpha$ is an arbitrary function. 

Consider the four-gluon tree amplitude with the numerator expressions in \eqref{eq:4ptcksolution}, one can perform the traditional GT for $\varepsilon_1$, and the numerators $N_{s,t,u}$ are not invariant but shifted as:
\begin{equation}
    N_{s}\rightarrow N_{s}+\alpha \delta_{s},  \quad \text{ with }\delta_{s}= N_{s}\big|_{\varepsilon_1\shortrightarrow p_1} \,,
\end{equation}
and so are $N_{t,u}$.
An explicit calculation shows that the deformations $\delta_{s,t,u}$ satisfy
\begin{equation}
\label{eq:tradGT4pt}
    \delta_s=-s \mathcal{V}_{234}, \quad \delta_t=t\mathcal{V}_{234}, \quad \delta_{u}=u\mathcal{V}_{234}\,,
\end{equation}
with $\mathcal{V}_{234}=\left(\varepsilon_{2}\cdot {\bf p}_{34}\right)\mathcal{E}_{34}+\left(\varepsilon_{3}\cdot {\bf p}_{42}\right)\mathcal{E}_{42}+\left(\varepsilon_{4}\cdot {\bf p}_{23}\right)\mathcal{E}_{23}$, following the notation in \eqref{eq:4ptcksolution}. 
It is easy to check that the full amplitude in \eqref{eq:4g-stu} is invariant.

The above transformation can be taken as a special case of generalized gauge transformation (GGT) \cite{Bern:2010ue}:
\begin{equation}
\label{eq:4g-generalizedGT}
    N_{s}\shortrightarrow N_s-s \Delta,\qquad  N_{t}\shortrightarrow N_t+t \Delta,\qquad  N_{u}\shortrightarrow N_u+u \Delta\,,
\end{equation}
where $\Delta$ can be an arbitrary function while the four-gluon amplitude still remains unchanged. 
The traditional GT in \eqref{eq:tradGT4pt} corresponds to the special choice $\Delta =\alpha\mathcal{V}_{234}$, but a general $\Delta$ does not necessarily originates from shifting polarization vectors and thus can be defined in more general cases. 

To be more explicit, let us consider embedding the four-point amplitudes into a general loop integrand like Figure~\ref{fig:ckstu}.
The following deformation of the corresponding loop-graph numerators leaves the loop integrand unchanged:
\begin{equation}
\label{eq:JacobiGT}
N_s \rightarrow N_s - l_s^2 \Delta \,, \quad  N_t \rightarrow N_t + l_t^2 \Delta \,, \quad N_u \rightarrow N_u + l_u^2 \Delta \,, 
\end{equation}
with $\Delta$ an arbitrary function, owing to the fact:\footnote{In this subsection, we omit possible loop-integration measure for simplicity.} 
\begin{equation}
    \frac{C_s\Delta_s}{l_s^2 \prod^{\prime}_{m}D_{m}}+\frac{C_t\Delta_t}{l_t^2 \prod^{\prime}_{m}D_{m}}+\frac{C_u\Delta_u}{l_u^2 \prod^{\prime}_{m}D_{m}}\ \propto \ (C_s-C_t-C_u)=0\,,
\end{equation}
where $\Delta_{s,t,u}=l_{s,t,u}^2 \Delta$ and $\prod^{\prime}_{a}$ denotes the product of common propagators other than $l^2_{s,t,u}$.
Such a deformation belongs to generalized gauge transformations at loop level.

\subsubsection*{Traditional GT for form factors and operator-induced GGT}
For form factors, one can conduct a parallel analysis. The simplest tree-level example is the three-point form factor of $\operatorname{tr}(\phi^2)$. It is easy to write down the following Feynman diagrams
\begin{equation}
\begin{aligned}
 \includegraphics[height=0.13\linewidth]{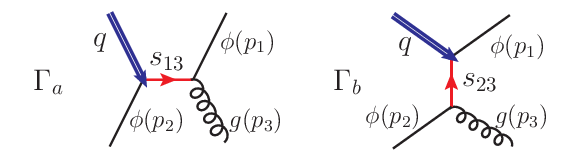}
\end{aligned}\,,
\end{equation}
so that the full-color form factor is
\begin{equation}
	\itbf{F}^{(0)}_{\operatorname{tr}(\phi^2),3}=\left(\frac{C_{s_{13}}N_{s_{13}}}{s_{13}}+\frac{C_{s_{23}}N_{s_{23}}}{s_{23}}\right)\,; \quad N_{s_{13}}=2\varepsilon_3\cdot p_1, \  N_{s_{23}}=-2\varepsilon_3\cdot p_2,
\end{equation} 
and $C_{s_{13}}=C_{s_{23}}=\tilde{f}^{a_1a_2a_3}$. Note that the color factor associated to the $q^2$ vertex is a $\delta$-function in color space. 
Performing a traditional GT similar to \eqref{eq:tragauge}, one has
\begin{equation}
N_{s_{13}}\rightarrow N_{s_{13}} + \alpha s_{13},\quad  N_{s_{23}}\rightarrow N_{s_{23}} - \alpha s_{23},
\end{equation}
which leaves the form factor invariant.

\begin{figure}[t!]
    \centering
    \subfigure[]{
        \includegraphics[width=0.23\linewidth]{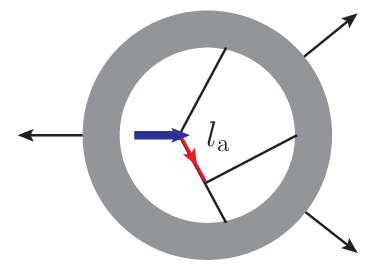}
         \label{fig:cka}
        }
    \centering
    \subfigure[]{
        \includegraphics[width=0.23\linewidth]{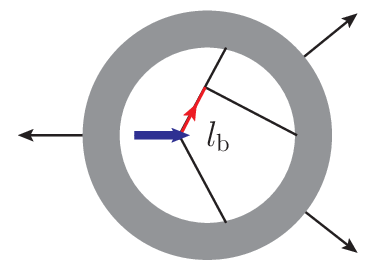}
        \label{fig:ckb}
        }
    \caption{Loop diagrams related by operator-induced relations for length-two operators. }
    \label{fig:ckab}
\end{figure}

As in the amplitude case, there are also generalized versions of GTs.
Let us consider the loop diagrams of form factor of  $\operatorname{tr}(\phi^2)$ operator in Figure~\ref{fig:ckab}. The two diagrams have same color factors $C_{\rm a} = C_{\rm b}$ since the $q^2$ vertex is a $\delta$-function.
One can perform a generalized GT as
\begin{equation}
\label{eq:generalGT2}
N_{\rm a}\rightarrow N_{\rm a} + l_{\rm a}^2 \Delta,\quad  N_{\rm b}\rightarrow N_{\rm b} - l_{\rm b}^2 \Delta\,,
\end{equation}
and it is easy to check that the full-color loop integrand is not altered for arbitrary $\Delta$:
\begin{equation}
\frac{C_{\rm a}l_{\rm a}^2\Delta}{l_{\rm a}^2\prod^{\prime}_{m}D_{m}}+\frac{C_{\rm b}l_{\rm b}^2\Delta}{l_{\rm b}^2\prod^{\prime}_{m}D_{m}} \ \propto \ (C_{\rm a} - C_{\rm b})=0\,.
\end{equation}

We see that for form factors, due to the insertion of local operators, a new type of generalized gauge transformations appears, where the ``operator-hopping transformation'' illustrated by Figure~\ref{fig:ckab} plays a fundamental role that is similar to the ``$s$-, $t$- and $u$-transformation''  in Figure~\ref{fig:ckstu}. Since the two types of generalized gauge transformations (GGTs),  \eqref{eq:JacobiGT} and \eqref{eq:generalGT2}, originate from different types of color relations, we assign different  names to distinguish them: 
\begin{align}
\eqref{eq:JacobiGT} : \qquad & \textrm{\emph{Jacobi-induced} generalized gauge transformation}, \nonumber\\
\eqref{eq:generalGT2} : \qquad & \textrm{\emph{operator-induced} generalized gauge transformation}. \nonumber
\end{align}

\begin{figure}[t]
    \centering
    \subfigure{\includegraphics[width=0.28\linewidth]{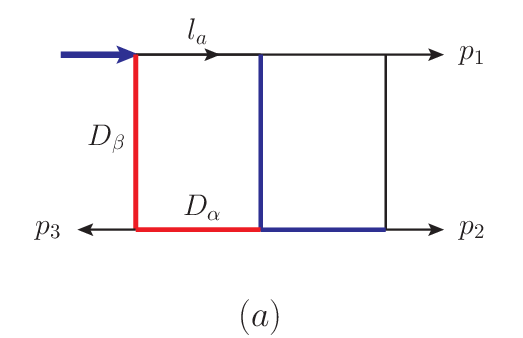}}
    \subfigure{\includegraphics[width=0.7\linewidth]{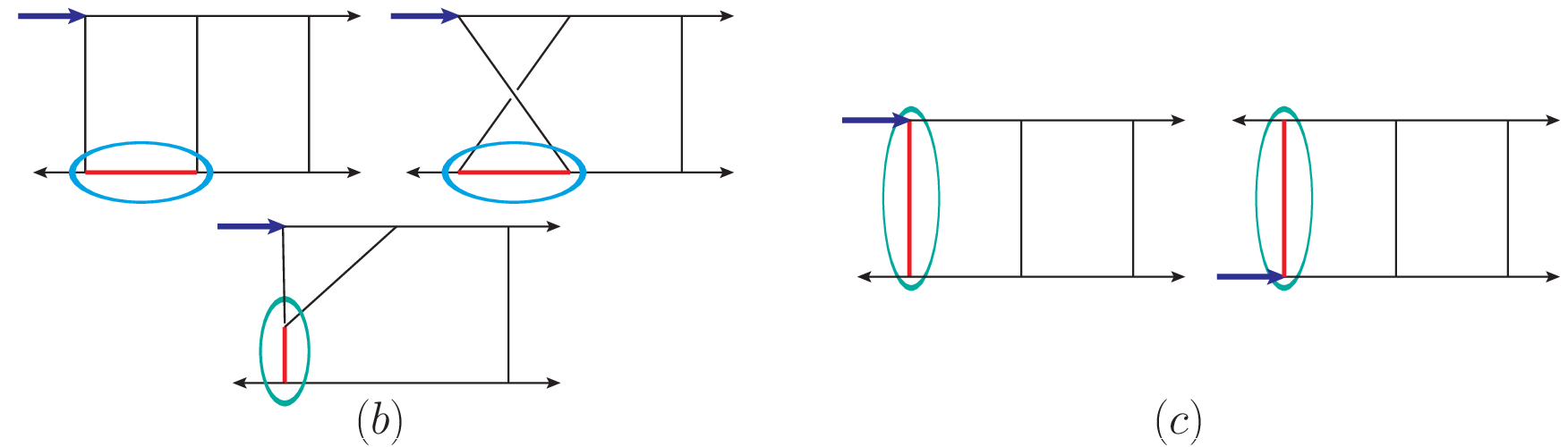}}
    \caption{Examples of generalized gauge transformations of ${\rm tr}(\phi^2)$ form factors. The momenta labelling is shown in Figure~(a). (b) and (c) are two types of generalized gauge transformations involved  in the cancellation of free parameter $c_1$. }
    \label{fig:GTff}
\end{figure}

\subsubsection*{CK-preserving deformations and cancellations of free parameters}

Since there are two types of GGTs for form factors, it is possible to get a large solution space, parametrized by the parameters in integrands, compared with the amplitude cases. The existence of the solution space is due to special CK-preserving deformations, which are formed based on the two types of GGTs.
Below we try to understand the roles played by these GGTs in the CK-preserving deformations and the cancellations of free parameters. 
For simplicity, we will consider the two-loop form factor as an example and the three-loop cases follow the same discussion. 

We first explain how the free parameters can be understood as GGTs. 
Let us consider the two-loop numerator solutions given in Appendix~\ref{ap:ck2loop} and focus on the free parameter $c_1$.
Without loss of generality, consider the numerator $N_4$ with topology shown in Figure~\ref{fig:GTff}(a), and the part proportional to $c_1$ (denoted as $N_4|_{c_1}$) is 
\begin{equation}\label{eq:n4c1}
	N_4\big|_{c_1}=c_1 s_{12}(s_{23}-s_{13})\big((l_a-p_1-p_2)^2-(l_a-q)^2\big)\,.
\end{equation}
One can observe that in $N_4|_{c_1}$, all terms are proportional to certain propagators. We explain now that they can be naturally generated from GGTs. 

To begin with, we notice that a Jacobi-induced GGT is illustrated in Figure~\ref{fig:GTff}(b), and an operator-induced GGT is shown in Figure~\ref{fig:GTff}(c), both of which involve the diagram shown in Figure~\ref{fig:GTff}(a). 
The first Jacobi-induced GGT leads to a deformation of $N_{4}$ proportional to $D_{\alpha}=(l_a-p_1-p_2)^2$ while the second operator-induced GGT gives the $D_{\beta}=(l_a-q)^2$ term. In this example, one can specifically set $\Delta=c_1s_{12}(s_{23}-s_{13})$ in \eqref{eq:JacobiGT} and \eqref{eq:generalGT2}, and the involved propagators are highlighted by red color in Figure~\ref{fig:GTff}. 
Thus, $N_4|_{c_1}$ in \eqref{eq:n4c1} can be put in a form as $(D_{\alpha}-D_{\beta})\times\Delta$, which originates from the two types of GGTs. One can check that the similar $c_1$ terms appear in the numerators of other topologies in Figure~\ref{fig:GTff}(b)--(c) and they cancel when adding together.\footnote{Note that the two diagrams in (c) belong to the same topology but with different external line permutations. They cancel because the factor $s_{12}(s_{23}-s_{13})$ is anti-symmetric when permuting $p_1$ and $p_2$.}

More interestingly, such a deformation also preserves the dual Jacobi relations. For example, one can inspect the $c_1$-relevant part in all $N_i$ given explicitly in Appendix~\ref{ap:ck2loop} and directly check that no dual Jacobi relations are spoiled by $c_i$ deformation. Here we would like to understand this point by studying the special structure of the deformations. 
Consider again the aforementioned $N_{4}|_{c_1}$ given by \eqref{eq:n4c1} as an example. One notes that $N_{4}|_{c_1}\propto (D_{\alpha}-D_{\beta})$ where both $D_{\alpha,\beta}$ are connected to the vertex associated to the massless $p_3$ leg. 
The similar structure also exists for all other $N_i$ in Appendix~\ref{ap:ck2loop}:
focusing for instance on the $c_1$ deformation and the $s_{12}(s_{23}-s_{13})\equiv S_{\rm p,3}$ terms in other numerators, one always finds the structure of $(P^2-(P-p_3)^2)$ where $P^2$ and  $(P-p_3)^2$ are two propagators connected to the $p_3$ leg. 

This structure is important for preserving dual Jacobi relations. 
Consider the dual Jacobi relation related to Figure~\ref{fig:GTff}(b), and 
extract the $s$-, $t$- and $u$-channel sub-diagrams as 
\begin{equation*}
\begin{aligned}
		\includegraphics[width=0.55\linewidth]{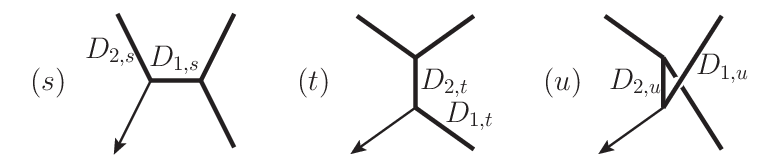}
	\end{aligned} \,.
\end{equation*}
Summing up the $c_1$-related deformation in the dual Jacobi relation, one finds
\begin{equation}\label{eq:jacobic1}
	\left(N_s-N_t-N_u\right)\big|_{c_1} \ \propto \ \left(D_{1,s}-D_{2,s}\right)-\left(D_{1,t}-D_{2,t}\right)-\left(D_{1,u}-D_{2,u}\right)\,.
\end{equation}
Using the on-shell condition of $p_3$ and momentum conservation, \eqref{eq:jacobic1} vanishes identically.

Finally, we comment that the operator-induced GGTs are indispensable in all the cancellations of free parameters for both two-loop and three-loop form factors, because they are involved in the cancellation of any one of the parameters. 
This shows the important role of the operator-induced GGTs for the existence of large CK-dual solution space, which also suggests that the method of constructing form factors via CK duality is promising for higher loops.

\section{Simplifying the Integrand}
\label{sec:simintegrand}

In this section, we implement some simplifications for the loop integrands of form factors so that they will be suitable for numerical evaluations and the cancellation of free parameters is also made manifest.
To begin with,  in Section~\ref{ssec:color}, we express the color factors of trivalent topologies in trace basis of color generators in $\SU{N_c}$, which divides the integrand into $N_c$-leading and $N_c$-subleading contributions.
Next, in Section~\ref{ssec:integrandstrategy}, we simplify the kinematical part of the integrand.
Finally,  in Section~\ref{ssec:dlog}, we discuss how to reorganize the results based on the $d\log$ forms, which are crucial to make the numerical evaluation more efficient in the next section.

\subsection{Color analysis}\label{ssec:color}

At three-loop level, the three-point form factors may have subleading contributions in the large $N_c$ expansion,
which requires a careful analysis on color structures.
Before going on, some terminologies need to be clarified.
The terminology ``planar form factors" refers to the $N_c$-leading component of full-color form factors. In contrast, the ``planar integrals" means strictly that the corresponding Feynman diagrams are planar.\footnote{A diagram is called planar if the diagram can be drawn on a plane with $p_1,p_2,p_3$ and $q$ aligned at infinity and no internal lines crossing with each other.} Similarly, ``non-planar form factors'' is  the $N_c$-subleading  component and should be distinguished from  ``non-planar integrals''. 
Both planar and non-planar form factors receive contributions from planar and non-planar integrals, see examples below. 

Because of the difference between color structures of length-two and length-three operators, we discuss them separately below.

\subsubsection*{Color decomposition for the form factor of ${\rm tr}(\phi^2)$}

At one- and two-loop levels, the three-point form factors of ${\rm tr}(\phi^2)$ in ${\cal N}=4$ SYM have no subleading color corrections, namely, the color factors contain only $N_c$-leading contributions. The reason is that for every trivalent diagram, an explicit calculation shows that its color factor has no $N_c$-subleading part. Therefore, 
for the three-point form factors of ${\rm tr}(\phi^2)$, the color factors can be taken as $N_c^\ell \tilde f^{a_1 a_2 a_3}$, where $\ell=1,2$ for one- and two-loop respectively. 

The $N_c$-subleading contribution starts to appear at three loops.
An explicit calculation of color factors associated with trivalent topologies of the $\operatorname{tr}(\phi^2)$ form factor shows that both color factors $N_c^3\tilde f^{a_1 a_2 a_3}$ and $N_c\tilde f^{a_1 a_2 a_3}$ appear. 
One can divide the $\operatorname{tr}(\phi^2)$ form factor into $N_c$-leading, denoted as PL representing planar, and $N_c$-subleading, denoted as NP representing non-planar, parts  as
\begin{equation}
\begin{aligned}
    \itbf{F}^{(3),\mathrm{PL}}_{\operatorname{tr}(\phi^2),3}&=N_{c}^{3}\tilde f^{a_1 a_2 a_3} \mathcal{F}^{(0)}_{\operatorname{tr}(\phi^2),3}\sum_{\sigma_e} {\sign{\sigma_e} } \sum_{i=1}^{29}  \int \prod_{j=1}^{3}\frac{ \mathrm{d}^{D}l_{j}}{(2\pi i)^{D}} \frac{c^{\text{PL}}_i}{S_i}\sigma_e \cdot \frac{ N_i}{\prod_{\alpha_i}P^2_{\alpha_i}}\,,\\
    \itbf{F}^{(3),\mathrm{NP}}_{\operatorname{tr}(\phi^2),3}&=12 N_{c}\tilde f^{a_1 a_2 a_3} \mathcal{F}^{(0)}_{\operatorname{tr}(\phi^2),3}\sum_{\sigma_e} {\sign{\sigma_e} } \sum_{i=1}^{29}  \int \prod_{j=1}^{3}\frac{ \mathrm{d}^{D}l_{j}}{(2\pi i)^{D}} \frac{c^{\text{NP}}_i}{S_i}\sigma_e \cdot \frac{ N_i}{\prod_{\alpha_i}P^2_{\alpha_i}}\,,
\end{aligned}
\end{equation}
where $c_i^{\text{PL}}$ and $c^{\text{NP}}_{i}$ are pure numbers appearing in the decomposition of color factor $C_{i}=(c_i^{\text{PL}} N_c^{3} + 12 c_i^{\text{NP}} N_c )\tilde f^{a_1 a_2 a_3}$ and 12 is simply a normalization convention. Following the order of 29 diagrams in Figure~\ref{fig:phi2tops}, $c_i^{\text{PL}}$ and $c_i^{\text{NP}}$  can be given explicitly as:
\begin{align}
        c_i^{\text{PL}} &: \ \{-1, -1, -1, -1, 0, 0, 0, -1, 0, 0, 0, 0, 0, 4, -2, 1, 0, 0, -2, -1, 2, 0, -8, -2, -4, 4, -2, 2, 0 \}\,, \nonumber\\
        c_i^{\text{NP}} &: \ \{0, -1, 0, 0, 0, -1, -1, 0, -1, -1, 0, 0, 0, 0, 0, 1, 0, 1, 0, 0, 0, 0, 0, 0, 0, 0, 0, 0, 0 \}\,.
\end{align}
It is interesting to notice that a large number of trivalent graphs have zero color factors, even though their numerators are important in the CK-dual construction.
In particular, there are only seven (top-level) topologies contributing to the $N_c$-subleading part. Besides, some non-planar topologies, such as (11), contribute to the planar form factor, while some planar topologies, such as (2), contribute to the non-planar form factor.

\subsubsection*{Color decomposition for the form factor of ${\rm tr}(\phi^3)$}

Compared with the $\operatorname{tr}(\phi^2)$ case, the color factors in the $\operatorname{tr}(\phi^3)$ case show more novel features. Since there are two inequivalent ways of connecting three internal lines to a single-trace length-three operator, the diagrams can always be grouped in pairs, which are  represented exactly by the summation over $\sigma_i$ in \eqref{eq:phi3res2}. Moreover, for the operator $\operatorname{tr}(\phi^3)$,  color factors of the two diagrams in pair can be summed up to be proportional to $\tilde d^{1 2 3} $, because they share the same kinematics factor. As a result, up to three loops, detailed computations manifest that no $N_c$-subleading contributions emerges. We show in detail why these facts are true.

\begin{figure}[t!]
    \centering
    \subfigure[$\Gamma_{a}$]{ \includegraphics[width=0.25\linewidth]{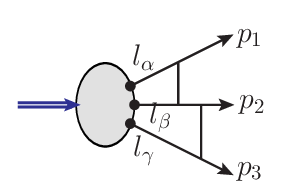}}
     \subfigure[$\Gamma_{b}$]{ \includegraphics[width=0.25\linewidth]{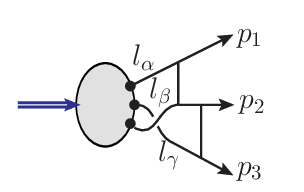}}
    \caption{Grouping diagrams in pair for length-three operators. $\Gamma_{a,b}$ have the same denominator and numerator but the color structures are different. The gray oval represent the trace of the operator.}
    \label{fig:phi3diagraminpair}
\end{figure}

Let us illustrate this feature by an example at two-loop level. Consider the two trivalent diagrams $\Gamma_{a,b} $ in Figure~\ref{fig:phi3diagraminpair}. The operator ${\rm tr}(\phi^3)$ carries a color factor $\operatorname{tr}(T^{\alpha}T^{\beta} T^{\gamma})$, and there are two inequivalent way of their connecting to the loop correction part, as shown by the figure. On one hand, their kinematical parts, \emph{i.e.}~the numerators $N$ and denominators $D$(as products of propagators), are identical:
\begin{equation}
    N(\Gamma_{a})=N(\Gamma_{b}),\qquad  D(\Gamma_{a})=D(\Gamma_{b})\,.
\end{equation}
On the other hand, their color factors $C_{a,b} $ are different and take the form
\begin{equation}\label{eq:phi3cab}
    \begin{aligned}
    C(\Gamma_a) &= \operatorname{tr}(T^{\alpha}T^{\beta} T^{\gamma})\tilde f^{\alpha a_1 d_1}\tilde f^{d_1d_2 \beta}\tilde f^{d_2 a_2 d_3}\tilde f^{d_3 a_3 \gamma}=N_c^2 \operatorname{tr}(T^{a_1}T^{a_2}T^{a_3})\,,\\
    C(\Gamma_b) &= \operatorname{tr}(T^{\alpha}T^{\gamma} T^{\beta})\tilde f^{\alpha a_1 d_1 }\tilde f^{d_1d_2 \gamma}\tilde f^{d_2 a_2 d_3}\tilde f^{d_3 a_3 \beta} =N_c^2 \operatorname{tr}(T^{a_1}T^{a_3}T^{a_2})  \neq C(\Gamma_{a})\,,
\end{aligned}
\end{equation}
Since $\Gamma_{a,b}$ have the same kinematical part, it is meaningful to directly sum up their color factors and get a color factor proportional to $ \tilde d^{a_1 a_2 a_3}$
\begin{equation}
    C(\Gamma_{a})+C(\Gamma_{b})=N_c^2 \tilde d^{a_1 a_2 a_3}\,.
\end{equation}
Similar analyses can be performed for all two-loop diagrams for three-point $\operatorname{tr}(\phi^3)$ form factor such that the color factor of $\itbf{F}^{(2)}_{\operatorname{tr}(\phi^3),3}$ is proportional to $N_c^2 \tilde d^{a_1 a_2 a_3} $.

At three-loop order, one can organize integrals into similar pairs as the above example, and we find all the pairs are summed to be color factors proportional to $N_c^3\tilde d^{a_1 a_2 a_3}$. 
In particular, there is no $N_c$-subleading contribution, which can also be justified by direct computations.\footnote{Interestingly, at three-loop level, color factors similar to $C_{a,b}$ in \eqref{eq:phi3cab} contain $N_c$-subleading parts, such as $N_c \tilde{f}^{a_1 a_2 a_3}$. Only after summing up $C_{a}$ and $C_{b}$ does the subleading parts cancel.}
Therefore, the full form factor take the form:\footnote{Note that in \eqref{eq:phi3res2}, the summation over $\sigma$ includes $\sigma_i\in S_3/\mathbb{Z}_3$. Here we are only left with $\sigma_e \in S_3$, since the summation on $\sigma_i$ has been taken into account in organizing integrals in pairs to get the $N_c^3 {\tilde d}^{a_1 a_2 a_3}$ color factor.}
\begin{equation}
    \itbf{F}^{(3)}_{\operatorname{tr}(\phi^3),3}=N_c^{3}\tilde d^{a_1 a_2 a_3} \sum_{\sigma_e} \sum_{i=1}^{26}  \int \prod_{j=1}^{3}\frac{ \mathrm{d}^{D}l_{j}}{(2\pi i)^{D}} \frac{c_i}{S_i}\sigma_e \cdot \frac{ N_i}{\prod_{\alpha_i}P^2_{\alpha_i}}\,,
\end{equation}
where $c_i$ are pure numbers in $C_i = c_i N_c^{3}\tilde d^{a_1 a_2 a_3}$ and can be given according to the order of Figure~\ref{fig:phi3tops} as 
\begin{equation}
    c_i: \ \{0,-1,1,-1,1,0,0,0,-1,0,1,0,0,0,0,0,0,0,0,0,1,-1,0,0,0,0\}\,.
\end{equation}
Again, there are a large number of diagrams with vanishing color factors.

With the analyses on color structures above, we can then focus on the kinematical parts to perform further simplification.

\subsection{Strategy for simplifying the kinematics part}\label{ssec:integrandstrategy}

For both leading and subleading contributions,  we get a schematic form of the kinematics part (after stripping off the color factor) as:
\begin{equation}\label{eq:kinematics_part_of_integrand}
    \sum_{\sigma_{e}} \sigma_{e}\cdot \sum_{\Gamma_i} \int \prod_{j=1}^{3} \frac{ \mathrm{d}^{D}l_{j}}{(2\pi i)^{D}} \frac{v_{i} N_i}{\prod_{\alpha_i} P^2_{\alpha_i}}\,,
\end{equation}
where $\sigma_e\in S_3 $ permutes three external on-shell legs, $v_{i}$ are some numbers arising from color and symmetry factors, and $N_i$ here still represent the numerators obtained from CK duality that contain free parameters. To check whether these parameters indeed drop out of final results as well as to get a simple form for integral evaluation, we need to further simplify \eqref{eq:kinematics_part_of_integrand}, which is the goal of this subsection.

\subsubsection*{General procedures}

To be more specific, the goal of the simplification is to rewrite \eqref{eq:kinematics_part_of_integrand} into the following form
\begin{equation}\label{eq:sim-target}
  \eqref{eq:kinematics_part_of_integrand}=\sum_{\sigma_{e}} \sigma_{e}\cdot \sum_{\Gamma} \int \prod_{j=1}^{3} \frac{ \mathrm{d}^{D}l_{j}}{(2\pi i)^{D}} \frac{\mathbb{N}(\Gamma)}{\mathbb{D}(\Gamma)}\,.
\end{equation}
The major specialty on the RHS is that the diagrams $\Gamma$ include not only top-level topologies (as in the CK-representation), but also sub-topologies obtained by shrinking propagators of the top-level topologies. 
$\mathbb{N}(\Gamma)$ is a polynomial of irreducible numerators of $\Gamma$, and  $\mathbb{D}(\Gamma)$ denotes the product of all the propagators. 
In this way, we expand \eqref{eq:kinematics_part_of_integrand} on a  \textsl{linear independent basis} of the integrand. 
 
To achieve this, we start from \eqref{eq:kinematics_part_of_integrand} which contains only top-level topologies. The numerators are obtained from CK duality, thus
they in general contain both irreducible numerators and (reducible) propagators. 
The part consisting solely of irreducible numerators can be directly mapped to the RHS of \eqref{eq:sim-target}, 
while in the remaining part we can shrink propagators and generate sub-topologies. 
By performing similar operations iteratively for sub-topologies, one can realize the simplification.

Below we take a three-loop integral from  the $\operatorname{tr}(\phi^3)$ form factor as an example to illustrate this procedure. 
Consider a $9$-propagator topology $\Gamma_{9:9}$
\begin{equation}
\begin{aligned}
    \Gamma_{9:9}&: \begin{aligned}
        \includegraphics[width=0.22 \linewidth]{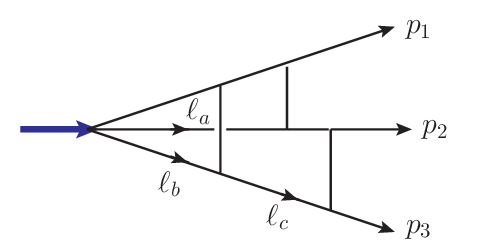}
    \end{aligned}\,.
\end{aligned}
\end{equation}
For $\Gamma_{r:i}$, the subscript $r$ refers to an $r$-propagator topology, and $i$ is a labelling (the enumeration for $9$-propagator topologies is the same as in Figure~\ref{fig:phi3tops}). For example, $\Gamma_{9:9}$ is the ninth topology with 9 propagators. In Appendix~\ref{ap:integrand} we also use such a notation.
 
We first introduce a basis of Lorentz products for topology $\Gamma_{9:9}$:
\begin{align}\label{eq:gamma99basis}
    \nonumber
    \{& \ell_a^{2},\ell_{b}^2,\ell_c^2,(\ell_b-\ell_c)^2,(\ell_c-p_3)^2 , (\ell_c-p_2-p_3)^2,(\ell_a+\ell_c-p_2-p_3)^2,(\ell_a+\ell_b-q)^2,(\ell_a+\ell_c-q)^2,\\
    &(\ell_a-q)^2,(\ell_c-q)^2,(\ell_a+\ell_c-p_3)^2, (\ell_b-p_3)^2,  (\ell_b-p_2-p_3)^2,(\ell_b+\ell_a-p_2-p_3)^2\}\,, 
\end{align}
where the first line gives the nine propagators of $\Gamma_{9:9}$, and the second line includes the irreducible numerators.
Given this basis, we can divide the numerator of $\Gamma_{9:9} $ into two parts.
One part consists only of irreducible numerators, which will be denoted as $\mathbb{N}(\Gamma_{9:9})$, or simply $\mathbb{N}_{9:9}$.
The other part of the numerator is composed of terms proportional to propagators, and they lead to sub-topologies with less propagators and thus contribute to numerators  $\mathbb{N}(\Gamma_{r \leq 8:i})$.
For example, a numerator $\ell_c^2(\ell_a-q)^2$ can shrink a propagator and generate an 8-propagator sub-topology:
\begin{equation}\label{eq:shrinkexample1}
\begin{aligned}
        \includegraphics[width=0.2\linewidth]{figure/Gamma99.eps}
    \end{aligned} \times \ell_c^2(\ell_a-q)^2
    \quad \xrightarrow{\rm shrink} \quad
\begin{aligned}
        \includegraphics[width=0.2\linewidth]{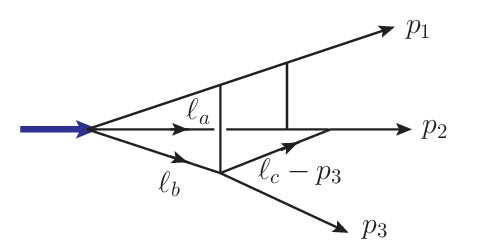}
    \end{aligned} \times (\ell_a-q)^2\,.
\end{equation}

Next we can go on to compute the numerators $\mathbb{N}(\Gamma_{r \leq 8:i})$ for sub-topologies. 
Since the same sub-topology may originate from different top-level topologies,
one needs to carefully take all possible contributions into account.
For example, another top-level topology $\Gamma_{9:11}$ generates an 8-propagator topology which is isomorphic to the one in \eqref{eq:shrinkexample1}:
\begin{equation}\label{eq:shrinkexample2}
\begin{aligned}
        \includegraphics[width=0.2\linewidth]{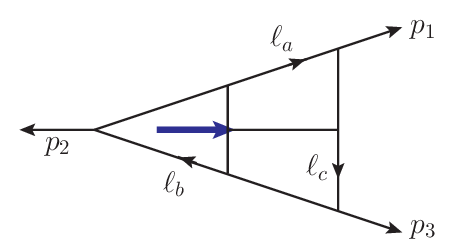}
    \end{aligned} \hskip -7pt \times (\ell_c-p_3)^2 (\ell_a\cdot p_2)
    \quad \xrightarrow{\rm shrink} 
\begin{aligned} \quad
        \includegraphics[width=0.2\linewidth]{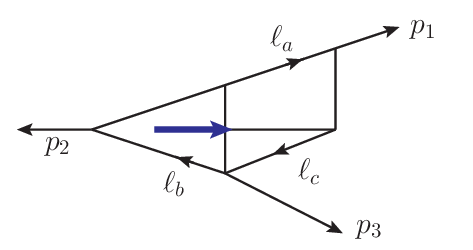}
    \end{aligned}\hskip -8pt \times (\ell_a\cdot p_2)\,.
\end{equation}
The numerators of sub-topologies, such as \eqref{eq:shrinkexample1} and \eqref{eq:shrinkexample2}, in general can have different momentum labelings. To combine them together, it is necessary to choose a uniform set of momentum labeling.
For example, choosing the momentum basis according to \eqref{eq:shrinkexample1}, one can map \eqref{eq:shrinkexample2} as
\begin{equation}
\begin{aligned}
        \includegraphics[width=0.2\linewidth]{figure/Gamma911s.eps}
    \end{aligned}\hskip -7pt \times (\ell_a\cdot p_2)
    \quad \xrightarrow{\rm map}  \quad
\begin{aligned}
        \includegraphics[width=0.2\linewidth]{figure/Gamma99s.eps}
    \end{aligned} \hskip -4pt \times (\ell_a+\ell_c-p_2-p_3)\cdot p_2 
    \label{eq:shrinkexample4}\,.
\end{equation} 
In this way, one can combine all contributions to the same 8-propagator topology. 

Now one can regard the 8-propagator topology as the ``top-topology'' and perform a simplification that is totally parallel to the operation above: (1) choose a set of Lorentz product basis, (2) divide the numerator into an irreducible part $\mathbb{N}(\Gamma_{8:i})$ and another part with terms proportional to propagators, and (3) shrink the latter part to sub-topologies and then combine the contributions to the numerator of the same topology together.
Such a procedure can be repeated iteratively, until one reaches the sub-topologies with minimal number of propagators.

Through the above procedures, one can simplify \eqref{eq:kinematics_part_of_integrand} into the form of \eqref{eq:sim-target}. 
Here we point out that there is a refinement about the choice of  irreducible numerators, which we explain in more detail below.
We also remark that the free parameters should cancel in such a simplification, which can be verified by explicit computations.

\subsubsection*{Select irreducible numerators} 

In the above procedure, we frequently encounter selecting a set of  irreducible numerators for a given topology. 
Different choices of the basis will lead to different final expressions (which are of course equivalent to each other). Here we briefly explain a few rules we use on selecting irreducible numerators:
\begin{enumerate}
\item For planar topologies, use zone variables to identify the set of irreducible numerators;
    \item For non-planar topologies, choose irreducible numerators 
    based on rung-rule \cite{Bern:1997nh};
    \item Symmetrize irreducible numerators if the topology bears some symmetries.\footnote{Such a requirement is consistent with our target form in \eqref{eq:sim-target}, where the summation over different permutations acts only on external legs.}
\end{enumerate}

\begin{figure}[t!]
    \centering
    \includegraphics[width=0.75\linewidth]{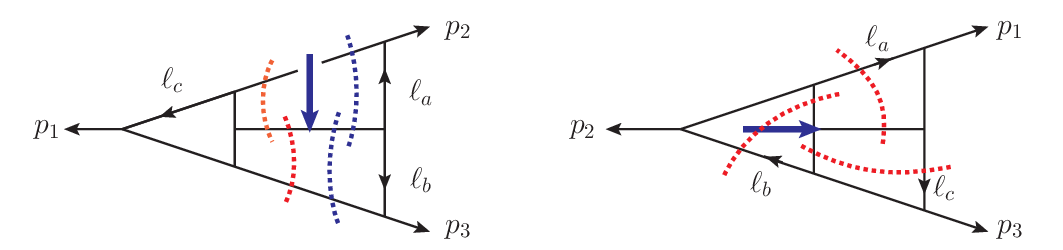}
    \caption{Irreducible-numerator candidates for two topologies. The dashed curves basically refers to rung-rule numerators. Note that the two blue lines in the left diagram refers to one (anti-symmetric) basis, and we only list half of the basis for the second diagram.}
    \label{fig:chooseEffPropCandidatesExample1}
\end{figure}

We use two examples to briefly illustrate these principles.
The first example is shown on the left of Figure~\ref{fig:chooseEffPropCandidatesExample1}. 
Considering rung-rule, some of the irreducible-numerator candidates can be picked out as (indicated by the dashed curves in the figure)
\begin{equation}\label{eq:effpropcandidate}
    (p_1+p_3-\ell_b)^2, \quad (p_1+p_2-\ell_a)^2, \quad (\ell_b+p_2)^2, \quad (\ell_a+p_3)^2 .
\end{equation}
These candidates together with the set of propagators are linearly dependent.
Thus, we may choose the following independent irreducible numerators as
\begin{equation}
    (p_1+p_3-\ell_b)^2, \quad (p_1+p_2-\ell_a)^2, \quad (\ell_b+p_2)^2-(\ell_a+p_3)^2\,,
\end{equation}
which also reflect the symmetry of the topology.\footnote{Note that we choose $(\ell_b+p_2)^2-(\ell_a+p_3)^2$ but not $(\ell_b+p_2)^2+(\ell_a+p_3)^2$  because the latter is not linearly independent with others.}

The second example is shown by the right figure in Figure~\ref{fig:chooseEffPropCandidatesExample1}. It appears as a trivalent topology in the $\operatorname{tr}(\phi^3)$ case and also as a sub-topology in the $\operatorname{tr}(\phi^2)$ case. Irreducible numerators can be chosen based on rung-rule as 
\begin{equation}
    (\ell_c-p_2-p_3)^2,\ (\ell_a+p_2)^2,\ (\ell_a-p_1-p_3)^2,\ (\ell_b+p_3)^2,\ (\ell_b-p_1-p_2)^2,\ (\ell_c+p_1)^2,
\end{equation}
which are also (partly) shown in Figure~\ref{fig:chooseEffPropCandidatesExample1}.
Note that these six candidates transform among each other under cyclic and reflection symmetries so that we do not need to spend extra effort symmetrizing them.

\subsubsection*{Simplified integrand result}\label{ssec:integrandresult}

Using the strategy described above, we obtain the simplified integrands summarized below.

For the form factor of $\operatorname{tr}(\phi^2)$, 
there are 32 topologies contributing to the $N_c$-leading part and 24 topologies to the $N_c$-subleading part. 
The explicit topologies and numerators are listed in Table~\ref{tab:integrandphi2Gp1}--\ref{tab:integrandphi2np} in Appendix~\ref{ap:integrand}, and the full-color form factor can be given as
\begin{equation}
\label{eq:simpInt2}
    \itbf{F}^{(3)}_{\operatorname{tr}(\phi^2),3}=\mathcal{F}^{(0)}_{\operatorname{tr}(\phi^2),3} \tilde f^{a_1 a_2 a_3} \big[ N_c^{3} \, \mathcal{I}^{(3),\pl}_{\operatorname{tr}(\phi^2)} + 12 N_c \,  \mathcal{I}^{(3),\np}_{\operatorname{tr}(\phi^2)} \big] \,,
\end{equation}
where
\begin{subequations}
    \begin{align}
        \mathcal{I}^{(3),\pl}_{\operatorname{tr}(\phi^2)} &= \sum_{\sigma_{e}\in S_3}\sigma_e\cdot \sum_{r=0}^{2}\sum_{j} \ \mathcal{I}(\Gamma^{\pl}_{(10-r):j},\mathbb{N}^{\pl}_{(10-r):j})\,, 
        \label{eq:simpInt2pl}\\
        \mathcal{I}^{(3),\np}_{\operatorname{tr}(\phi^2)} &= \sum_{\sigma_{e}\in S_3}\sigma_e\cdot \sum_{r=0}^{2}\sum_{j} \ \mathcal{I}(\Gamma^{\np}_{(10-r):j},\mathbb{N}^{\np}_{(10-r):j})\,.\label{eq:simpInt2np}
    \end{align}
\end{subequations}
The $\sigma_e$ permutes  three external momenta, and $\mathcal{I}(\Gamma,\mathbb{N})$ refers to a loop integral with denominator defined by $\Gamma$ and numerator given by $\mathbb{N}$. 
The numerators all take simple forms. We can also check that all the planar integrals are dual conformal invariant (DCI) \cite{Drummond:2006rz} for form factors \cite{Bork:2010wf,Brandhuber:2014ica}.

For $\operatorname{tr}(\phi^3)$, there are 20 distinct topologies contributing to the final form factor. The numerators are very simple and all the planar integrals are also DCI.
Topologies and numerators are listed in Table~\ref{tab:integrandphi3} in Appendix~\ref{ap:integrand}, and we summarize the result as 
\begin{equation}
\label{eq:simpInt3}
    \itbf{F}^{(3)}_{\operatorname{tr}(\phi^3),3}=N_c^{3}\tilde d^{a_1 a_2 a_3} \mathcal{I}^{(3)}_{\operatorname{tr}(\phi^3)}, \qquad \mathcal{I}^{(3)}_{\operatorname{tr}(\phi^3)} =  \sum_{\sigma_{e}\in S_3}\ \sigma_e \cdot \sum_{r=0}^{2}\sum_{j} \mathcal{I}(\Gamma_{(9-r):j},\mathbb{N}_{(9-r):j})\,.
\end{equation}

\subsection{$d\log$ form and an alternative organization of  the integrand}
\label{ssec:dlog}

In the last subsection, we have obtained the integrands in a compact form, which should be convenient for numerical evaluations. However, for the three-loop integrals, in particular for some non-planar topologies, such a computation turns out to be still very challenging.  
To overcome this problem, we observe that some improved basis of integrals can substantially improve the efficiency of the calculations, and we use them to further reorganize the integrands.

As an illuminating example, let us consider the following two integrals of the same topology|a scalar integral $I_1$ and a rank-two tensor integral $I_2$ as
\begin{equation}\label{eq:i1i2}
    I_1=\begin{aligned}
     \includegraphics[width=0.2\linewidth]{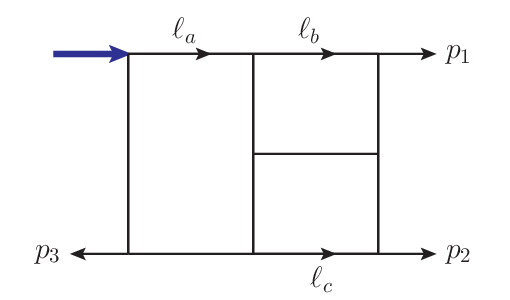}
    \end{aligned} \hspace{-7pt} \times 1, \qquad 
    I_2=\begin{aligned}
     \includegraphics[width=0.2\linewidth]{figure/plint1.eps}
    \end{aligned}  \hspace{-7pt}  \times {\big[ (\ell_a-p_1)^2 \big]^2}\,.
\end{equation}
Naively, one may expect the scalar integral $I_1$ to be much simpler than the rank-two tensor integral $I_2$. However, a test computation using FIESTA 4.2 \cite{Smirnov:2015mct} shows that the integration time of $I_1$ is more than 8000 seconds, while the time for $I_2$ is less than 300 seconds, almost 30 times faster than the scalar one.\footnote{In this example we have used 12 CPU cores (2.7GHz) in the HPC Cluster of ITP-CAS and the Vegas integrator with its default precision in FIESTA 4.2.} The setup and more details about the numerical evaluation will be given in Section~\ref{ssec:numerical}.

The speciality of ${I}_2$ is that its integrand can be written as a special $d\log$ form which presents simple pole structures. 
A $d\log$ integral refers to a loop integral whose integrand $\Omega$ (with measure) can be written as the following $d\log$ form \cite{Arkani-Hamed:2014via, Bern:2014kca}:
\begin{equation}\label{eq:dlog}
\Omega=\sum_{k} c_{k} \, d \log g_{1}^{(k)} \wedge d \log g_{2}^{(k)} \wedge \ldots \wedge d \log g_{n}^{(k)}\,,
\end{equation}
where $g_{i}^{(k)}$ are functions of both loop and external momenta while $c_{k}$ are called leading singularities independent of loop (integration) variables. In this paper, the leading singularities $c_{k}$ are not necessarily pure numbers but simple (rational) function of Mandelstam variables, which are sometimes denoted as mixed leading singularities, see \emph{e.g.}~\cite{Wasser:2018qvj}.

It is non-trivial that the integrand of an integral can take the $d\log$ form \eqref{eq:dlog}|it needs to have certain special numerator, for example, the numerator $[(\ell_a-p_1)^2]^2$ in \eqref{eq:i1i2}.
Searching these numerators, nevertheless, is in general not a trivial task. Below we briefly explain the strategies employed in this paper. 

A straightforward way is to parametrize the loop momenta with proper parametrization and try to re-express the loop integrand into the $d\log$ form in \eqref{eq:dlog}. 
Such an algorithm and codes (the Mathematica package $\texttt{DlogBasis}$) have been developed in \cite{Wasser:2018qvj, Henn:2020lye}. The basic strategy is that one can make an ansatz for the numerator and fix the parameters by requiring that a $d\log$ form can be made; relevant operations like partial fractions and rationalizing square roots can be automatically carried out in the package $\texttt{DlogBasis}$. For the purpose of the current paper, we will apply some simple ansatz with power-counting constraints (such that loop-momentum power is no more than $[(\ell+p)^2]^2$) and also allow mixed singularities. These conditions are practically useful when dealing with complicated topologies.  

In many situations, another  convenient way to get the $d\log$ numerator is to apply the unitarity-cut picture.
The idea of the cut-based method is to separate some well-studied parts of the integrals, such as one-loop sub-box, via unitarity cuts, and thus to transform the problem to a simpler one \cite{Bern:2014kca}. 
As an example, let us consider the three-loop integral $I_{11}$ in Appendix~\ref{ap:ut} with numerators
\begin{equation}
\label{eq:dlogN11}
  N^{d\log}_{11}= s_{12}^2s_{23}\big[ (\ell_a-p_1)^2-(\ell_a-\ell_b)^2-(\ell_b-p_1)^2-(\ell_a-p_1-p_2)^2+(\ell_b-p_1-p_2)^2\big] \,.
\end{equation}
Under the double cut $\ell_a^2=(\ell_a-p_1-p_2)^2=0$, the three-loop integral factorizes as the product of two integrals: a one-loop box and a two-loop non-planar box. The $d\log$ numerator \eqref{eq:dlogN11}, together with the Jacobian factor $(\ell_a-p_1)^2$ from the cut,\footnote{For more details, see \cite{Bern:2014kca}.} also factorizes as the product of $s_{12}s_{23}$ and $s_{12}(\ell_a-p_1)^2 \big[ (\ell_a-p_1)^2-(\ell_a-\ell_b)^2-(\ell_b-p_1)^2+(\ell_b-p_1-p_2)^2\big]$, which are precisely the $d\log$ numerators of the one-loop box integral and the two-loop non-planar box integral respectively. 
Conversely, starting from the simpler building blocks, one may reconstruct the three-loop $d\log$ numerators.  
 
By combining the above strategies, we get the $d\log$ integrals listed in Table~\ref{tab:utintlist} in Appendix~\ref{ap:ut}. Topics related to integrations will be covered in Section~\ref{ssec:numerical}, and some remarks and details will also be given in Appendix~\ref{ap:ut}.

\section{IR exponentiation and finite remainder functions}\label{sec:ir}

In this section, we consider the integration of three-loop form factors and study their infrared divergences and finite remainders. 
We first review the (planar and non-planar) IR structures and finite remainders in Section~\ref{ssec:ir}. 
Then we describe the numerical integration in Section~\ref{ssec:numerical}, where the properties of the integrated form factors are also discussed. Some technical details are included in Appendix~\ref{ap:ut} and \ref{ap:ir}. 

\subsection{Structure of IR divergences and finite remainders}\label{ssec:ir}
Amplitudes and form factors in massless gauge theories contain IR divergences that originate from phase space configurations where loop momenta become soft or collinear.
The BPS form factors have only IR divergence because of the protection of supersymmetry. 
The structure of IR singularities, coming from long-distance dynamics, is universal and depends only on the external on-shell legs but not on the local operators. Therefore, the match of IR-divergence structures is a very strong check of the correctness of our results. 

Before considering the full-color structure, we first review the planar (\emph{i.e.}~$N_c$-leading) contribution.
In this case, due to the planarity all the internal lines are confined to the wedges formed by two adjacent hard lines, and the IR structure has a relatively simple form as (see \emph{e.g.}~\cite{Sterman:2002qn, Bern:2005iz})
\begin{equation}
\label{eq:planarlog}
\log {\mathcal{I}_n}=-\sum_{\ell=1}^{\infty} g^{2 \ell}\left[\frac{\gamma_{\mathrm{cusp}}^{(\ell)}}{( \ell \epsilon)^{2}}+\frac{\mathcal{G}_{\mathrm{coll}}^{(\ell)}}{ \ell \epsilon}\right] \sum_{i=1}^{n}\left(-s_{i i+1}\right)^{-\ell \epsilon}+{O}\left(\epsilon^{0}\right) ,
\end{equation}
where $\mathcal{I}_n=\mathcal{F}_n/\mathcal{F}_n^{(0)}$ with a loop expansion $\mathcal{I}_n=1+\sum_{\ell=1}^{\infty} g^{2\ell}\mathcal{I}_n^{(\ell)}$, $\gamma_{\rm cusp}$ is the cusp anomalous dimension \cite{Korchemsky:1985xj, Korchemsky:1988si}
and $\mathcal{G}_{\rm coll}$ is the collinear anomalous dimension (see \emph{e.g.}~\cite{Cachazo:2007ad}). These anomalous dimensions can be fixed by the calculation of Sudakov form factors \cite{Mueller:1979ih,Collins:1980ih,Sen:1981sd, Magnea:1990zb}. 

A convenient way to rewrite \eqref{eq:planarlog} is to express the $\epsilon^{-2,-1}$ poles on the RHS with one-loop corrections. In ${\cal N}=4$ SYM, this is given by the so-called BDS ansatz \cite{Bern:2005iz} (see also \cite{Anastasiou:2003kj}):
\begin{equation}
\label{eq:bds}
\log {\mathcal{I}_n}=\sum_{\ell=1}^{\infty} g^{2 \ell}\left(f^{(\ell)}(\epsilon)\mathcal{I}_n^{(1)}(\ell \epsilon)+C^{(\ell)}+\mathcal{R}_n^{(\ell)}+O(\epsilon^{1})\right),
\end{equation}
where $f^{(\ell)}(\epsilon)$ takes the form 
\begin{equation}
\label{eq:fell-form}
f^{(\ell)}(\epsilon) = a_{0}^{(\ell)}+a_{1}^{(\ell)}\epsilon+a_{2}^{(\ell)}\epsilon^{2} \,,
\end{equation}
$C^{(\ell)}$ is a pure number, and $\mathcal{R}_n^{(\ell)}$ is the $\ell$-loop finite remainder function which has a nice behavior of ${\cal R}_n^{(\ell)} \rightarrow {\cal R}_{n-1}^{(\ell)}$ in the collinear limit (more details will be given later). 
The $a_0^{(\ell)}$ and $a_1^{(\ell)}$ in \eqref{eq:fell-form} are completely fixed by the cusp and collinear anomalous dimensions in \eqref{eq:planarlog}, while $a_2^{(\ell)}$ and $C^{(\ell)}$ are determined by collinear properties of $\mathcal{R}_n^{(\ell)}$, as will be discussed shortly.

To generalize the planar picture to full color,  one encounters more complicated structures because of the entanglement between color and space-time degrees of freedom.  
At two-loop order, a useful factorization form was introduced by Catani \cite{Catani:1998bh} in the late 90s. And generalized higher-loop structures were conjectured based on the \emph{dipole-formula} in \cite{Becher:2009cu, Gardi:2009qi}. 
Later, concrete three-loop computations revealed that the {dipole-formula} is not complete and new non-dipole contributions begin to appear at three-loop order \cite{Almelid:2015jia,Gardi:2016ttq,Almelid:2017qju}.

Specifically, a convenient way to represent the full-color IR singularity structure is to use the fact that IR divergences can be mapped to ultraviolet (UV) divergences of Wilson loops  \cite{Korchemsky:1985xj,Korchemsky:1987wg}. Such an IR-UV correspondence leads to a factorization formula of IR divergences in a way similar to the renormalization constant as  
\begin{equation}\label{eq:fullcolorir}
    \itbf{F}(p_i,a_i,\epsilon)={\itbf{Z}}(p_i, \epsilon)\itbf{F}^{\rm fin}(p_i,a_i,\epsilon)\,,
\end{equation}
where the ${\itbf{Z}}$ factor captures all IR divergences,  and $\itbf{F}^{\rm fin}$ denotes the finite ``hard" form factor. 
Note that the form factor $\itbf{F}$ is regarded as a tensor in color space carrying indices $a_i$ and ${\itbf{Z}}$ plays the role of an operator in color space. 

The ${\itbf{Z}}$ factor originated from the Wilson line calculation can be written as a solution of a renormalization-group equation as \cite{Magnea:1990zb}
\begin{equation}\label{eq:zfactorwilson}
    {\itbf{Z}}(p_i,\epsilon)=\mathcal{P} \exp\left\{-\frac{1}{2}\int_{0}^{\mu^2}\frac{\mathrm{d}\lambda^2}{\lambda^2}\boldsymbol{\Gamma}(p_i,\lambda,\bar\alpha_{\rm s}(\lambda^2))\right\}\,,
\end{equation}
where $\mathcal{P}$ is the path ordering for Wilson lines and $\boldsymbol{\Gamma}$ is the soft anomalous dimension matrix. 
One can divide $\boldsymbol{\Gamma}$ into dipole and non-dipole terms  \cite{Almelid:2015jia}, where the dipole terms contribute to the planar IR divergences mentioned above and involve two external legs at the same time, and non-dipole terms are the remaining parts which are due to multi-leg and non-planar corrections. 

For $\mathcal{N}=4$ SYM in $4-2\epsilon$ dimension, the ${\itbf{Z}}$ factor can be given in a more explicit form as (see \emph{e.g.}~\cite{Henn:2016jdu})
\begin{equation}\label{eq:zfactor}
    \itbf{Z}(p_i,\epsilon)=\mathcal{P} \exp\left\{\sum_{\ell=1}^{\infty}g^{2\ell}\left[\frac{\gamma_{\rm cusp}^{(\ell)}}{(\ell\epsilon)^2}\mathbf{D}_0-\frac{\gamma_{\rm cusp}^{(\ell)}}{\ell \epsilon}\mathbf{D}-n\frac{\mathcal{G}_{\rm coll}^{(\ell)}}{\ell \epsilon}\mathbf{1}+\frac{1}{\ell \epsilon} \mathbf{\Delta}^{(\ell)}\right]\right\}\,,
\end{equation}
where we keep the $\mathcal{P}$ to stress the one should keep the order when computing the exponential color operators, since the operators in color space do not commute in general.
$\mathbf{1}$ is the identity operator in color space, and the explicit expressions of operators $\mathbf{D}_0$ and $\mathbf{D}$ in color space are given by\footnote{Note that there is a $1/N_c$ factor in the definition, which is consistent with the our convention of anomalous dimensions, see \eqref{eq:defCuAD}-\eqref{eq:defCoAD} in Appendix \ref{ap:ir} for a detailed definition.}
\begin{equation}\label{eq:D0Ddef}
    \mathbf{D}_0=\frac{1}{N_c}\sum_{i \neq j}\mathbf{T}_i\cdot\mathbf{T}_j, \qquad \mathbf{D}=\frac{1}{N_c}\sum_{i \neq j}\mathbf{T}_i\cdot \mathbf{T}_j \log\left(\frac{-s_{ij}}{\mu^2}\right)\,,
\end{equation}
where the action of $\mathbf{T}_i^{a}$ goes as $\mathbf{T}_i^{a}T^{a_i}=-i {f}^{a a_i \mathrm{x}}T^{\rm x}$. The first three terms in \eqref{eq:zfactor} correspond to the dipole terms. And the last $\mathbf{\Delta}^{(\ell)}$ describes the term beyond the dipole form|it involves three or more particles at the same time. $\mathbf{\Delta}^{(\ell)}$ vanishes when  $\ell=1,2$. Starting at three-loops, the $\mathbf{\Delta}^{(\ell)}$ encodes non-trivial contributions that can only have kinematics dependence on the cross ratios.

In our consideration of three-point form factors, it is impossible to write down a cross ratio because there are only three external on-shell legs. However, there is still a non-trivial kinematics-independent non-dipole correction given as \cite{Almelid:2015jia}
\begin{equation}\label{eq:3loopdelta}
    \mathbf{\Delta}_{3}^{(3)}=-8 (\zeta_5+2\zeta_2\zeta_3) \sum_{i}\sum_{\substack{ j<k\\j,k\neq i}}  {f}^{abe}{f}^{cde} (\mathbf{T}^{a}_i\mathbf{T}^{d}_i+\mathbf{T}^{d}_i\mathbf{T}^{a}_i)\mathbf{T}^{b}_{j}\mathbf{T}^{c}_{k}\,.
\end{equation}
A more detailed derivation from \eqref{eq:zfactorwilson} to \eqref{eq:3loopdelta} is given in Appendix~\ref{ap:ir}.

We briefly comment on the finite remainders in the end. The planar BDS form should be consistent with the full-color IR divergence in \eqref{eq:fullcolorir} at $N_c$-leading order.  The definitions of finite remainders, however, can be different in these two IR-subtraction schemes. Of course, they can be easily translated from one another via a scheme change.

With the above general discussions, we consider below concrete applications at both two and three loops.

\subsubsection*{Two-loop form factors (a review)}

We first consider the two-loop $\operatorname{tr}(\phi^2)$ form factor as a warm-up example.
We begin with the BDS form in the planar limit. 
Expanding \eqref{eq:bds} to the $g^4$ order, one gets \cite{Brandhuber:2012vm}
\begin{equation}
\label{eq:bds-2loop}
{\mathcal{I}}_n^{(2)}(\epsilon)-\frac{1}{2}\left({\mathcal{I}}^{(1)}_n(\epsilon)\right)^{2}
= f^{(2)}(\epsilon) {\mathcal{I}}_n^{(1)}(2 \epsilon)+ C^{(2)} + \mathcal{R}_n^{(2)}+O(\epsilon)\,,
\end{equation}
where
\begin{equation}
f^{(2)}(\epsilon)= - 2 \big( \zeta_2 + \zeta_3 \epsilon + \zeta_4 \epsilon^2 \big),
\qquad C^{(2)} = 4 \zeta_4\,.
\end{equation}
The terms $\zeta_2+\zeta_3\epsilon$ in $f^{(2)}$ are determined by the two-loop cusp and collinear anomalous dimensions.
The term $\zeta_{4}\epsilon^2$ in $f^{(2)}$ and the constant $C^{(2)}$  can be fixed by requiring the remainder to satisfy the collinear limit behavior
\begin{equation}\label{eq:twoloopcollinear}
{\cal R}_n^{(2)}  \ \xlongrightarrow[\mbox{}]{\mbox{$p_i \parallel p_{i+1}$}} \ {\cal R}_{n-1}^{(2)} \,.
\end{equation}
Specifically, for the two-point (\emph{i.e.}~$n$=2 Sudakov form factor) and three-point cases, \eqref{eq:twoloopcollinear} requires:
\begin{equation}
{\cal R}_2^{(2)} = 0\,, \qquad {\cal R}_3^{(2)}  \xlongrightarrow[\mbox{}]{\mbox{$p_i \parallel p_{i+1}$}} 0 \,.
\end{equation} 
These two relations are enough to determine $C^{(2)}$ and the term $\zeta_{4}\epsilon^2$ in $f^{(2)}$, see more details in \cite{Brandhuber:2012vm}.

Next, we consider the full-color IR structure. We focus on the $n$=3 case with $\itbf{F}_3^{(0)}=\tilde f^{a_1 a_2 a_3}\mathcal{F}^{(0)}_{\operatorname{tr}(\phi^2),3}$. Expanding  \eqref{eq:fullcolorir} up to $g^4$, one has
\begin{equation}
  \begin{aligned}\label{eq:fullcolorir-2loop}
    \itbf{F}_3^{(1)}=&\mathfrak{D}^{(1)} \itbf{F}_3^{(0)}+\itbf{F}_3^{(1),\rm fin}\,, \\
    \itbf{F}_3^{(2)}=&\frac{1}{2}(\mathfrak{D}^{(1)})^2 \itbf{F}_3^{(0)} +\mathfrak{D}^{(1)} \itbf{F}_3^{(1),\rm fin}  + \mathfrak{D}^{(2)} \itbf{F}_3^{(0)} +\itbf{F}_3^{(2),\rm fin} \,,
  \end{aligned}
\end{equation}
where $\mathfrak{D}^{(\ell)}$ is the $\ell$-loop dipole part given by
\begin{equation}
  \mathfrak{D}^{(\ell)}=\frac{\gamma_{\rm cusp}^{(\ell)}}{(\ell\epsilon)^2}\mathbf{D}_0-\frac{\gamma_{\rm cusp}^{(\ell)}}{\ell \epsilon}\mathbf{D}-n\frac{\mathcal{G}_{\rm coll}^{(\ell)}}{\ell \epsilon}\mathbf{1}\,.
\end{equation}
One can  show that: (1) there is no $N_c$-subleading contribution up to two loops due to the following color identity 
\begin{equation}\label{eq:threepointdipole}
    \mathbf{T}_i\cdot \mathbf{T}_j f^{a_1 a_2 a_3}=-\frac{1}{2}N_c f^{a_1 a_2 a_3} \,,\quad \forall i,j \,,
\end{equation}
so that the action of every $\mathbf{D}$ and $\mathbf{D}_0$ does not give any additional $N_c$ power and all the $N_c$ powers come from the anomalous dimensions; (2) the color degrees of freedom can be factorized out such that:\footnote{Note that in the definition of $\mathbf{D}_0$, any $(i,j)$ pair is actually counted twice, which cancels the 1/2 factor in the above equation. }
\begin{align}
  \itbf{F}_3^{(1)}=N_c \tilde{f}^{a_1 a_2 a_3}\mathcal{F}^{(0)}_{\operatorname{tr}(\phi^2),3} &\bigg\{ \Big(\sum_{i<j} -\frac{1}{\epsilon^2}+\frac{1}{\epsilon}\log(-s_{ij})\Big) +\mathcal{I}_3^{(1),\rm fin}\bigg\} \,, \\
  \itbf{F}_3^{(2)}=N_c^2 \tilde{f}^{a_1 a_2 a_3}\mathcal{F}^{(0)}_{\operatorname{tr}(\phi^2),3}&\bigg\{\frac{1}{2}\Big(\sum_{i<j} -\frac{1}{\epsilon^2}+\frac{1}{\epsilon}\log(-s_{ij})\Big)^2 + \Big(\sum_{i<j} - \frac{1}{\epsilon^2} + \frac{1}{\epsilon}\log({-s_{ij}})\Big)\mathcal{I}_3^{(1),\rm fin} \nonumber\\ 
    & \ +\Big(\sum_{i<j}\frac{\zeta_2}{2\epsilon^2}   -\frac{\zeta_2}{\epsilon}\log({-s_{ij}})+\frac{\zeta_3}{\epsilon}\Big) + \mathcal{I}_3^{(2),\rm fin}\bigg\} \,,
\end{align}
where $\mathcal{I}_3^{(\ell),\mathrm{fin}}$ with $\ell=1,2$ are loop corrections after subtracting IR divergences.

One can compare the above $\mathcal{I}_3^{\mathrm{fin}}$ with the previous planar remainder function $\mathcal{R}_3$ defined in the BDS subtraction \eqref{eq:bds-2loop}, so that
\begin{equation}
    \mathcal{I}_3^{(2),\mathrm{fin}}-\frac{1}{2}\Big(\mathcal{I}_3^{(1),\rm fin}\Big)^2 -\mathcal{R}^{(2)}_{3}=(\text{logs and } \zeta_n \textrm{ terms}) +O(\epsilon) \,,
\end{equation}
where the logs and $\zeta_n$ terms refer to a linear combination of simple terms $\{\zeta_2\log^2, \zeta_3\log, \zeta_4\}$, which come from the finite part of $f^{(2)}(\epsilon)\mathcal{I}_3^{(1)}(2\epsilon)+C^{(2)}$ in \eqref{eq:bds-2loop}. This provides a concrete example of scheme changes mentioned previously. 

\subsubsection*{The three-loop form factors}
We now turn to our main object, the three-loop case.

We discuss the planar contribution first and clarify some details in the BDS ansatz. The three-loop BDS ansatz is 
\begin{equation}
\label{eq:bds-3loop}
{\mathcal{I}}_n^{(3)}(\epsilon)
= -\frac{1}{3}\left({\mathcal{I}}_n^{(1)}(\epsilon)\right)^{3} + {\mathcal{I}}_n^{(2)}(\epsilon) {\mathcal{I}}_n^{(1)}(\epsilon)+f^{(3)}(\epsilon) {\mathcal{I}}_n^{(1)}(3 \epsilon)+C^{(3)}+\mathcal{R}_n^{(3)}+O(\epsilon)\,,
\end{equation}
where
\begin{equation}\label{eq:f3}
    f^{(3)}(\epsilon)=4\Big(\frac{11}{2}\zeta_4 +(6 \zeta_5+5\zeta_2\zeta_3)\epsilon+(c_1\zeta_6+c_2\zeta_3^2)\epsilon^2\Big)\,.
\end{equation}
The exact number of $c_{1,2}$ in $f^{(3)}$, as well as $C^{(3)}$, are not yet available. 
The estimated of $f^{(3)}$ has been computed numerically in \cite{Spradlin:2008uu} based on three-loop five-point amplitudes as
\begin{equation}
\label{eq:c12value}
(c_1\zeta_6+c_2\zeta_3^2) = 85.263 \pm 0.004 \,.
\end{equation}
Since $f^{(3)}$ is also understood by the iteration of the splitting amplitude $r_{S}$ that obeys  \cite{Bern:2005iz}
\begin{equation}
r_{S}^{(3)}= -\frac{1}{3}\big( r_{S}^{(1)}(\epsilon) \big)^{3} + r_{S}^{(1)}(\epsilon) r_{S}^{(2)}(\epsilon) + f^{(3)}(\epsilon) r_{S}^{(1)}(3 \epsilon) + O(\epsilon)\,,
\end{equation}
it also applies to the form factor considered here.
The constant $C^{(3)}$, similar to the two-loop case, can be fixed by using the three-loop Sudakov result \cite{Gehrmann:2011xn}:\footnote{We point out  that $f^{(3)}$ for form factors and amplitudes are the same (up to overall normalization) but $C^{(3)}$ can be different (this is also the case for the previous two-loop discussion). Given the same $f^{(3)}$, the difference between $C^{(3)}$ for form factors and amplitudes is equivalent to the difference between the remainders (defined via a BDS form without $C^{(3)}$) of Sudakov form factors and four-point amplitudes, which in general does not vanish.}
\begin{equation}
{\cal R}_2^{(3)} = 0 \quad \Rightarrow \quad {11 \pi^6 \over 270} - {8 \over 9} (c_1\zeta_6+c_2\zeta_3^2) + C^{(3)} = 8 \left( - {13\over 9} \zeta_3^2 - {193 \pi^6 \over 25515} \right)  \,,
\end{equation}
which by plugging in the value of \eqref{eq:c12value} leads to 
\begin{equation}
C^{(3)} = -38.252 \pm 0.004 \,.
\end{equation}
Knowing the value of $f^{(3)}$ and $C^{(3)}$ will be necessary for comparing our form factor result with the remainder result derived from the bootstrap method \cite{Dixon:2020bbt}.

Next, we move on to the full-color structure and study the more interesting non-planar IR divergences. The $g^6$ order of \eqref{eq:fullcolorir} reads
\begin{equation}\label{eq:fullcolorir-3loop}
\begin{aligned}
    \itbf{F}^{(3)}_3=&\frac{1}{6} (\mathfrak{D}^{(1)})^{3} \itbf{F}^{(0)}_3 + \frac{1}{2} (\mathfrak{D}^{(1)})^2 \itbf{F}^{(1),\rm fin}_3 +\mathfrak{D}^{(1)}\itbf{F}^{(2),\rm fin}_3 + \frac{1}{2}\left(\mathfrak{D}^{(2)}\mathfrak{D}^{(1)}+\mathfrak{D}^{(1)}\mathfrak{D}^{(2)}\right)\itbf{F}^{(0)}_3\\
    & + \mathfrak{D}^{(2)}\itbf{F}^{(1),\rm fin}_3 + \mathfrak{D}^{(3)}\itbf{F}^{(0)}_3 +  \frac{1}{3\epsilon} \boldsymbol{\Delta}^{(3)}\itbf{F}^{(0)}_3+\itbf{F}^{(3),\rm fin}_3\,.
\end{aligned}
\end{equation}
Similar to the two-loop discussion, dipole terms generate only $N_c$-leading contributions, since acting the dipole part $\mathfrak{D} $ (consisting of $\mathbf{D}_0$ and $\mathbf{D}$) on $\itbf{F}^{(0)}_3$ gives solely leading $N_c$ powers. 
The only source to generate the $N_c$-subleading IR contribution is the $\boldsymbol{\Delta}^{(3)}$ term.
Applying \eqref{eq:3loopdelta} to both $\operatorname{tr}(\phi^2)$ and $\operatorname{tr}(\phi^3)$ form factors, one finds their $N_c$-subleading IR divergences are
\begin{equation}\label{eq:non-planarir}
    \frac{1}{3\epsilon}\boldsymbol{\Delta}_{3}^{(3)}\itbf{F}_{\operatorname{tr}(\phi^L),3}^{(0)}=\left\{\begin{array}{cc}
     12N_{\rm c}\times   (-2)\epsilon^{-1}(\zeta_5+2\zeta_2\zeta_3)  \tilde f^{a_1 a_2 a_3}\,, & \text{ for} \operatorname{tr}(\phi^2) \,, \\ \\
         0 \,, & \text{ for} \operatorname{tr}(\phi^3) \,.
    \end{array}\right. 
\end{equation}
Regarding this equation, we mention that the vanishing of subleading contributions for the $\operatorname{tr}(\phi^3)$ case is a direct consequence of the color identity $\boldsymbol{\Delta}_{3}\tilde d^{a_1 a_2 a_3}=0$, which is consistent with the fact that $\itbf{F}^{(3)}_{\operatorname{tr}(\phi^3),3}$ has merely $N_c$-leading part from Section~\ref{ssec:color}. 
The $\operatorname{tr}(\phi^2)$ case, on the other hand, shoud have $1/\epsilon$ pole in the $N_c$-subleading part. 

We finally remark on the minimal form factor of $\operatorname{tr}(\phi^3)$. Since in this case there is no $N_c$-subleading contribution, it is enough to apply the BDS subtraction. Another difference to notice is that, while the $f^{(\ell)}$ factor is the same as in the $\operatorname{tr}(\phi^2)$ case, there is no need to introduce the $C^{(\ell)}$ term for the $\operatorname{tr}(\phi^3)$ form factor in BDS ansatz. 

\subsection{Numerical integrations and results}\label{ssec:numerical}

In this subsection we consider the evaluation of three-loop integrals and then perform the IR subtraction and obtain finite remainders. 
Since the analytic results are only known for some planar three-loop integrals  \cite{DiVita:2014pza}
and a major part in our problem are the unknown non-planar three-loop integrals (which include the most challenging ones), 
in this work we will compute the form factors numerically. 

\subsubsection*{Numerical integrations}
We take the approach of sector decomposition \cite{Binoth:2000ps},
for which several pubic codes are available, including FIESTA 
\cite{Smirnov:2008py,Smirnov:2015mct}, and pySecDec \cite{Borowka:2017idc} based on  SecDec \cite{Carter:2010hi}.
Our computation is mainly based on FIESTA 4.2 and pySecDec 1.4.5.\footnote{Recently, pySecDec develops some interesting new functions \cite{Heinrich:2021dbf} like expansion by regions and automatic summation in the version 1.5, which was not applied in our calculation. Also, the new FIESTA5 has been published in \cite{Smirnov:2021rhf}.}

We first clarify some basic settings. For the strategy of performing sector decomposition, we choose the `geometric' method when using pySecDec, while for FIESTA 4.2, we employ `STRATEGY$\_$X' (and occasionally `STRATEGY$\_$KU'). 
As for the integrator, the quasi-Monte-Carlo (QMC) \cite{dick_kuo_sloan_2013, Li:2015foa} integrator in pySecDec \cite{Borowka:2018goh} is suitable for a  high-precision integration;  in FIESTA 4.2, we use the VEGAS algorithm \cite{lepage1980vegas} implemented in the CUBA library \cite{Hahn:2005pf}. We would like to mention that for most integrals, pySecDec with QMC integrator are typically much faster than FIESTA, 
including all those appearing in $N_c$-leading contributions.
However, for some most complicated non-planar integrals, such as integrals for topology (10) in Figure~\ref{fig:phi2tops}, pySecDec 1.4.5 
was not efficient enough to fulfill the required calculations, and the computations were performed with massive parallelization (typically thousands cores in a few days) using FIESTA 4.2.\footnote{It would be interesting to test these computations in the latest version pySecDec 1.5 which contains several important improvements.}

As already discussed  in Section~\ref{ssec:dlog}, for our computations, a crucial point to fulfill the numerical integration is to use $d\log$ integrals.
The use of $d\log$ integrals can often improve the efficiency of the computation by orders of magnitude, as discussed in the beginning of Section~\ref{ssec:dlog}. Below we give some arguments for the reason of the improvement, in company with concrete examples in our calculation.

A special property of the $d\log$ integrals is that they have only logarithm singularities and are free of double poles in certain parametrization forms. 
One may expect that the similar simplicity should be reflected in the sector decomposition algorithms via Feynman parametrization. In these algorithms, there exists a step separating divergences in all sectors, for example the ``pole resolution" step in  FIESTA 4.2. The number of terms after this step can be taken as an index reflecting the complexity of the pole structure of an integral. 
For instance, we consider further details for the example \eqref{eq:i1i2} in Section~\ref{ssec:dlog}. The numbers of terms after the step of ``pole resolution" are about $1.7\times 10^5$ and $4.6\times 10^4$  for $I_1$ and $I_2$ respectively. 
This shows that the $d\log$ integral $I_2$ indeed has fewer terms, which may be expected from the simpler pole structure mentioned above. 
Moreover, many other examples of the non-planar topologies also corroborate this point: 
using $d\log$ integrals can often reduce the number of terms after pole resolution by two to three times. For example, let us consider the integral $I_{12} $ in Table~\ref{tab:utintlist} in Appnedix~\ref{ap:ut}. It has a special numerator containing $[(\ell_a-p_1)^2]^2$ and other terms proportional to propagators. The special linear combination is vitally important because if we consider only $[(\ell_a-p_1)^2]^2$ as the numerator, the number of terms increases by two times and the total computational time increases by nearly 10 times. 

With the above discussion, it is reasonable for us to first reorganize the integrand based on $d\log$ integrals, and then perform numerical integrations accordingly. We would like to also mention that similar improvement was observed and the strategies were taken in the numerical computation of the four-loop non-planar Sudakov form factors \cite{Boels:2017ftb}, and our results provide further support to these observations. 
As a side remark, in our problem the three-loop non-planar integrals at top level, \emph{i.e.}~with 10 propagators, are the hardest tasks for numerical evaluation, and other integrals with planar topologies or shrunk propagators are much faster.  Thus, practically it is sufficient to  only deal with the hard part using $d\log$ integrals and leave the remaining part invariant, so that we can get a balance between a minor modification of the integrand and a more efficient integration. 

\subsubsection*{Results and checks}

We summarize the final numerical results of form factors in Table~\ref{tab:numres}, which is computed at a special kinematics point $s_{12}=s_{23}=s_{13}=-2$. To reach the given precisions, a large amount of computational resources are needed, and nearly  $O(10^7)$ CPU core hours were used in total. 
Sample numerical results for a few individual $d\log$ integrals are also given in Appendix~\ref{ap:ut}.

Since we take numerical methods, a few words about the credibility and estimated error in the numerical computation are in order here. 
First, for the available analytic planar integrals \cite{DiVita:2014pza}, we check that our numerical computations always match perfectly. 
Second, most integrals have been cross checked by using both FIESTA and pySecDec and they are consistent within error bars. 
Third, for many integrals, we have checked that the errors converge by increasing evaluating points.
Finally, as discussed in more details below, for the planar form factors of both ${\rm tr}(\phi^2)$ and ${\rm tr}(\phi^3)$, the differences between three-loop results and the BDS ansatz predictions are all within the error range, and the planar remainder of ${\rm tr}(\phi^2)$ also reproduces the bootstrap result, and more non-trivially, the non-planar IR divergence matches the prediction as well considering the estimated errors; all these confirms that the estimated errors produced by the programs are trustworthy.

\begin{table}[t!]
    \centering
    \caption{Numerical results for the three-loop three-point form factors on the kinematics point $(s_{12},s_{23},s_{13}) = (-2,-2,-2)$.} 
    \vskip .5 cm
    \begin{tabular}{|c|c|c|c|c|c|c|c|}
    \hline 
 &
\multicolumn{7}{c|}{$\mathcal{I}^{(3),\text{PL}}_{\operatorname{tr}(\phi^2)}$} \\\hline
        & $\epsilon^{-6}$ & $\epsilon^{-5}$ & $\epsilon^{-4}$ & $\epsilon^{-3}$ & $\epsilon^{-2}$ & $\epsilon^{-1}$ & $\epsilon^{0}$ \\\hline
      ${\cal I}^{(3)}$  & -4.5 & $9.35749 $ & -22.6136 & $55.8891 $ & -77.252 & $92.943 $ & -336.51 \\\hline
      error & $6\times10^{-7} $ & $2.3\times10^{-5} $ & $3.3\times10^{-4} $ & $0.0021 $ & $0.012 $ & $0.078 $ & $0.59 $ \\\hline\hline
&
\multicolumn{7}{c|}{$\mathcal{I}^{(3),\text{NP}}_{\operatorname{tr}(\phi^2)}$} \\\hline
        & $\epsilon^{-6}$ & $\epsilon^{-5}$ & $\epsilon^{-4}$ & $\epsilon^{-3}$ & $\epsilon^{-2}$ & $\epsilon^{-1}$ & $\epsilon^{0}$ \\\hline
       ${\cal I}^{(3)}$  & -2.3$\times 10^{-7}$ & $5.8\times10^{-6}$ & $3.8\times10^{-5}$ & $5.6\times10^{-4}$ & -0.001 & -9.989 & -265.31 \\\hline
      error & $1.2\times 10^{-6}$ & $2.4\times 10^{-5}$ & $3.0\times 10^{-4}$ & $2.5\times10^{-3}$ & $0.02$ & $0.185$ & $1.76$\\\hline\hline
 &
\multicolumn{7}{c|}{$\mathcal{I}^{(3)}_{\operatorname{tr}(\phi^3)}$} \\\hline
        & $\epsilon^{-6}$ & $\epsilon^{-5}$ & $\epsilon^{-4}$ & $\epsilon^{-3}$ & $\epsilon^{-2}$ & $\epsilon^{-1}$ & $\epsilon^{0}$ \\\hline
        ${\cal I}^{(3)}$  & -4.5 & $9.357488$ & -6.02807 & 31.5028 & 19.5617 & 123.565 & 217.11 \\\hline
      error & $2\times 10^{-7}$ & $9.4\times 10^{-6}$ & $8.1\times 10^{-5}$ & $5.4\times 10^{-4}$ & 0.0035 & 0.023 & 0.21\\\hline
    \end{tabular}
    \label{tab:numres}
\end{table}
Given these numerical form factor results, we can study their properties  and compare them with known structures and results: 
\begin{itemize}
\item[1.] At $N_c$-leading order, we compare the divergent part of our results with the BDS ansatz. This also requires high order of $\epsilon$-expansion of $\mathcal{I}^{(\ell)}$ with $\ell=1,2$. Concretely, to get $O(\epsilon^0)$ of  \eqref{eq:bds-3loop}, results of $\mathcal{I}^{(1)}$ up to $O(\epsilon^4)$ and $\mathcal{I}^{(2)}$ up to $O(\epsilon^2)$ are necessary. Most of the required expression are known analytically \cite{DiVita:2014pza}, and one unknown non-planar master integral for the $\operatorname{tr}(\phi^2)$ form factor are evaluated numerically. 
Considering the BDS form for $\operatorname{tr}(\phi^3)$ form factors as an example, the epsilon expansion is 
\begin{align}
    &-\frac{1}{3}\left({\mathcal{I}}^{(1)}(\epsilon)\right)^{3} + {\mathcal{I}}^{(2)}(\epsilon) {\mathcal{I}}^{(1)}(\epsilon)+f^{(3)}(\epsilon) {\mathcal{I}}^{(1)}(3 \epsilon)\\
    \nonumber
    &\hskip .5cm=-\frac{9}{2\epsilon^{6}}+\frac{9.3574869}{\epsilon^{5}}-\frac{6.028071}{\epsilon^4}+\frac{31.50306}{\epsilon^3}+\frac{19.5639}{\epsilon^2}+\frac{123.580}{\epsilon}+56.63+O(\epsilon)\,,
\end{align}
which shows a perfect match with the result in Table~\ref{tab:numres}. 

\item[2.]As for the $N_c$-subleading IR divergences, the prediction in \eqref{eq:non-planarir} tells us that $\epsilon^{-6}$ to $\epsilon^{-2}$ poles should cancel and the residue of $\epsilon^{-1}$ pole should be $-2(\zeta_5+2\zeta_2\zeta_3)=-9.983$. We observe that  the non-planar result in Table~\ref{tab:numres} matches the prediction. It should be mentioned that at $\epsilon^{-1}$ order, the value of a single integral is typically a number of ${\cal O}(10^2)$ or even ${\cal O}(10^3)$ (see some integral data in Appendix~\ref{ap:ut}), but these large numbers successfully cancel to get a correct small number. 

\item[3.]Furthermore, we can compare our numerical result of $\operatorname{tr}(\phi^2)$ three-point form factor at planar level with the remainder result obtained recently via a totally different method, \emph{i.e.}~bootstrap \cite{Dixon:2020bbt} utilizing input from form factor operator product expansion (OPE) \cite{Sever:2020jjx, Sever:2021nsq}. More concretely, using the remainder result ${\cal R}^{(3), \textrm{planar}}_{\operatorname{tr}(\phi^2),3} = -8.372$ in Table~6 of \cite{Dixon:2020bbt} and together with the three-loop BDS ansatz \eqref{eq:bds-3loop}, one gets a prediction that  the $\operatorname{tr}(\phi^2)$ form factor at $O(\epsilon^0)$ should be $-336.71$, which is totally consistent with our value $-336.51\pm 0.59$. 

\item[4.]
Finally, we also obtain some previously unknown results. We can apply three-loop BDS subtraction for the form factor of $\operatorname{tr}(\phi^3)$ and obtain the finite remainder as 
\begin{equation}
  \mathcal{R}^{(3)}_{\operatorname{tr}(\phi^3),3}=160.48\pm 0.22 \,.
\end{equation}
We also obtain non-planar ($N_c$-subleading) three-loop finite remainder for $\operatorname{tr}(\phi^2)$, which is read directly from the $O(\epsilon^0)$ order result in  Table~\ref{tab:numres} as $-265.31\pm 1.76$.\footnote{Here we just define the $O(\epsilon^0)$ term as the finite remainder. A refined definition would include a subtraction of the three-loop splitting function such that the remainder has a simple collinear behavior, as for the planar remainder function.}
These results may be useful for further discussions on both numerical and analytical analyses of three-loop three-point form factors in future studies.

\end{itemize}

\section{Summary and Outlook}\label{sec:summary}

In this paper, we discuss the detailed construction of the full-color three-loop three-point form factors in ${\cal N}=4$ SYM based on the the color-kinematics duality and generalized unitarity methods. 
These results provide a concrete step of extending the application scope of color-kinematics duality.
An intriguing and surprising finding is that the ``simplest''-type CK-dual integrand solutions still contain a large number of free parameters, considering that it is usually not easy to find high-loop solutions for amplitudes. 
These large solution spaces originate from a new type of generalized gauge transformations induced by the insertion of local operators in form factors. 
The solution spaces as well as the new generalized gauge transformations strongly imply the constructibility of CK-dual integrand solutions at four or even higher loops \cite{Lin:2021lqo} and suggest that the form-factor-type quantities may be an ideal arena for applying color-kinematics duality.  
Based on CK-dual solutions, the double copy construction and its physical meaning in gravity are natural problems to explore.
Recently, color-kinematics duality and the corresponding double copy have been studied for amplitudes in AdS space \cite{Armstrong:2020woi,Albayrak:2020fyp,Alday:2021odx,Diwakar:2021juk,Zhou:2021gnu} which are equivalent to the correlation functions of boundary operators. Form factors also contain local operators and it is interesting to explore their double copy, which we leave to another work \cite{Lin:2021pne}.

Having the compact integrand results, we have performed numerical integrations and study the IR divergences and finite remainders, which further confirm the correctness of the results and also provide new data of non-planar remainders. 
In the numerical calculation, we have constructed some $d\log$ integrals.
To construct a complete set of $d\log$ basis and to obtain the analytical expressions via the method of differential equation \cite{Henn:2013pwa} for these three-loop integrals are certainly important topics for further investigations. 
It is also promising to consider efficient numerical methods which may provide high-precision results, such as the recently developed auxiliary mass flow method \cite{Liu:2017jxz, Liu:2021wks}. 
Given these advances, it is hopeful that the three-loop integration problem can be solved in the near future. 

Finally, we mention that phenomenologically it is worthwhile to calculate similar three-point form factors in QCD at three-loop level. 
The two-loop calculations \cite{Gehrmann:2011aa, Brandhuber:2012vm, Banerjee:2016kri, Jin:2018fak,Jin:2019ile,Jin:2019opr,Brandhuber:2018xzk,Brandhuber:2018kqb} have shown remarkable connections between analytic expressions in various supersymmetric and non-supersymmetric theories, and it would be of great interest to examine similar connections at three loops, in particular for the non-planar corrections.

\acknowledgments

We would like to thank Yuchen Ding, Yuanhong Guo, Song He and Yanqing Ma for discussions. 
This work is supported in part by the National Natural Science Foundation of China (Grants No.~11822508, 11935013, 12047502, 12047503, 11947301), and by the Key Research Program of the Chinese Academy of Sciences, Grant NO. XDPB15.
We also thank the support of the HPC Cluster of ITP-CAS and CAS Xiandao-1 computing environment.

\appendix

\section{CK-dual solution of the two-loop form factor}\label{ap:ck2loop}

In this appendix, we provide various factors for the CK-dual integrand \eqref{eq:ff2loop} of the two-loop three-point form factor of ${\rm tr}(\phi^2)$.
For simplicity, we introduce two functions of Mandelstam variables as $S_{\rm p,1}=s_{23}(s_{12}-s_{13})$ and  $S_{\rm s,1}=s_{12}^2-s_{13}^2$ while other $S_{{\rm p},i}$ and $S_{{\rm s}, j}$ can be obtained by cyclic permutations. 
\begin{longtable}{|c|c|c|c|}
    \caption{CK-dual factors of the two-loop form factor.}
    \label{tab:resphi2-2loop}
    \endfirsthead
    \endhead
    \hline 
    \begin{tabular}{cc}   \\ ~  \end{tabular} $\Gamma_{i,123} $ &  $N_{i,123} $ & $C_{i,123} $ & $S_i$  \\
    \hline 
    \begin{tabular}{cc}    
        \includegraphics[width=0.20\linewidth]{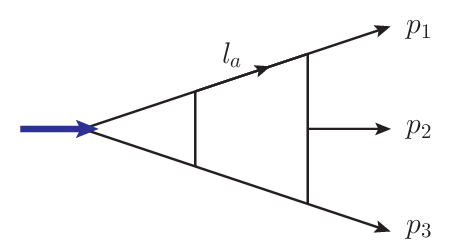}  \\
        (1)  
    \end{tabular}
    &
    $\begin{aligned}
          & q^2 s_{12} s_{23}/2 \\
        - & c_1 
        \begin{aligned}[t]
            &\bigl(S_{\rm p,3}\big((l_a-p_1-p_2)^2-(l_a-q)^2\big)\\
               & + S_{\rm p,2}\big((l_a-p_1)^2-(l_a-p_1-p_2)^2\big)\\
               & +S_{\rm p,1}(l_a^2-(l_a-p_1)^2\big)\bigr)
        \end{aligned} \\
        - & c_2 
         \begin{aligned}[t]
            &\bigl(S_{\rm s,3}\big((l_a-p_1-p_2)^2-(l_a-q)^2\big)\\
               & + S_{\rm s,2}\big((l_a-p_1)^2-(l_a-p_1-p_2)^2\big)\\
               & +S_{\rm s,1}(l_a^2-(l_a-p_1)^2\big)\bigr)
        \end{aligned}
    \end{aligned} $   
    & $2N_c^2 \tilde f^{a_1 a_2 a_3} $ & 2  \\
    \hline 
    \begin{tabular}{cc}    
        \includegraphics[width=0.20\linewidth]{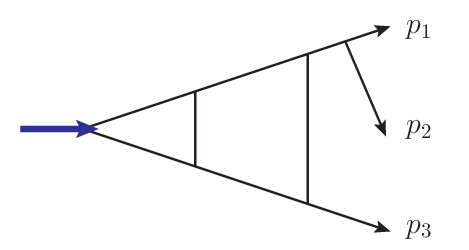}  \\
        (2)  
    \end{tabular}
    &
    $\begin{aligned}
          & q^2 s_{12} (s_{13}+s_{23})/2 \\
        + & c_1 
        \begin{aligned}[t]
            \bigl(S_{\rm p,1}-S_{\rm p,2}\bigr) s_{12}
        \end{aligned} \\
        + & c_2
        \begin{aligned}[t]
            \bigl(S_{\rm s,1}-S_{\rm s,2}\bigr) s_{12}
        \end{aligned}
    \end{aligned} $
    & $4N_c^2 \tilde f^{a_1 a_2 a_3} $ & 2 \\
    \hline 
    \begin{tabular}{cc}    
        \includegraphics[width=0.20\linewidth]{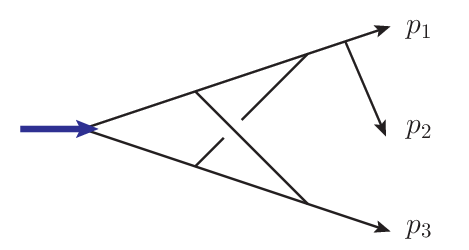}  \\
        (3)  
    \end{tabular} 
    &
    $\begin{aligned}
          & q^2 s_{12} (s_{13}+s_{23})/2 \\
        + & c_1 
        \begin{aligned}[t]
            \bigl(S_{\rm p,1}-S_{\rm p,2}\bigr) s_{12}
        \end{aligned} \\
        + & c_2
        \begin{aligned}[t]
            \bigl(S_{\rm s,1}-S_{\rm s,2}\bigr) s_{12}
        \end{aligned}
    \end{aligned} $
    & $2N_c^2 \tilde f^{a_1 a_2 a_3} $ & 4 \\
    \hline 
    \begin{tabular}{cc}    
        \includegraphics[width=0.20\linewidth]{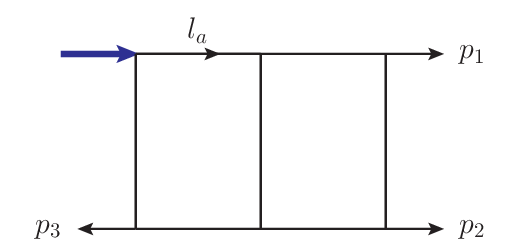}  \\
        (4)  
    \end{tabular} 
    & 
    $\begin{aligned}
          & s_{12}(s_{23}(l_{a}\cdot p_2)-s_{13}(l_{a}\cdot p_1)) \\
        + & c_1
        \begin{aligned}[t]
            S_{\rm p,3} \bigl((l_a-p_1-p_2)^2 - (l_a-q)^2 \bigr)
        \end{aligned} \\
        + & c_2
        \begin{aligned}[t]
            S_{\rm s,3} \bigl((l_a-p_1-p_2)^2 - (l_a-q)^2 \bigr)
        \end{aligned} \\
        + & c_3
        \begin{aligned}[t]
            S_{\rm p,3} \bigl((l_a-q)^2 - l_a^2 \bigr)
        \end{aligned} \\
        + & c_4
        \begin{aligned}[t]
         S_{\rm s,3} \bigl((l_a-q)^2-l_a^2 \bigr)
        \end{aligned}
    \end{aligned} $
    & $N_c^2 \tilde f^{a_1 a_2 a_3} $ & 1 \\
    \hline 
    \begin{tabular}{cc}    
        \includegraphics[width=0.20\linewidth]{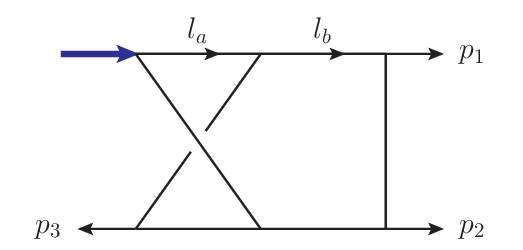}  \\
        (5)  
    \end{tabular} 
    & 
    $\begin{aligned}
          & s_{12} (s_{23}q^2/2+s_{13}(l_a\cdot p_1)-s_{23}(l_a\cdot p_2)) \\
        + & c_1
        \begin{aligned}[t]
            &\bigl( S_{\rm p,2} \big( (l_b-p_1)^2- (l_b-p_1-p_2)^2\big)  \\
                 &+ S_{\rm p,1} \big( (l_b)^2- (l_b-p_1)^2\big) \\
                 &+ S_{\rm p,3} \big( (l_a-l_b)^2- (l_a-l_b-p_3)^2\big) \bigr)
        \end{aligned} \\
        + & c_2
        \begin{aligned}[t]
            &\bigl( S_{\rm s,2} \big( (l_b-p_1)^2- (l_b-p_1-p_2)^2\big)  \\
                 &+ S_{\rm s,1} \big( (l_b)^2- (l_b-p_1)^2\big) \\
                 &+ S_{\rm s,3} \big( (l_a-l_b)^2- (l_a-l_b-p_3)^2\big) \bigr)
        \end{aligned} \\
        - & c_3
        \begin{aligned}[t]
            S_{\rm p,3} ((l_a-q)^2 - l_a^2)
        \end{aligned} \\
        - & c_4
        \begin{aligned}[t]
           S_{\rm s,3} ((l_a-q)^2 - l_a^2)
        \end{aligned}
    \end{aligned} $
    & $N_c^2 \tilde f^{a_1 a_2 a_3} $ & 2 \\
    \hline 
    \begin{tabular}{cc}    
        \includegraphics[width=0.20\linewidth]{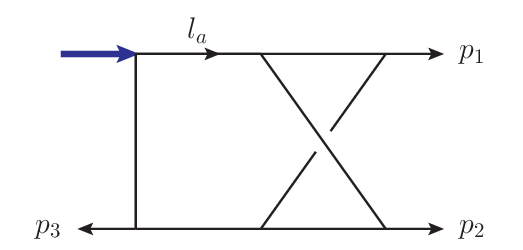}  \\
        (6)  
    \end{tabular} 
    & 
    $\begin{aligned}
          & s_{12} (s_{13}(l_a\cdot p_1)-s_{23}(l_a\cdot p_2))\\
        - & c_1
        \begin{aligned}[t]
            S_{\rm p,3} ((l_a-p_1-p_2)^2-(l_a-q)^2)
        \end{aligned} \\
        - & c_2
        \begin{aligned}[t]
            S_{\rm s,3} ((l_a-p_1-p_2)^2-(l_a-q)^2)
        \end{aligned} \\
        - & c_3
        \begin{aligned}[t]
            S_{\rm p,3} ((l_a-q)^2-l_a^2)
        \end{aligned} \\
        - & c_4
        \begin{aligned}[t]
           S_{\rm s,3} ((l_a-q)^2-l_a^2)
        \end{aligned}
    \end{aligned} $
    & $0 $ & 2 \\
    \hline
\end{longtable}

\section{Simplified three-loop Integrands}\label{ap:integrand}

In this appendix, we provide explicit results for the simplified integrands discussed in Section~\ref{ssec:integrandstrategy}. They are collected in three tables, corresponding to the topologies and numerators in \eqref{eq:simpInt2pl}, \eqref{eq:simpInt2np}, and \eqref{eq:simpInt3}, respectively.

\begin{longtable}{|c|c|}
\caption{$N_c$-leading $\operatorname{tr}(\phi^2)$ form factor results in \eqref{eq:simpInt2pl}.}
    \label{tab:integrandphi2Gp1}
\endfirsthead
\endhead
    \hline diagram &  numerator  \\
    \hline $\Gamma^{\pl}_{10:14}=\begin{aligned}
        \includegraphics[width=0.23\linewidth]{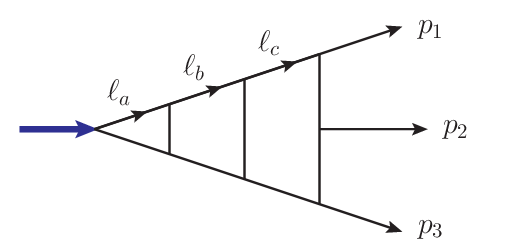}
    \end{aligned}$
    & $\mathbb{N}^{\pl}_{10:14}=-s_{12}s_{23}(q^2)^2 $ \\
    \hline $\Gamma^{\pl}_{10:15}=\begin{aligned}
        \includegraphics[width=0.23\linewidth]{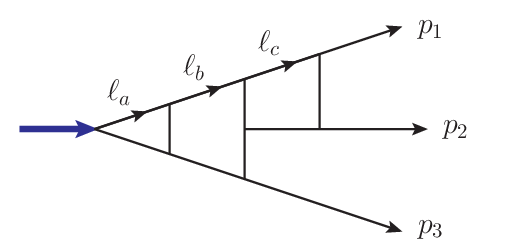}
    \end{aligned}$
    & $\mathbb{N}^{\pl}_{10:15}=-s_{12}q^2(s_{12}s_{23}+(s_{13}+s_{23})(\ell_b-p_1)^2) $ \\
    \hline $\Gamma^{\pl}_{10:1}=\begin{aligned}
        \includegraphics[width=0.23\linewidth]{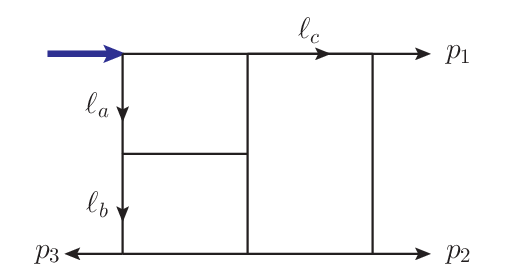}
    \end{aligned}$
    & { ${ \mathbb{N}^{\pl}_{10:1}}=-\frac{1}{2}s_{12}\begin{aligned}[t]
        \bigl( & s_{23}(\ell_a-p_3)^2(q^2-(l_c-q)^2) \\
             + & (\ell_c-q)^2(s_{12}s_{23}+(s_{13}+s_{23})(\ell_a-p_2-p_3)^2) \bigr)
    \end{aligned} $} \\
    \hline $\Gamma^{\pl}_{10:2}=\begin{aligned}
        \includegraphics[width=0.23\linewidth]{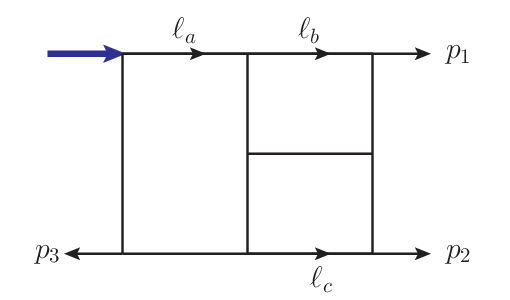}
    \end{aligned}$
    & $\mathbb{N}^{\pl}_{10:2}=-\frac{1}{2}s_{12}(\ell_a-p_1)^2(s_{12}s_{23}+(s_{13}+s_{23})(\ell_a-p_1)^2) $ \\
    \hline $\Gamma^{\pl}_{10:3}=\begin{aligned}
        \includegraphics[width=0.23\linewidth]{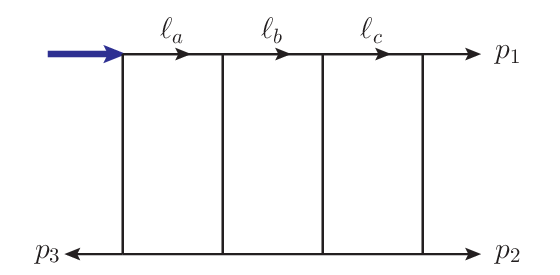}
    \end{aligned}$
    & $\mathbb{N}^{\pl}_{10:3}=-\frac{1}{2}s_{12}^2(s_{12}s_{23}+(s_{13}+s_{23})(\ell_a-p_1)^2) $ \\
    \hline $\Gamma^{\pl}_{10:16}=\begin{aligned}
        \includegraphics[width=0.23\linewidth]{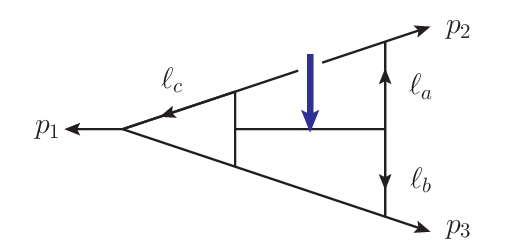}
    \end{aligned}$
    & \hskip 2pt {\small $\begin{aligned}
        \mathbb{N}^{\pl}_{10:16}= & -\frac{1}{4}s_{13}(s_{23}+2s_{12})((\ell_a-p_1-p_2)^2+s_{23})(\ell_a-p_1-p_2)^2 \\
                  & -\frac{1}{8}(4s_{12}s_{13}+s_{23}(s_{12}+s_{13}))(\ell_a-p_1-p_2)^2(\ell_b-p_1-p_3)^2 \\
                  & +\frac{1}{2}s_{13}s_{23}(\ell_a-p_1-p_2)^2(\ell_a+\ell_b)\cdot(p_2-p_3)
    \end{aligned} $} \hskip 2pt \\
    \hline $\Gamma^{\pl}_{10:19}=\begin{aligned}
        \includegraphics[width=0.23\linewidth]{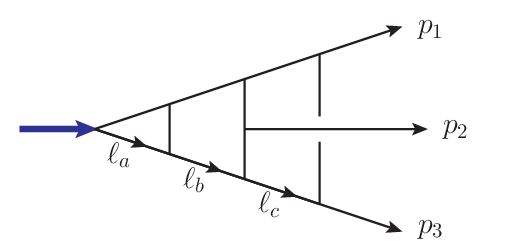}
    \end{aligned}$
    & $\begin{aligned}
        \mathbb{N}^{\pl}_{10:19}=s_{13}s_{23}q^2((\ell_b-p_3)^2-\frac{1}{2}s_{12})
    \end{aligned} $ \\
    \hline $\Gamma^{\pl}_{10:20}=\begin{aligned}
        \includegraphics[width=0.23\linewidth]{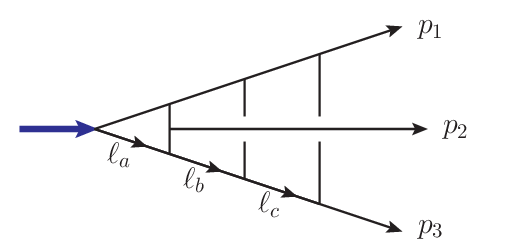}
    \end{aligned}$
    & $\begin{aligned}
        \mathbb{N}^{\pl}_{10:20}=-\frac{1}{4}s_{13}\begin{aligned}[t]
            \bigl( & s_{12}s_{13}s_{23}+(s_{12}+s_{23})q^2(\ell_b-p_3)^2 \\
                 - & 2s_{13}s_{23}(\ell_a-p_3)^2 \bigr)
        \end{aligned}
    \end{aligned} $ \\
    \hline $\Gamma^{\pl}_{10:21}=\begin{aligned}
        \includegraphics[width=0.23\linewidth]{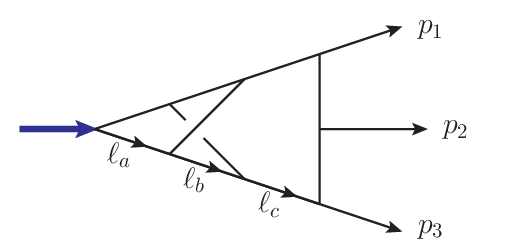}
    \end{aligned}$
    & $\begin{aligned}
        \mathbb{N}^{\pl}_{10:21}=-\frac{1}{4}s_{12}s_{23}(q^2)^2
    \end{aligned} $ \\
    \hline $\Gamma^{\pl}_{10:4}=\begin{aligned}
        \includegraphics[width=0.23\linewidth]{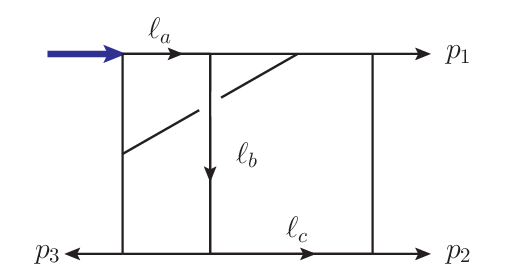}
    \end{aligned}$
    & {\small $\begin{aligned}
        \mathbb{N}^{\pl}_{10:4}= & -\frac{1}{2}s_{12}s_{23}\begin{aligned}[t]
            \bigl( & (s_{12}-(\ell_a-p_1-p_2)^2)(q^2-(\ell_c+p_3)^2) \\
                 + & q^2(\ell_a+\ell_c-p_1-p_2)^2 \bigr)
        \end{aligned} \\
                   & -\frac{1}{2}s_{12}(s_{13}+s_{23})(q^2(\ell_b-\ell_c+p_2)^2-(\ell_a-p_1)^2(\ell_c+p_3)^2)
    \end{aligned} $ }\\
    \hline $\Gamma^{\pl}_{10:8}=\begin{aligned}
        \includegraphics[width=0.23\linewidth]{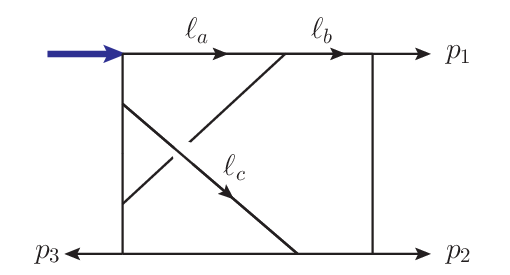}
    \end{aligned}$
    & {\small $\begin{aligned}
        \mathbb{N}^{\pl}_{10:8}=\frac{1}{2}s_{12}\begin{aligned}[t]
            \bigl( & s_{23}((\ell_a-p_1-p_2)^2(q^2-(\ell_b-q)^2)+s_{12}(\ell_b-q)^2) \\
                 + & (s_{13}+s_{23})(\ell_a-p_1)^2(\ell_b-q)^2 \bigr)
        \end{aligned}
    \end{aligned} $ }\\
    \hline $\Gamma^{\pl}_{9:1}=\begin{aligned}
        \includegraphics[width=0.23\linewidth]{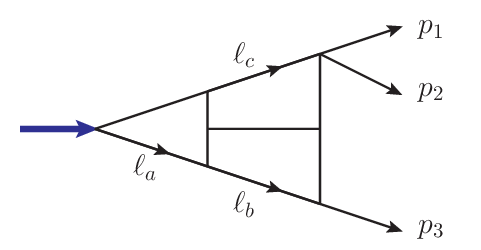}
    \end{aligned}$
    & $\begin{aligned}
        \mathbb{N}^{\pl}_{9:1}=-\frac{1}{2}(s_{13}+s_{23})q^2(\ell_a-p_3)^2
    \end{aligned} $ \\
    \hline $\Gamma^{\pl}_{9:2}=\begin{aligned}
        \includegraphics[width=0.23\linewidth]{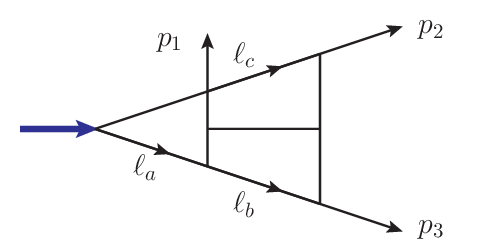}
    \end{aligned}$
    & $\begin{aligned}
        \mathbb{N}^{\pl}_{9:2}=\frac{1}{2}s_{12}s_{23}(\ell_a-p_3)^2
    \end{aligned} $ \\
    \hline $\Gamma^{\pl}_{9:3}=\begin{aligned}
        \includegraphics[width=0.23\linewidth]{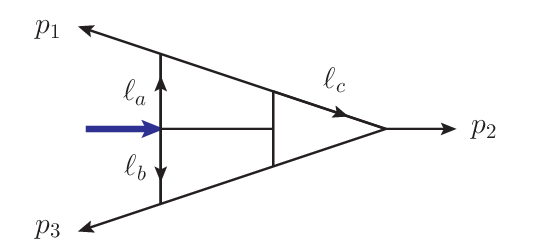}
    \end{aligned}$
    & $\begin{aligned}
        \mathbb{N}^{\pl}_{9:3}=\frac{1}{4}s_{13}s_{23}(\ell_a-p_1-p_2)^2
    \end{aligned} $ \\
    \hline $\Gamma^{\pl}_{9:4}=\begin{aligned}
        \includegraphics[width=0.23\linewidth]{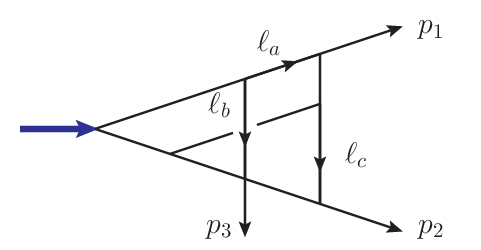}
    \end{aligned}$
    & $\begin{aligned}
        \mathbb{N}^{\pl}_{9:4}=\frac{1}{4}\bigl( & s_{12}(s_{13}+2s_{23})(\ell_b-p_2-p_3)^2 \\
        + & s_{23}(s_{13}+2s_{12})(\ell_a-p_1-p_2)^2+2s_{12}s_{23}q^2 \bigr)
    \end{aligned} $ \\
    \hline $\Gamma^{\pl}_{9:5}=\begin{aligned}
        \includegraphics[width=0.23\linewidth]{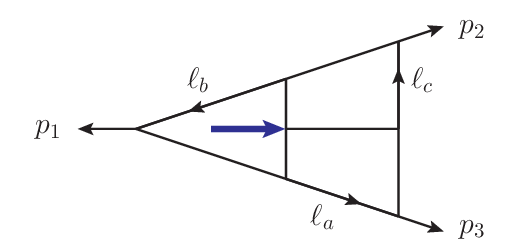}
    \end{aligned}$
    & $\begin{aligned}
        \mathbb{N}^{\pl}_{9:5}=\frac{1}{4}s_{12}((s_{13}+2s_{23})(\ell_a-p_2-p_3)^2+\frac{1}{3}s_{13}s_{23})
    \end{aligned} $ \\
    \hline $\Gamma^{\pl}_{9:6}=\begin{aligned}
        \includegraphics[width=0.23\linewidth]{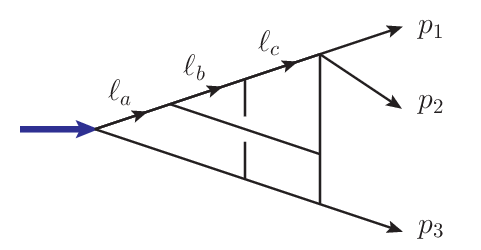}
    \end{aligned}$
    & $\begin{aligned}
        \mathbb{N}^{\pl}_{9:6}=\frac{1}{2}(s_{13}+s_{23})q^2(\ell_a-p_1-p_2)^2
    \end{aligned} $ \\
    \hline $\Gamma^{\pl}_{9:7}=\begin{aligned}
        \includegraphics[width=0.23\linewidth]{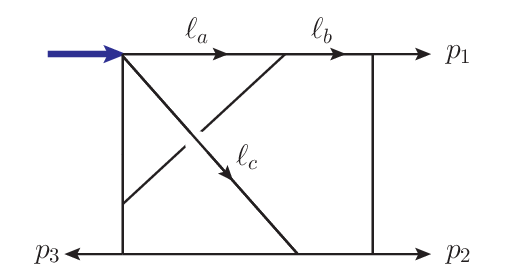}
    \end{aligned}$
    & $\begin{aligned}
        \mathbb{N}^{\pl}_{9:7}=-\frac{1}{2}s_{12}s_{23}(\ell_b-q)^2
    \end{aligned} $ \\
    \hline $\Gamma^{\pl}_{9:8}=\begin{aligned}
        \includegraphics[width=0.23\linewidth]{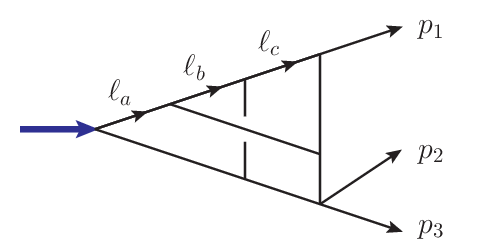}
    \end{aligned}$
    & $\begin{aligned}
        \mathbb{N}^{\pl}_{9:8}=\frac{1}{2}(s_{12}+s_{13})q^2(\ell_a-p_1)^2
    \end{aligned} $ \\
    \hline $\Gamma^{\pl}_{9:9}=\begin{aligned}
        \includegraphics[width=0.23\linewidth]{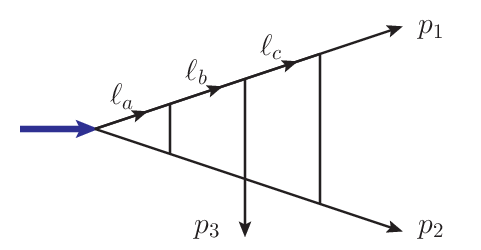}
    \end{aligned}$
    & $\begin{aligned}
        \mathbb{N}^{\pl}_{9:9}=s_{12}s_{23}q^2
    \end{aligned} $ \\
    \hline $\Gamma^{\pl}_{9:10}=\begin{aligned}
        \includegraphics[width=0.23\linewidth]{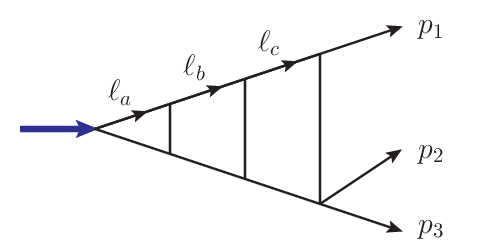}
    \end{aligned}$
    & $\begin{aligned}
        \mathbb{N}^{\pl}_{9:10}=-2(s_{12}+s_{13})(q^2)^2
    \end{aligned} $ \\
    \hline $\Gamma^{\pl}_{9:11}=\begin{aligned}
        \includegraphics[width=0.23\linewidth]{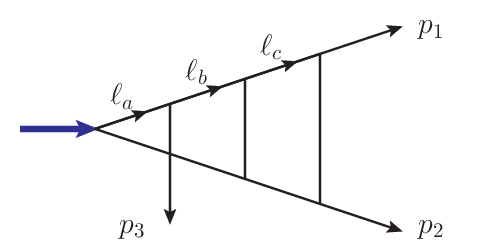}
    \end{aligned}$
    & $\begin{aligned}
        \mathbb{N}^{\pl}_{9:11}=\frac{1}{2}s_{12}^2s_{23}
    \end{aligned} $ \\
    \hline $\Gamma^{\pl}_{9:12}=\begin{aligned}
        \includegraphics[width=0.23\linewidth]{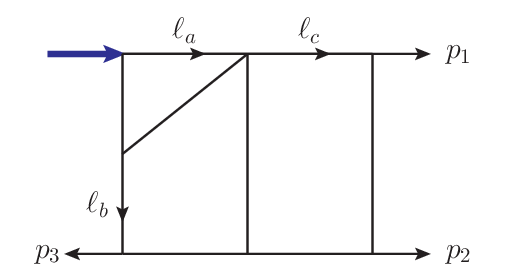}
    \end{aligned}$
    & $\begin{aligned}
        \mathbb{N}^{\pl}_{9:12}=\frac{1}{2}s_{12}s_{23}q^2
    \end{aligned} $ \\
    \hline $\Gamma^{\pl}_{9:13}=\begin{aligned}
        \includegraphics[width=0.23\linewidth]{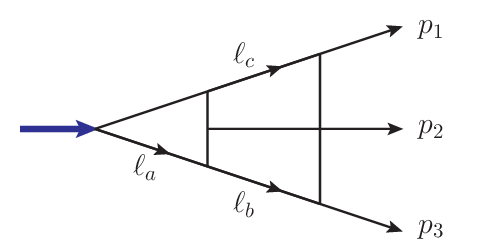}
    \end{aligned}$
    & $\begin{aligned}
        \mathbb{N}^{\pl}_{9:13}=\frac{1}{4}s_{12}s_{23}q^2
    \end{aligned} $ \\
    \hline $\Gamma^{\pl}_{9:14}=\begin{aligned}
        \includegraphics[width=0.23\linewidth]{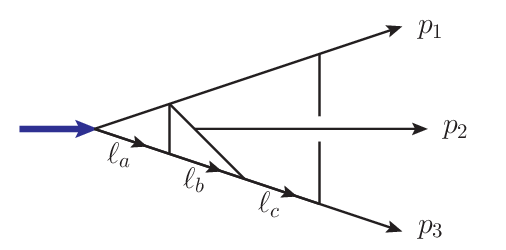}
    \end{aligned}$
    & $\begin{aligned}
        \mathbb{N}^{\pl}_{9:14}=-\frac{1}{2}s_{13}s_{23}q^2
    \end{aligned} $ \\
    \hline $\Gamma^{\pl}_{9:15}=\begin{aligned}
        \includegraphics[width=0.23\linewidth]{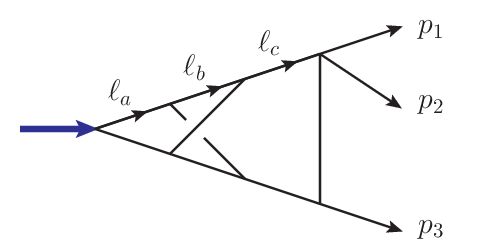}
    \end{aligned}$
    & $\begin{aligned}
        \mathbb{N}^{\pl}_{9:15}=-\frac{1}{2}(s_{13}+s_{23})(q^2)^2
    \end{aligned} $ \\
    \hline $\Gamma^{\pl}_{9:16}=\begin{aligned}
        \includegraphics[width=0.23\linewidth]{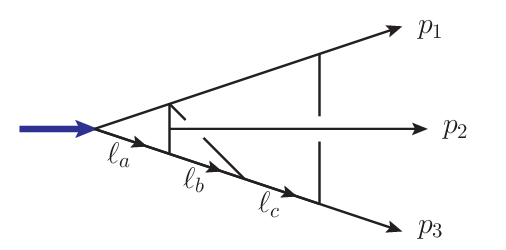}
    \end{aligned}$
    & $\begin{aligned}
        \mathbb{N}^{\pl}_{9:16}=\frac{1}{2}s_{13}(s_{12}+s_{23})q^2
    \end{aligned} $ \\
    \hline $\Gamma^{\pl}_{9:17}=\begin{aligned}
        \includegraphics[width=0.23\linewidth]{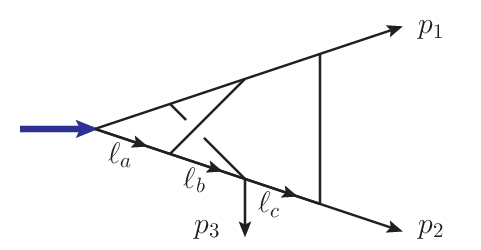}
    \end{aligned}$
    & $\begin{aligned}
        \mathbb{N}^{\pl}_{9:17}=\frac{1}{2}s_{13}s_{23}q^2
    \end{aligned} $ \\
    \hline $\Gamma^{\pl}_{9:18}=\begin{aligned}
        \includegraphics[width=0.23\linewidth]{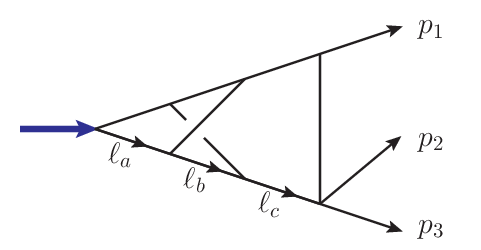}
    \end{aligned}$
    & $\begin{aligned}
        \mathbb{N}^{\pl}_{9:18}=-\frac{1}{2}(s_{12}+s_{13})(q^2)^2
    \end{aligned} $ \\
    \hline $\Gamma^{\pl}_{9:19}=\begin{aligned}
        \includegraphics[width=0.23\linewidth]{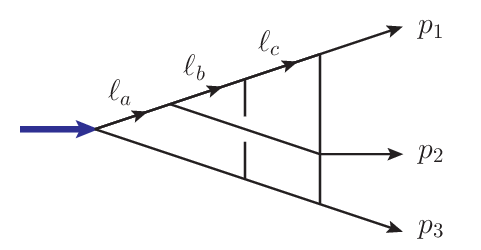}
    \end{aligned}$
    & $\begin{aligned}
        \mathbb{N}^{\pl}_{9:19}=-\frac{1}{2}s_{12}s_{23}q^2
    \end{aligned} $ \\
    \hline $\Gamma^{\pl}_{8:1}=\begin{aligned}
        \includegraphics[width=0.23\linewidth]{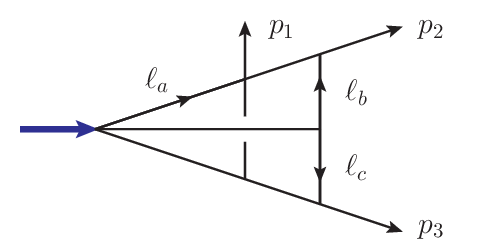}
    \end{aligned}$
    & $\begin{aligned}
        \mathbb{N}^{\pl}_{8:1}=-\frac{1}{2}s_{12}s_{23}
    \end{aligned} $ \\
    \hline $\Gamma^{\pl}_{8:2}=\begin{aligned}
        \includegraphics[width=0.23\linewidth]{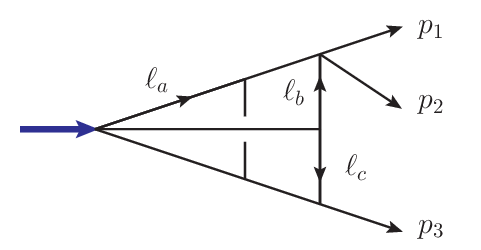}
    \end{aligned}$
    & $\begin{aligned}
        \mathbb{N}^{\pl}_{8:2}=-\frac{1}{2}(s_{13}+s_{23})q^2
    \end{aligned} $ \\
    \hline
\end{longtable}


\begin{longtable}{|c|c|}
\caption{$N_c$-subleading $\operatorname{tr}(\phi^2)$ form factor results in  \eqref{eq:simpInt2np}.}
    \label{tab:integrandphi2np}
\endfirsthead
\endhead
    \hline diagram &  numerator  \\
    \hline $\Gamma^{\np}_{10:2}=\begin{aligned}
        \includegraphics[width=0.23\linewidth]{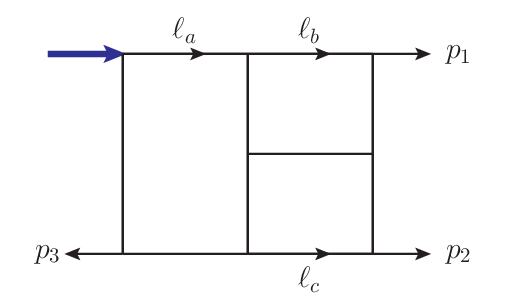}
    \end{aligned}$
    & $\begin{aligned}
        \mathbb{N}^{\np}_{10:2}=-\frac{1}{2}s_{12} (\ell_a-p_1)^2 \bigl((s_{13}+s_{23})(\ell_a-p_1)^2+s_{12}s_{23} \bigr)
    \end{aligned} $ \\
    \hline $\Gamma^{\np}_{10:16}=\begin{aligned}
        \includegraphics[width=0.23\linewidth]{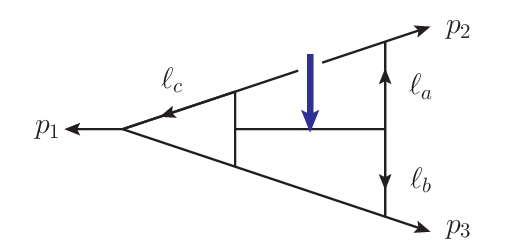}
    \end{aligned}$
    & $\begin{aligned}
        \mathbb{N}^{\np}_{10:16}= & -\frac{1}{4}s_{13}(s_{23}+2s_{12})((\ell_a-p_1-p_2)^2+s_{23})(\ell_a-p_1-p_2)^2 \\
                  & -\frac{1}{8}(4s_{12}s_{13}+s_{23}(s_{12}+s_{13}))(\ell_a-p_1-p_2)^2(\ell_b-p_1-p_3)^2 \\
                  & +\frac{1}{2}s_{13}s_{23}(\ell_a-p_1-p_2)^2(\ell_a+\ell_b)\cdot(p_2-p_3)
    \end{aligned} $ \\
    \hline $\Gamma^{\np}_{10:10}=\begin{aligned}
        \includegraphics[width=0.23\linewidth]{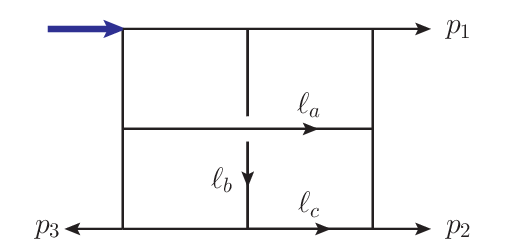}
    \end{aligned}$
    & $\begin{aligned}
        \mathbb{N}^{\np}_{10:10}=\begin{aligned}[t]
            & -\frac{1}{2}s_{23}(s_{12}+s_{13})\begin{aligned}[t]
                \bigl( & (\ell_a-\ell_b+\ell_c)^2 \\
                + & (\ell_c-p_1-p_2)^2 \bigr)(\ell_a-\ell_b+\ell_c)^2
            \end{aligned} \\
            & +\frac{1}{2}s_{12}s_{23}\bigl((\ell_a-\ell_b+\ell_c-p_2)^2+(\ell_c+p_3)^2 \bigr)(\ell_a-\ell_b+\ell_c)^2 \\
            & -s_{13}s_{23}\bigl((\ell_a-\ell_b+\ell_c)^2-\frac{1}{2}s_{12} \bigr)\ell_a\cdot\ell_b \\
            & +\frac{1}{2}s_{12}s_{23}\begin{aligned}[t]
                \bigl( & (s_{12}+s_{13}-s_{23})(\ell_a-\ell_b+\ell_c)^2 \\
                - & s_{12}(\ell_c+p_3)^2+\frac{1}{2}s_{12}s_{23} \bigr)
            \end{aligned}
        \end{aligned}
    \end{aligned} $ \\
    \hline $\Gamma^{\np}_{10:18}=\begin{aligned}
        \includegraphics[width=0.23\linewidth]{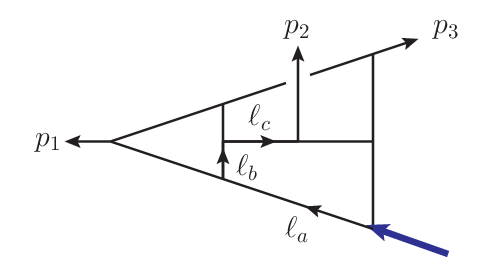}
    \end{aligned}$
    & $\begin{aligned}
        \mathbb{N}^{\np}_{10:18}=\begin{aligned}[t]
             &\frac{1}{2}s_{12}s_{23}\bigl((\ell_a-p_1)^2-(\ell_a-p_1-p_2)^2 \bigr)(\ell_a-\ell_c)^2 \\
            -&\frac{1}{2}s_{13}s_{23}\bigl((\ell_a-p_1)^2+(\ell_a-\ell_c)^2 \bigr)(\ell_a-p_1-p_2)^2 \\
            -&\frac{1}{2}s_{12}\begin{aligned}[t]
                \Bigl( & s_{13}(\ell_a-p_1)^2\bigl((\ell_a-p_1)^2-s_{23} \bigr) \\
                + & s_{23}^2(\ell_a-\ell_c)^2 \Bigr)
            \end{aligned}
        \end{aligned}
    \end{aligned} $ \\
    \hline $\Gamma^{\np}_{10:6}=\begin{aligned}
        \includegraphics[width=0.23\linewidth]{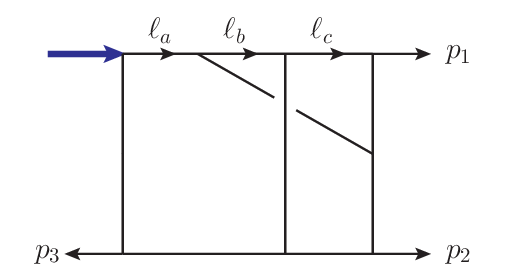}
    \end{aligned}$
    & $\begin{aligned}
        \mathbb{N}^{\np}_{10:6}=-\frac{1}{2}s_{12}\bigl((s_{13}+s_{23})(\ell_a-p_1)^2+s_{12}s_{23} \bigr)(\ell_a-p_1)^2
    \end{aligned} $ \\
    \hline $\Gamma^{\np}_{10:7}=\begin{aligned}
        \includegraphics[width=0.23\linewidth]{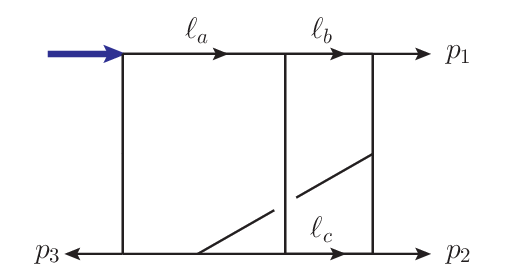}
    \end{aligned}$
    & $\begin{aligned}
        \mathbb{N}^{\np}_{10:7}=-\frac{1}{2}s_{12}\bigl((s_{13}+s_{23})(\ell_a-p_1)^2+s_{12}s_{23} \bigr)(\ell_a-p_1)^2
    \end{aligned} $ \\
    \hline $\Gamma^{\np}_{10:9}=\begin{aligned}
        \includegraphics[width=0.23\linewidth]{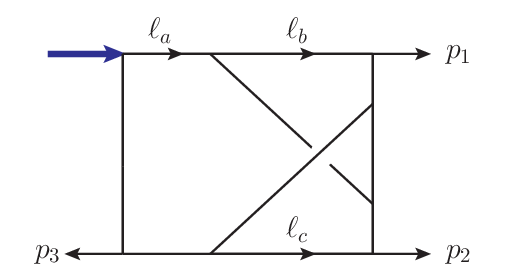}
    \end{aligned}$
    & $\begin{aligned}
        \mathbb{N}^{\np}_{10:9}=-\frac{1}{2}s_{12}\bigl((s_{13}+s_{23})(\ell_a-p_1)^2+s_{12}s_{23} \bigr)(\ell_a-p_1)^2
    \end{aligned} $ \\
    \hline $\Gamma^{\np}_{9:1}=\begin{aligned}
        \includegraphics[width=0.23\linewidth]{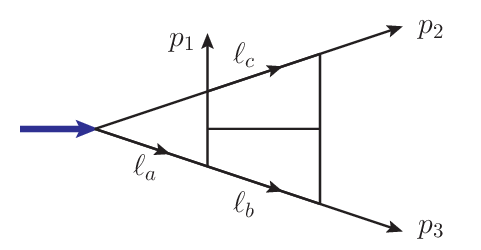}
    \end{aligned}$
    & $\begin{aligned}
        \mathbb{N}^{\np}_{9:1}=\frac{1}{2}s_{12}s_{23}(\ell_a-p_3)^2
    \end{aligned} $ \\
    \hline $\Gamma^{\np}_{9:2}=\begin{aligned}
        \includegraphics[width=0.23\linewidth]{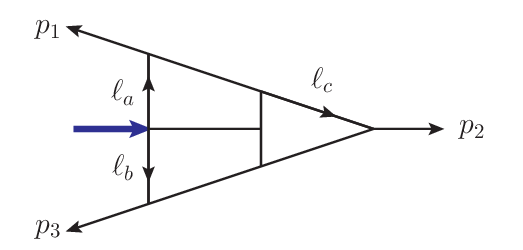}
    \end{aligned}$
    & $\begin{aligned}
        \mathbb{N}^{\np}_{9:2}=\frac{1}{4}s_{13}s_{23}(\ell_a-p_1-p_2)^2
    \end{aligned} $ \\
    \hline $\Gamma^{\np}_{9:3}=\begin{aligned}
        \includegraphics[width=0.23\linewidth]{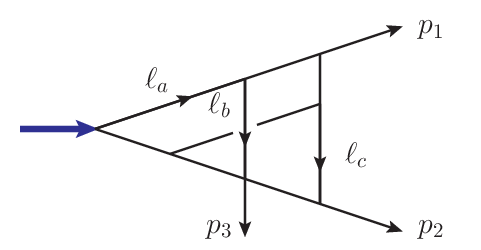}
    \end{aligned}$
    & $\begin{aligned}
        \mathbb{N}^{\np}_{9:3}=-\frac{1}{4} \begin{aligned}[t]
            \Bigl(
             &s_{23}(s_{13}+2s_{12})\bigl((\ell_b-p_3)^2-2(\ell_a-\ell_b-p_1-p_2)^2 \\
            -&\bigl(s_{13}(s_{12}+s_{23})+4s_{12}s_{23} \bigr)(\ell_b-p_2-p_3)^2 \bigr) \\
            +&s_{13}s_{23}\ell_a\cdot(p_1-p_3) \\
            +&s_{23}\bigl(s_{13}^2+2(s_{12}+s_{13})(s_{12}+s_{23}) \bigr) \Bigr)
        \end{aligned}
    \end{aligned} $ \\
    \hline $\Gamma^{\np}_{9:4}=\begin{aligned}
        \includegraphics[width=0.23\linewidth]{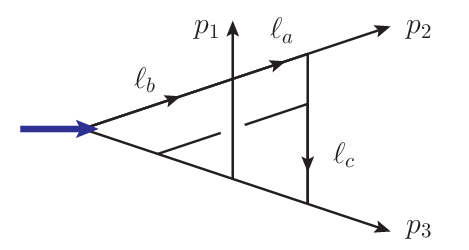}
    \end{aligned}$
    & $\begin{aligned}
        \mathbb{N}^{\np}_{9:4}=-\frac{1}{4}s_{12}s_{23}\bigl((\ell_a-\ell_b+p_1+p_3)^2+(\ell_a-p_2-p_3)^2 \bigr)
    \end{aligned} $ \\
    \hline $\Gamma^{\np}_{9:5}=\begin{aligned}
        \includegraphics[width=0.23\linewidth]{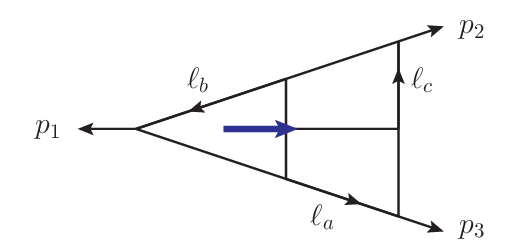}
    \end{aligned}$
    & $\begin{aligned}
        \mathbb{N}^{\np}_{9:5}=\frac{1}{2}s_{12}\bigl((s_{13}+2s_{23})(\ell_a-p_2-p_3)^2+\frac{1}{3}s_{13}s_{23} \bigr)
    \end{aligned} $ \\
    \hline $\Gamma^{\np}_{9:6}=\begin{aligned}
        \includegraphics[width=0.23\linewidth]{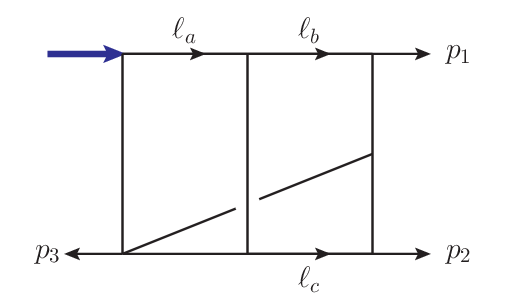}
    \end{aligned}$
    & $\begin{aligned}
        \mathbb{N}^{\np}_{9:6}=-\frac{1}{2}s_{12}\bigl(s_{23}(\ell_a-p_1-p_2)^2-(s_{13}+2s_{23})(\ell_a-p_1)^2-s_{12}s_{23} \bigr)
    \end{aligned} $ \\
    \hline $\Gamma^{\np}_{9:7}=\begin{aligned}
        \includegraphics[width=0.23\linewidth]{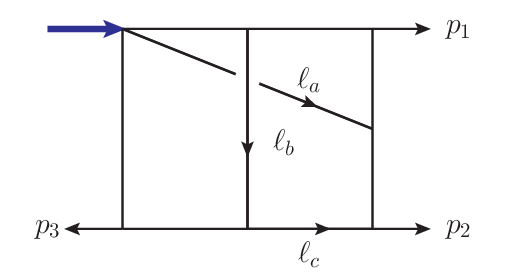}
    \end{aligned}$
    & $\begin{aligned}
        \mathbb{N}^{\np}_{9:7}=\frac{1}{2}s_{12}s_{13}\bigl((\ell_a-\ell_b+\ell_c-p_2)^2+2\ell_a\cdot\ell_b \bigr)
    \end{aligned} $ \\
    \hline $\Gamma^{\np}_{9:8}=\begin{aligned}
        \includegraphics[width=0.23\linewidth]{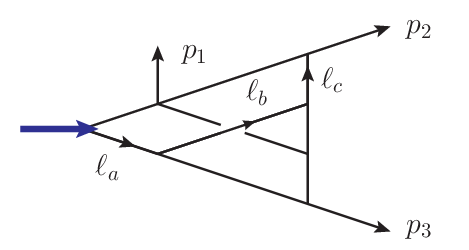}
    \end{aligned}$
    & $\begin{aligned}
        \mathbb{N}^{\np}_{9:8}=\frac{1}{2}s_{12}s_{23}(\ell_a-p_3)^2
    \end{aligned} $ \\
    \hline $\Gamma^{\np}_{9:9}=\begin{aligned}
        \includegraphics[width=0.23\linewidth]{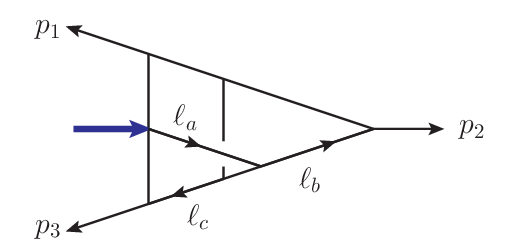}
    \end{aligned}$
    & $\begin{aligned}
        \mathbb{N}^{\np}_{9:9}=\frac{1}{2}s_{13}s_{23}(\ell_c-\ell_a)^2
    \end{aligned} $ \\
    \hline $\Gamma^{\np}_{9:10}=\begin{aligned}
        \includegraphics[width=0.23\linewidth]{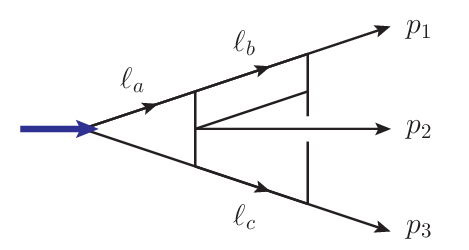}
    \end{aligned}$
    & $\begin{aligned}
        \mathbb{N}^{\np}_{9:10}=\frac{1}{2}s_{12}s_{13}\bigl((\ell_a-p_1)^2-\frac{1}{2} s_{23} \bigr)
    \end{aligned} $ \\
    \hline $\Gamma^{\np}_{8:1}=\begin{aligned}
        \includegraphics[width=0.23\linewidth]{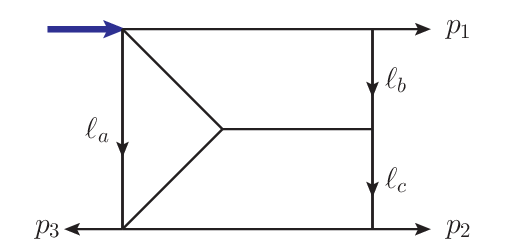}
    \end{aligned}$
    & $\begin{aligned}
        \mathbb{N}^{\np}_{8:1}=-\frac{1}{4}s_{13}s_{23}
    \end{aligned} $ \\
    \hline $\Gamma^{\np}_{8:2}=\begin{aligned}
        \includegraphics[width=0.23\linewidth]{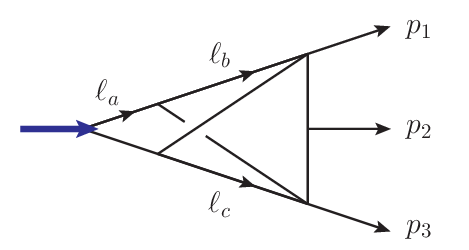}
    \end{aligned}$
    & $\begin{aligned}
        \mathbb{N}^{\np}_{8:2}=-\frac{1}{8}\bigl(4s_{12}s_{23}+s_{13}(s_{12}+s_{23}) \bigr)
    \end{aligned} $ \\
    \hline $\Gamma^{\np}_{8:3}=\begin{aligned}
        \includegraphics[width=0.23\linewidth]{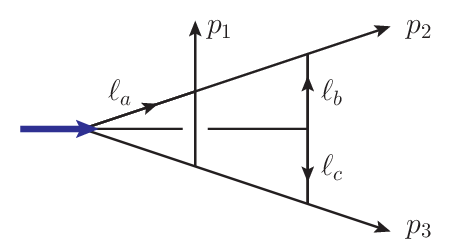}
    \end{aligned}$
    & $\begin{aligned}
        \mathbb{N}^{\np}_{8:3}=\frac{1}{4}s_{12}s_{13}
    \end{aligned} $ \\
    \hline $\Gamma^{\np}_{8:4}=\begin{aligned}
        \includegraphics[width=0.23\linewidth]{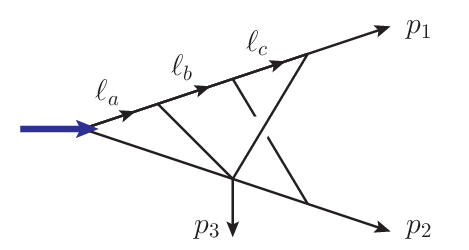}
    \end{aligned}$
    & $\begin{aligned}
        \mathbb{N}^{\np}_{8:4}=\frac{1}{4}s_{12}(s_{13}+s_{23})
    \end{aligned} $ \\
    \hline $\Gamma^{\np}_{8:5}=\begin{aligned}
        \includegraphics[width=0.23\linewidth]{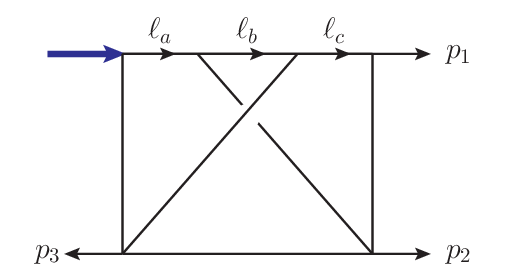}
    \end{aligned}$
    & $\begin{aligned}
        \mathbb{N}^{\np}_{8:5}=\frac{1}{2}s_{12}s_{23}
    \end{aligned} $ \\
    \hline $\Gamma^{\np}_{8:6}=\begin{aligned}
        \includegraphics[width=0.23\linewidth]{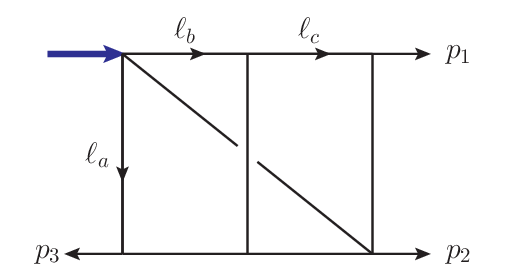}
    \end{aligned}$
    & $\begin{aligned}
        \mathbb{N}^{\np}_{8:6}=\frac{1}{2}s_{12}(s_{13}+s_{23})
    \end{aligned} $ \\
    \hline $\Gamma^{\np}_{8:7}=\begin{aligned}
        \includegraphics[width=0.23\linewidth]{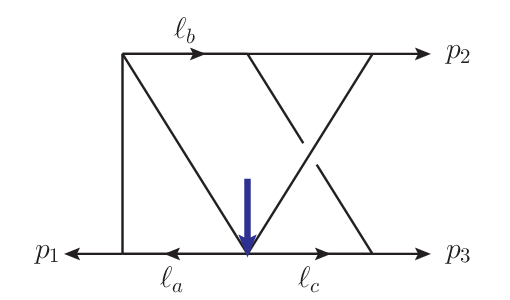}
    \end{aligned}$
    & $\begin{aligned}
        \mathbb{N}^{\np}_{8:7}=\frac{1}{2}s_{23}(s_{12}+s_{13})
    \end{aligned} $ \\
    \hline
\end{longtable}


\begin{longtable}{|c| c| c| }
\caption{Results of $\operatorname{tr}(\phi^3)$ form factor in \eqref{eq:simpInt3}.}
    \label{tab:integrandphi3}
\endfirsthead
\endhead
    \hline
       diagram &  numerator  \\\hline
    
      $\Gamma_{9:2}=\begin{aligned}
        \includegraphics[width=0.23\linewidth]{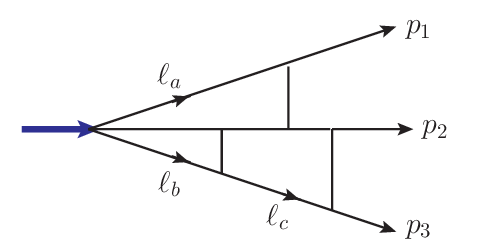}
    \end{aligned}$
    & $\mathbb{N}_{9:2}=(\ell_a-q)^2(\ell_c-q^2)s_{23}$   \\\hline
      $\Gamma_{9:3}=\begin{aligned}
        \includegraphics[width=0.23\linewidth]{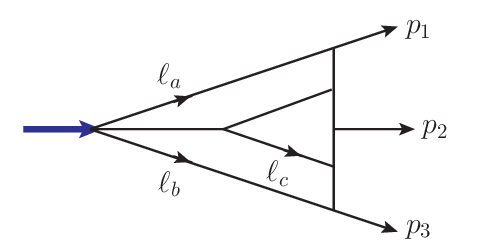}
    \end{aligned}$
    &  $\mathbb{N}_{9:3}=(\ell_a-q)^2(\ell_b-p_2-p_3)^2s_{12}-\frac{1}{2}(\ell_a-q)^2(\ell_b-q)^2q^2$   \\\hline
     
     $\Gamma_{9:4}=\begin{aligned}
        \includegraphics[width=0.23\linewidth]{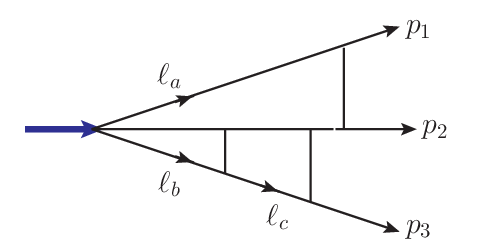}
    \end{aligned}$
    &  $\mathbb{N}_{9:4}=\left[(\ell_a-q)^2\right]^2 s_{12}$   \\\hline
    
     $\Gamma_{9:5}=\begin{aligned}
        \includegraphics[width=0.23\linewidth]{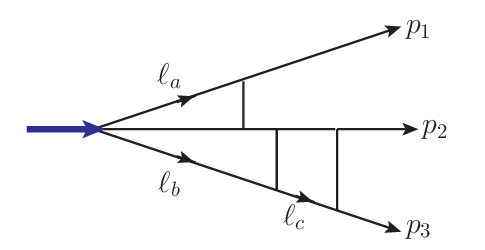}
    \end{aligned}$
    & $\mathbb{N}_{9:5}=(\ell_b-q)^2 s_{23}^2 $     \\\hline
      
     $\Gamma_{9:9}=\begin{aligned}
        \includegraphics[width=0.23\linewidth]{figure/Gamma99.eps}
    \end{aligned}$
    & $\mathbb{N}_{9:9}=(\ell_a-q)^2(\ell_c-q)^2s_{23}$    \\\hline
   
     $\Gamma_{9:11}=\begin{aligned}
        \includegraphics[width=0.23\linewidth]{figure/Gamma911.eps}
    \end{aligned}$
    &  $\begin{aligned}
        \mathbb{N}_{9:11}=&4(\ell_a\cdot p_2)(\ell_c \cdot p_1)s_{23}+2(p_3\cdot \ell_b)(p_3\cdot (p_1-\ell_a))s_{12}\\
        &-2(p_1\cdot \ell_c)s_{12}s_{13}+\frac{1}{6}s_{12}s_{13}s_{23}
    \end{aligned}
    $  \\\hline
      $\Gamma_{9:20}=\begin{aligned}
        \includegraphics[width=0.23\linewidth]{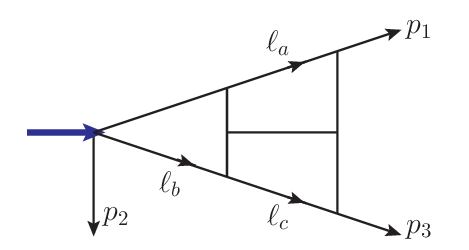}
    \end{aligned}$
    & $\mathbb{N}_{9:20}=(\ell_b-p_3)^2 s_{13}^2$    \\\hline
   
      $\Gamma_{9:21}=\begin{aligned}
        \includegraphics[width=0.23\linewidth]{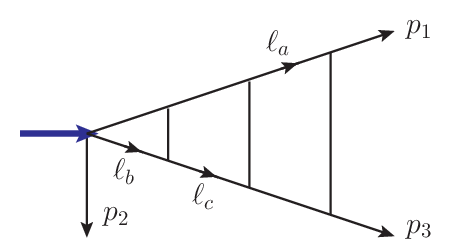}
    \end{aligned}$
    & $\mathbb{N}_{9:21}=-s_{13}^3$    \\\hline
     
      $\Gamma_{8:1}=\begin{aligned}
        \includegraphics[width=0.23\linewidth]{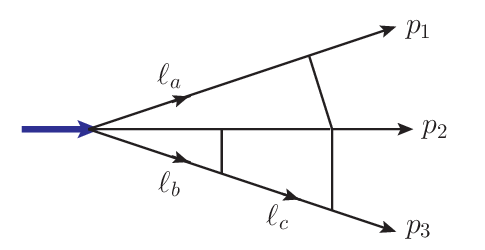}
    \end{aligned}$
    & $\mathbb{N}_{8:1}=-(\ell_a-q)^2q^2$    \\\hline
       
      $\Gamma_{8:2}=\begin{aligned}
        \includegraphics[width=0.23\linewidth]{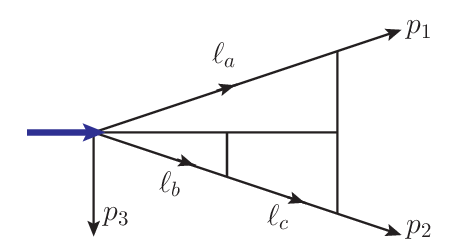}
    \end{aligned}$
    &  $\mathbb{N}_{8:2}=(\ell_a-p_1-p_2)^2s_{12}$    \\\hline
       
      $\Gamma_{8:3}=\begin{aligned}
        \includegraphics[width=0.23\linewidth]{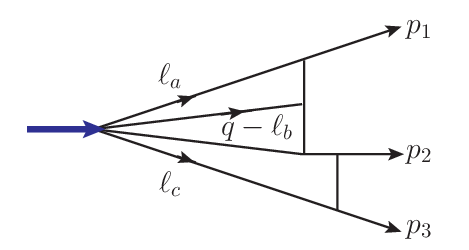}
    \end{aligned}$
    &  $\mathbb{N}_{8:3}=(\ell_c-q)^2s_{23}$   \\\hline
         
     $\Gamma_{8:4}=\begin{aligned}
        \includegraphics[width=0.23\linewidth]{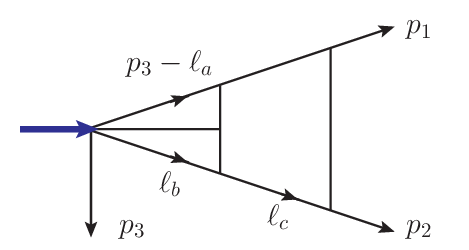}
    \end{aligned}$
    &  $\mathbb{N}_{8:4}=\frac{1}{2}s_{12}^2$   \\\hline
         
     $\Gamma_{8:5}=\begin{aligned}
        \includegraphics[width=0.23\linewidth]{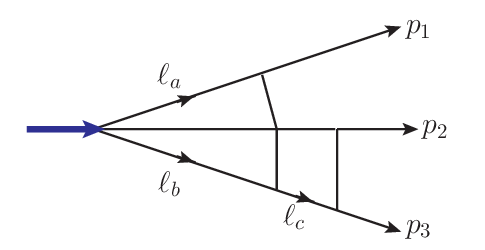}
    \end{aligned}$
    &  $\mathbb{N}_{8:5}=-s_{12}q^2$   \\\hline
      
     $\Gamma_{8:6}=\begin{aligned}
        \includegraphics[width=0.23\linewidth]{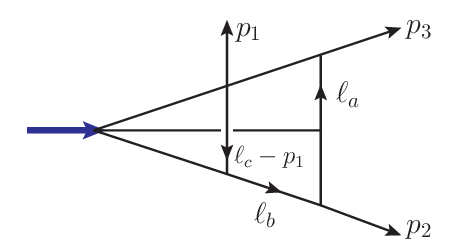}
    \end{aligned}$
    &  $\begin{aligned}
        \mathbb{N}_{8:6}=
        &2\left(\ell_b\cdot(\ell_a - p_3)\right)s_{12}+2\left(\ell_b\cdot p_1\right) s_{23}
    \end{aligned}$   \\\hline
       
     $\Gamma_{8:7}=\begin{aligned}
        \includegraphics[width=0.23\linewidth]{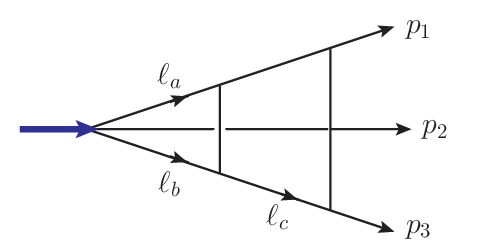}
    \end{aligned}$
    &  $\mathbb{N}_{8:7}=-\frac{1}{2}(\ell_a+\ell_b)^2 q^2$   \\\hline
       
     $\Gamma_{8:8}=\begin{aligned}
        \includegraphics[width=0.23\linewidth]{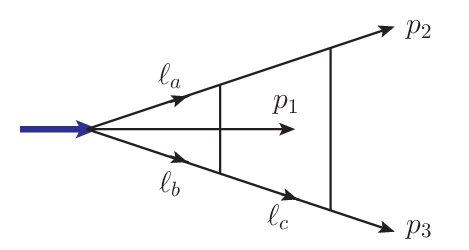}
    \end{aligned}$
    &  $\mathbb{N}_{8:8}=-\frac{1}{2}s_{23} q^2$   \\\hline
       
     $\Gamma_{7:1}=\begin{aligned}
        \includegraphics[width=0.23\linewidth]{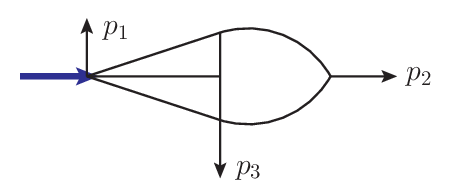}
    \end{aligned}$
    &  $\mathbb{N}_{7:1}=-s_{23}$   \\\hline
       
     $\Gamma_{7:2}=\begin{aligned}
        \includegraphics[width=0.23\linewidth]{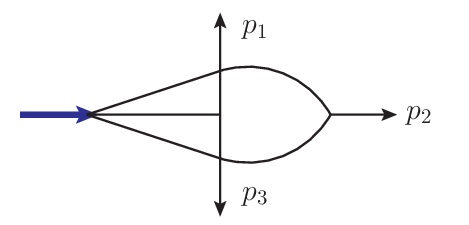}
    \end{aligned}$
    &  $\mathbb{N}_{7:2}=-\frac{1}{2}s_{13}$   \\\hline
       
     $\Gamma_{7:3}=\begin{aligned}
        \includegraphics[width=0.23\linewidth]{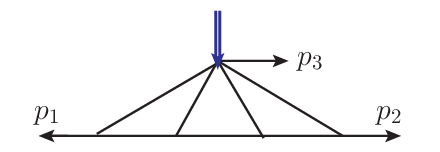}
    \end{aligned}$
    &  $\mathbb{N}_{7:3}=\frac{1}{2}s_{12}$   \\\hline
              
     $\Gamma_{7:4}=\begin{aligned}
        \includegraphics[width=0.23\linewidth]{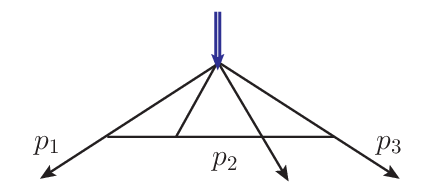}
    \end{aligned}$
    &  $\mathbb{N}_{7:4}=-s_{23}+s_{12}$   \\\hline
\end{longtable}

\section{$d\log$ integrals}\label{ap:ut}

In this appendix we give some details of the $d\log$ integrals that are used to improve numerical evaluations. 

We first collect the $d\log$ integrals that we used in the non-planar calculation in Table~\ref{tab:utintlist}. For the purpose of numerical computation, we are satisfied with $d\log$ integrals with mixed leading singularities.
We would like to point out that the methods and tools employed in this paper, especially with the \texttt{DlogBasis} package, are expected to be able to get a complete pure $d\log$-integral expansion of our form factor integrands and we leave it for future study.

\begin{longtable}{|c| c| c| c | }
\caption{$d\log$ integrals for non-planar $\operatorname{tr}(\phi^2)$ form factor. }
    \label{tab:utintlist}
\endfirsthead
\endhead
    \hline
     &  topology &  numerator & leading singularity  \\\hline
    $I_1$ & $\begin{aligned}
        \includegraphics[width=0.23\linewidth]{figure/phi2NplGamma10to4.eps}
    \end{aligned}$ & $
    \begin{aligned}
        s_{12}s_{23}\big(&(\ell_a-p_1)^2-s_{23}+\ell_a^2\\
       &-(\ell_a-p_1-p_2)^2 \big)(\ell_a-\ell_c)^2
    \end{aligned}
     $ & $\left\{1\right\}$ \\\hline
     $I_2$ & $\begin{aligned}
        \includegraphics[width=0.23\linewidth]{figure/phi2NplGamma10to4.eps}
    \end{aligned}$ & $
    \begin{aligned}
   s_{12} s_{13}(\ell_a-p_1)^2\big(&(\ell_a-p_1-p_2)^2\\
    &-(p_1+p_3-\ell_a+\ell_c)^2\big)
    \end{aligned}
    $ & $\left\{\frac{s_{12}}{ q^2},\frac{s_{12}}{s_{13}+s_{12}},\frac{s_{12}}{s_{23}+s_{12}}\right\}$ \\\hline
     $I_3$ & $\begin{aligned}
        \includegraphics[width=0.23\linewidth]{figure/phi2NplGamma10to4.eps}
    \end{aligned}$ & $\begin{aligned}
    s_{13}s_{23}(\ell_a-\ell_c)^2\big(&(\ell_a-p_1-p_2)^2\\
    &-(p_1+p_3-\ell_a+\ell_c)^2\big)
    \end{aligned}$ & $\left\{1\right\}$ \\\hline
      $I_4$ & $\begin{aligned}
        \includegraphics[width=0.23\linewidth]{figure/phi2NplGamma10to4.eps}
    \end{aligned}$ & $\begin{aligned}
   s_{23} s_{13}(\ell_a-p_1)^2\big(&(\ell_a-p_1)^2\\
    &-(p_1+p_3-\ell_a+\ell_c)^2\big)
    \end{aligned}$ & $\left\{\frac{s_{23}}{q^2},\frac{s_{23}}{s_{13}+s_{23}},\frac{s_{23}}{s_{12}+s_{23}}\right\}$ \\\hline
    $I_5$ & $\begin{aligned}
        \includegraphics[width=0.23\linewidth]{figure/phi2NplGamma10to4.eps}
    \end{aligned}$ & $
    s_{12}s_{23}s_{13}(\ell_a-p_1)^2$ & {\small 
   $ \begin{array}{l}
    \Big\{1,\frac{s_{23}}{q^2},\frac{s_{12}}{q^2},  \frac{s_{12}}{s_{12}+s_{23}}\Big\}
    \end{array}$}
     \\\hline
      $I_{6}$ & $\begin{aligned}
        \includegraphics[width=0.23\linewidth]{figure/phi2NplGamma10to4.eps}
    \end{aligned}$ & 
    $\begin{aligned}
    s_{23}s_{12}&(\ell_a-\ell_c)^2\big(s_{23}\\
    &-(\ell_a-p_1-p_2-p_3)^2\big)
    \end{aligned}$
    & {\small 
   $\begin{array}{l}
    \Big\{1,\frac{s_{12}}{s_{13}},\frac{s_{12}}{q^2}\Big\}
    \end{array}$}
     \\\hline
     $I_7$ & $\begin{aligned}
        \includegraphics[width=0.23\linewidth]{figure/phi2NplGamma10to5.eps}
    \end{aligned}$ & $\begin{aligned}
    s_{12}&(s_{13}+s_{23})(\ell_a-p_1)^2\big((\ell_a-p_1)^2\\
    &-(\ell_b-\ell_c)^2-(\ell_a+\ell_c-\ell_b-p_1)^2\big)
    \end{aligned}$ & $\left\{1,\frac{s_{13}+s_{23}}{s_{13}}\right\}$ \\\hline
    $I_8$ & $\begin{aligned}
        \includegraphics[width=0.23\linewidth]{figure/phi2NplGamma10to5.eps}
    \end{aligned}$ & $\begin{aligned}
    s_{12}^2 s_{23}\big(&(\ell_a-p_1)^2-(\ell_b-\ell_c)^2\\
    &-(\ell_a+\ell_c-\ell_b-p_1)^2\big)
    \end{aligned}$ & $\left\{1,\frac{s_{23}}{q^2},\frac{s_{23}}{s_{13}}\right\}$ \\\hline
    $I_9$ & $\begin{aligned}
        \includegraphics[width=0.23\linewidth]{figure/phi2NplGamma10to7.eps}
    \end{aligned}$ & $\begin{aligned}
    s_{12}s_{13}\big(&(\ell_a-p_1)^2\big)^2
    \end{aligned}$ & $\left\{1\right\}$ \\\hline
    $I_{10}$ & $\begin{aligned}
        \includegraphics[width=0.23\linewidth]{figure/phi2NplGamma10to7.eps}
    \end{aligned}$ & $\begin{aligned}
    s_{12}s_{23}(\ell_a-p_1)^2\big(&(\ell_a-p_1)^2+s_{12}\\
    &-\ell_a^2-(\ell_a-p_2-p_1)^2\big)
    \end{aligned}$ & $\left\{1\right\}$ \\\hline
     $I_{11}$ & $\begin{aligned}
        \includegraphics[width=0.23\linewidth]{figure/phi2NplGamma10to6.eps}
    \end{aligned}$ & $\begin{aligned}
    s_{12}^2s_{23}\big(&(\ell_a-p_1)^2-(\ell_a-\ell_b)^2\\
    &-(\ell_b-p_1)^2-(\ell_a-p_1-p_2)^2\\
    &+(\ell_b-p_1-p_2)^2\big)
    \end{aligned}$ & $\left\{1\right\}$ \\\hline
    $I_{12}$ & $\begin{aligned}
        \includegraphics[width=0.23\linewidth]{figure/phi2NplGamma10to6.eps}
    \end{aligned}$ & $\begin{aligned}
    s_{12}&(s_{13}+s_{23})(\ell_a-p_1)^2\\
    &\big((\ell_a-p_1)^2-(\ell_a-p_2-p_1)^2\\
    &\ \ -(\ell_a-\ell_b)^2-(\ell_b-p_1)^2\big)
    \end{aligned}$ & $\left\{1,\frac{s_{23}+s_{13}}{s_{13}}\right\}$ \\\hline
     $I_{13}$ & $\begin{aligned}
        \includegraphics[width=0.23\linewidth]{figure/phi2NplGamma10to6.eps}
    \end{aligned}$ & $\begin{aligned}
    s_{12}s_{23}(\ell_b-p_1-p_2)^2\big(&s_{12}-\ell_a^2\big)
    \end{aligned}$ & $\left\{1,\frac{s_{23}}{s_{13}},\frac{s_{23}}{q^2}\right\}$ \\\hline
\end{longtable}

Next we give explicitly some numerical integral results, and each of them consumes $O(10^5)$ CPU core hours using FIESTA 4.2.
\begin{align}
I_{2}=&+\frac{0.1373265\pm 2\times 10^{-7}}{\epsilon ^5}+\frac{0.527396\pm 2\times 10^{-6}}{\epsilon^4} \\
&-\frac{4.90339\pm 2\times 10^{-5}}{\epsilon^3}-\frac{59.4854\pm 0.0002}{\epsilon ^2}-\frac{470.454\pm 0.001}{\epsilon }-(3185.00\pm 0.01) \,,\nonumber\\
I_{4}=&+\frac{0.1018519}{\epsilon ^6}-\frac{0.3643800\pm 4\times 10^{-7}}{\epsilon^5}-\frac{2.380663\pm 4\times 10^{-6}}{\epsilon^4}\\
&-\frac{3.40443\pm 3\times 10^{-5}}{\epsilon^3}+\frac{33.4031\pm 0.0003}{\epsilon^2}+\frac{295.227\pm 0.003}{\epsilon }+(1827.06\pm 0.03) \,,\nonumber\\
I_{5}=&+\frac{0.2444444}{\epsilon ^6}+\frac{0.102032\pm 1\times 10^{-6}}{\epsilon ^5}-\frac{0.05261\pm 2\times 10^{-5}}{\epsilon^4}\\
&-\frac{0.254901\pm 0.00013}{\epsilon ^3}-\frac{8.65496\pm0.0013}{\epsilon ^2}-\frac{166.366\pm 0.013}{\epsilon}-(1738.61\pm 0.13) \,,\nonumber\\
I_{6}=&-\frac{0.3953704}{\epsilon ^6}-\frac{0.001810\pm 1\times 10^{-6}}{\epsilon ^5}+\frac{5.81108\pm 1\times 10^{-5}}{\epsilon^4}\\
&+\frac{29.1661\pm 0.0001}{\epsilon ^3}+\frac{145.274\pm 0.001}{\epsilon ^2}+\frac{908.068\pm 0.014}{\epsilon}+(6034.\pm 0.16) \,,\nonumber\\
I_{9}=&+\frac{0.0972222}{\epsilon ^6}+\frac{0.255587\pm 3\times 10^{-7}}{\epsilon^5}+\frac{0.176656\pm 4\times 10^{-6}}{\epsilon ^4}\\
&+\frac{2.78628\pm 4\times 10^{-5}}{\epsilon ^3}+\frac{51.3948\pm 0.0003}{\epsilon^2}+\frac{424.088\pm 0.003}{\epsilon}+(2740.39\pm 0.03) \,,\nonumber\\
I_{11}=&-\frac{0.0805556}{\epsilon ^6}-\frac{0.1803833\pm 5\times 10^{-7}}{\epsilon \
^5}-\frac{2.913784\pm 8\times 10^{-6}}{\epsilon^4}\\
&-\frac{15.5231\pm 7\times 10^{-5}}{\epsilon^3}-\frac{66.6142\pm 0.0007}{\epsilon ^2}-\frac{263.088\pm 0.008}{\epsilon }-(1140.57\pm 0.08) \,,\nonumber\\
I_{12}=&+\frac{0.0722222}{\epsilon ^6}+\frac{0.0329202\pm 6\times 10^{-7}}{\epsilon ^5}-\frac{3.164532\pm 8\times 10^{-6}}{\epsilon ^4}\\
&-\frac{9.9251\pm 0.0001}{\epsilon^3}-\frac{60.6163\pm 0.0011}{\epsilon^2}-\frac{506.186\pm 0.012}{\epsilon}-(4151.88\pm 0.13) \,,\nonumber\\
I_{13}=&+\frac{0.1555556}{\epsilon ^6}-\frac{0.152574\pm 7\times 10^{-7}}{\epsilon^5}-\frac{3.83705\pm 8\times 10^{-6}}{\epsilon ^4}\\
&-\frac{22.6628\pm 9\times 10^{-5}}{\epsilon^3}-\frac{141.16\pm 0.001}{\epsilon^2}-\frac{972.808\pm 0.012}{\epsilon }-(6694.44\pm 0.12) \,.\nonumber
\end{align}


\section{IR conventions and non-dipole terms}\label{ap:ir}
 
In this appendix we provide some details about IR divergences discussed in Section~\ref{ssec:ir}. 

We first setup the convention of anomalous dimensions. Following the discussion in Section~\ref{ssec:ir}, we use the cusp and collinear anomalous dimensions in the following expansion:
\begin{align}
& \sum_{l=1}^\infty \gamma_{\rm cusp}^{(l)} g^{2l} =  g^2 N_{\rm c} - 2 \zeta_2 g^4 N_{\rm c}^2 + 22 \zeta_4 g^6 N_{\rm c}^3 + {\cal O}(g^8)  \,, 
\label{eq:defCuAD}\\
& \sum_{l=1}^\infty \mathcal{G}_{\rm coll}^{(l)} g^{2l} = -\zeta_3 g^4 N_{\rm c}^2 + 8 \Big( \zeta_5 + {5\over 6} \zeta_2 \zeta_3 \Big) g^6 N_{\rm c}^3 + {\cal O}(g^8) \,.\label{eq:defCoAD}
\end{align}
Note that up to three-loop, only $N_{\rm c}$-leading powers appear. At four-loops, there is also $N_c$-subleading contribution to these anomalous dimensions, see \emph{e.g.}~\cite{Boels:2017skl,Moch:2018wjh,Henn:2019swt,Huber:2019fxe,Agarwal:2021zft,Lee:2021lkc}.
We can compare the above convention with the Sudakov form factor result in \cite{Boels:2017ftb}: 
\begin{equation}\label{eq:subdakovlog}
\begin{aligned}
\log F_{\rm sudakov} & = - \sum_{{l}} g^{2l}(-s_{12}^2)^{-l \epsilon}  \bigg[ \frac{\tilde\gamma_{\textrm{cusp}}^{({l})} }{(2 {l} \epsilon)^2} + \frac{\tilde{\cal G}_{\textrm{coll}}^{({l})} }{2 {l} \epsilon} + {\rm Fin}^{(l)} \bigg] + {\mathcal O}\left(\epsilon\right) + {\cal O}(g^8) \, , \\
\tilde\gamma_{\rm cusp}  &= \sum_{l=1}^\infty \tilde\gamma_{\rm cusp}^{(l)} g^{2l} =  8g^2 - 16 \zeta_2 g^4 + 176 \zeta_4 g^6  \,, \\
\tilde{\cal G}_{\rm coll} &= \sum_{l=1}^\infty \tilde{\cal G}_{\rm coll}^{(l)} g^{2l} = - 4\zeta_3 g^4 + 32\Big(  \zeta_5 + {5\over 6} \zeta_2 \zeta_3 \Big) g^6 + {\cal O}(g^8) \,,
\end{aligned}
\end{equation}
which can be checked to equal to the IR structure used in this paper, see also \eqref{eq:planarlog}:
\begin{equation}\label{eq:plir}
\log {\mathcal{I}}=-\sum_{\ell=1}^{\infty} g^{2 \ell}\left[\frac{\gamma_{\mathrm{cusp}}^{(\ell)}}{( \ell \epsilon)^{2}}+\frac{\mathcal{G}_{\mathrm{coll}}^{(\ell)}}{ \ell \epsilon}\right] \sum_{i=1}^{n}\left(-s_{i i+1}\right)^{-\ell \epsilon}+{O}\left(\epsilon^{0}\right)\,,
\end{equation}
by noting that: $\sum_{i=1}^{n}$ for $n=2$ provides a factor 2 in \eqref{eq:plir}, and there is an extra 2 before the $l\epsilon$ in the denominators in \eqref{eq:subdakovlog}.

Next, we provide some details on the calculation from \eqref{eq:zfactorwilson} to \eqref{eq:zfactor}. 
Let us reproduce  \eqref{eq:zfactorwilson} explicitly here:
\begin{equation}\label{eq:zfactorwilsonApp}
    {\itbf{Z}}(p_i,\epsilon)=\mathcal{P} \exp\left\{-\frac{1}{2}\int_{0}^{\mu^2}\frac{\mathrm{d}\lambda^2}{\lambda^2}\boldsymbol{\Gamma} \bigg(p_i,\lambda, \bar\alpha_s\Big({\lambda^2 \over \mu^2}, \alpha_s(\mu^2), \epsilon \Big) \bigg)\right\}\,,
\end{equation}
where $\bar\alpha_s$ is the $(4-2\epsilon)$-dimensional running coupling; since for $\mathcal{N}=4$ SYM, $\alpha_{\rm s}(\mu^2)={g_{\rm YM}^2 \over 4\pi}$ is a constant,  $\bar\alpha_s$ has trivial scale dependence as 
\begin{equation}
    \frac{\bar\alpha_s\Big({\lambda^2 \over \mu^2}, \alpha_s(\mu^2), \epsilon \Big)}{4\pi}=\left(\frac{\mu^2}{\lambda^2}\right)^{\epsilon}\frac{(4\pi e^{-\gamma_{E}})^{\epsilon}g_{\rm YM}^2}{(4\pi)^2}= \left(\frac{\mu^2}{\lambda^2}\right)^{\epsilon} g^2\,.
\end{equation}
One can separate $\boldsymbol{\Gamma}$ as \cite{Almelid:2015jia}
\begin{equation}
\label{eq:softmatrixdecomp}
    \boldsymbol{\Gamma}(p_i,\lambda,\bar\alpha_{\rm s}(\lambda^2))=\boldsymbol{\Gamma}^{\rm dip}(p_i,\lambda, \bar\alpha_{\rm s}(\lambda^2))+\text{non-dipole terms}\,,
\end{equation}
where the dipole terms, denoted as $\boldsymbol{\Gamma}^{\rm dip} $, contribute to the planar IR divergences, and non-dipole terms are the remaining parts. 
The $\boldsymbol{\Gamma}^{\rm dip}$ in \eqref{eq:softmatrixdecomp} is
\begin{equation}
    \boldsymbol{\Gamma}^{\rm dip}(p_i,\lambda, \bar\alpha_{\rm s}(\lambda^2))=-\frac{1}{2}\hat{\gamma}_{\rm cusp}(\bar\alpha_{\rm s})\sum_{i<j}\log\left(-\frac{s_{ij}}{\lambda^2}\right)\mathbf{T}_i\cdot \mathbf{T}_j + \sum_{i=1}^{n}\hat{\mathcal{G}}_{\mathrm{coll},i}(\bar\alpha_{\rm s})\,,
\end{equation}
where the action of operator $\mathbf{T}_i$ on the color space goes as $\mathbf{T}_i^{a}T^{a_i}=-i {f}^{a a_i \mathrm{x}}T^{\rm x}$. 
The anomalous dimensions can be perturbatively expanded as follows:\footnote{The normalization factor $1/4$ and $-1/2$ are defined to be consistent with our convention in \eqref{eq:planarlog}.}
\begin{equation}
    \hat{\gamma}_{\rm cusp}=\frac{1}{4}\sum_{\ell=1}^{\infty} \left(\frac{\bar\alpha_{\rm s}}{4\pi}\right)^{\ell} \gamma_{\rm cusp}^{(\ell)}, \qquad \hat{\mathcal{G}}_{\rm coll}=-\frac{1}{2}\sum_{\ell=1}^{\infty} \left(\frac{\bar\alpha_{\rm s}}{4\pi}\right)^{\ell} \mathcal{G}_{\rm coll}^{(\ell)}\,.
\end{equation}
Then we can perform the integration in \eqref{eq:zfactorwilsonApp}, where one meets following typical integrals:
\begin{equation}
    \int_{0}^{\mu^2}\frac{\mathrm{d}\lambda^2}{\lambda^2}\left(\frac{\mu^2}{\lambda^2}\right)^{\ell \epsilon}=-\frac{1}{\ell \epsilon}, \qquad 
    \int_{0}^{\mu^2}\frac{\mathrm{d}\lambda^2}{\lambda^2}\left(\frac{\mu^2}{\lambda^2}\right)^{\ell \epsilon}\log\left(-\frac{s_{ij}}{\lambda^2}\right)=\frac{1}{(\ell \epsilon)^2} - \frac{1}{\ell \epsilon} \log\left(-\frac{s_{ij}}{\mu^2}\right)\,.
\end{equation}
In this way, we finally get \eqref{eq:zfactor}:
\begin{equation}\label{eq:zfactor-ap}
    \itbf{Z}(p_i,\epsilon)=\mathcal{P} \exp\left\{\sum_{\ell=1}^{\infty}g^{2\ell}\left[\frac{\gamma_{\rm cusp}^{(\ell)}}{(\ell\epsilon)^2}\mathbf{D}_0-\frac{\gamma_{\rm cusp}^{(\ell)}}{\ell \epsilon}\mathbf{D}-n\frac{\mathcal{G}_{\rm coll}^{(\ell)}}{\ell \epsilon}\mathbf{1}+\frac{1}{\ell \epsilon} \mathbf{\Delta}^{(\ell)}\right]\right\}\,,
\end{equation}
with $\mathbf{D}_0,\mathbf{D}, \mathbf{\Delta}_{3}^{(3)}$ given in \eqref{eq:D0Ddef}--\eqref{eq:3loopdelta}.

Let us give some more explanations on factor $\itbf{Z}(p_i,\epsilon)$ of this form. 
\begin{itemize}
\item 
Firstly, we can compare our convention with the one given in \cite{Henn:2016jdu}. 
Taking into account the difference of the coupling convention $\alpha = 4 g^2$, this is the same as the anomalous dimensions given in (9)--(10) in \cite{Henn:2016jdu}. 

\item
Second, the action of $\mathbf{D}_0$ and $\mathbf{D}$ usually gives a minus sign, explaining the sign difference  between cusp anomalous dimensions $\gamma_{\rm cusp}$ in planar \eqref{eq:plir} and full-color  \eqref{eq:zfactor-ap} IR divergences.

\item
Finally, confining \eqref{eq:zfactor-ap} to three-point form factors, the dipole terms contribute only to the $N_{\rm c}$-leading part which has been given in Section~\ref{ssec:ir}. 
The non-dipole term contributes to the $N_{\rm c}$-subleading part; when $n=3$, one has $\mathbf{\Delta}_{3}^{(3)}$ given as
\begin{align}
\mathbf{\Delta}_{3}^{(3)} & = - 8(\zeta_5+2\zeta_2\zeta_3 ) \sum_{i=1}^3\sum_{\substack{ j<k, j,k\neq i}}  {f}^{abe} {f}^{cde} (\mathbf{T}^{a}_i\mathbf{T}^{d}_i+\mathbf{T}^{d}_i\mathbf{T}^{a}_i)\mathbf{T}^{b}_{j}\mathbf{T}^{c}_{k}\,.
\end{align}
A direct calculation shows:
\begin{align}
& \Big[ {f}^{abe} {f}^{cde} (\mathbf{T}^{a}_1\mathbf{T}^{d}_1+\mathbf{T}^{d}_1\mathbf{T}^{a}_1)\mathbf{T}^{b}_{2}\mathbf{T}^{c}_{3} \Big] {\tilde f}^{a_1 a_2 a_3}
= 3 {N_c} {\tilde f}^{a_1 a_2 a_3} \\
& \Rightarrow  \ \frac{1}{3\epsilon}\mathbf{\Delta}_{3}^{(3)} \tilde{f}^{a_1 a_2 a_3}  = -\frac{24}{\epsilon} N_c (\zeta_5+2\zeta_2\zeta_3) \tilde{f}^{a_1 a_2 a_3} = 12 N_c {-2 (\zeta_5+2\zeta_2\zeta_3) \over \epsilon} \tilde{f}^{a_1 a_2 a_3}\,. \nonumber
\end{align}
Comparing with our expansion \eqref{eq:simpInt2} of the form factor 
\begin{align}  
\itbf{F}_{\operatorname{tr}(\phi^2), 3}^{(3)} = {\cal F}_{\operatorname{tr}(\phi^2), 3}^{(0)} {\tilde f}^{a_1 a_2 a_3} \Big( N_c^3 {\mathcal{I}}^{(3),{\rm PL}}_{\operatorname{tr}(\phi^2)} + 12 N_c {\mathcal{I}}_{\operatorname{tr}(\phi^2)}^{(3), {\rm NP}} \Big) \,,
\end{align}
the non-planar divergent part of $\mathcal{I}_{\operatorname{tr}(\phi^2)}^{(3), {\rm NP}}$ is $-2(\zeta_5+2\zeta_3\zeta_2)/\epsilon$. 

\end{itemize}

\providecommand{\href}[2]{#2}\begingroup\raggedright\endgroup

\end{document}